\documentclass[10pt,journal,twoside]{IEEEtran}
\usepackage{stfloats}
\usepackage{amsmath}
\usepackage{graphicx} 
\usepackage{amssymb}
\usepackage{amsfonts}
\usepackage{amsthm}
\usepackage{cite}
\usepackage{booktabs}
\usepackage{bm,comment}
\usepackage{algorithm}
\usepackage{algpseudocode}

\usepackage{subeqnarray}
\usepackage{subfigure}
\usepackage{diagbox}
\usepackage{slashbox}
\usepackage{stfloats}
\usepackage{float}
\usepackage{color} 
\usepackage{cases}
\usepackage{setspace}
\usepackage{threeparttable} 
\usepackage{longtable}
\usepackage{type1cm}
\usepackage{moresize}
\usepackage{anyfontsize}
\usepackage{t1enc}
\usepackage{multirow}
\usepackage{array}
\usepackage{makecell}

\usepackage[table,xcdraw]{xcolor}
\newcommand{\beizhu}[1]{}

\newcommand{\tabincell}[2]{\begin{tabular}{@{}#1@{}}#2\end{tabular}}

\newtheorem{remark}{Remark}

\newtheorem{lemma}{Lemma}

\usepackage{geometry}
\floatname{algorithm}{Algorithm}

\graphicspath{{surfigs/}}

\begin{document}
 \title{Beamforming Technologies for Ultra-Massive MIMO in Terahertz Communications}
 \author{Boyu Ning, \IEEEmembership{Member, IEEE}, Zhongbao Tian, \IEEEmembership{Student Member, IEEE},\\ Weidong Mei, \IEEEmembership{Member, IEEE},  Zhi Chen, \IEEEmembership{Senior Member, IEEE},  Chong Han, \IEEEmembership{Member, IEEE},  \\ Shaoqian Li, \IEEEmembership{Fellow, IEEE}, Jinhong Yuan, \IEEEmembership{Fellow, IEEE} and Rui Zhang, \IEEEmembership{Fellow, IEEE} \vspace{-6pt}
 \thanks{This work was supported in part by the National Key R$\&$D Program of China under Grant 2018YFB1801500. \emph{(Boyu Ning and Zhongbao Tian are co-first authors.) (Corresponding author: Zhi Chen.)}}
 \thanks{Boyu Ning, Zhongbao Tian, Weidong Mei,  Zhi Chen, and Shaoqian Li are with the National Key Laboratory of Science and Technology on Communications, University of Electronic Science and Technology of China, Chengdu 611731, China (e-mails: boydning@outlook.com; vincent11231@outlook.com; mwduestc@gmail.com;
 chenzhi@uestc.edu.cn; lsq@uestc.edu.cn).}
 \thanks{Chong Han is with the Terahertz Wireless Communications (TWC) Laboratory,
Shanghai Jiao Tong University, Shanghai 200240, China (e-mail:
chong.han@sjtu.edu.cn).}
 \thanks{Jinhong Yuan is with the University of New South Wales, Australia (e-mail:
j.yuan@unsw.edu.au).}
\thanks{ R. Zhang is with the Chinese University of Hong Kong, Shenzhen, and Shenzhen Research Institute of Big Data, Shenzhen, China 518172 (e-mail: rzhang@cuhk.edu.cn). He is also with the Department of Electrical and Computer Engineering, National University of Singapore, Singapore 117583 (e-mail: elezhang@nus.edu.sg). }

 }

\maketitle
\begin{abstract}
Terahertz (THz) communications with a frequency band $0.1-10$ THz are envisioned as a promising solution to future high-speed wireless communication. Although with tens of gigahertz available bandwidth, THz signals suffer from severe free-spreading loss and molecular-absorption loss, which limit the wireless transmission distance. To compensate for the propagation loss, the ultra-massive multiple-input-multiple-output (UM-MIMO) can be applied to generate a high-gain directional beam by beamforming technologies. In this paper, a review of beamforming technologies for THz UM-MIMO systems is provided. Specifically, we first present the system model of THz UM-MIMO and identify its channel parameters and architecture types. Then, we illustrate the basic principles of beamforming via UM-MIMO and discuss the far-field and near-field assumptions in THz UM-MIMO. Moreover, an important beamforming strategy in THz band, i.e., beam training, is introduced wherein the beam training protocol and codebook design approaches are summarized. The intelligent-reflecting-surface (IRS)-assisted joint beamforming and multi-user beamforming in THz UM-MIMO systems are studied, respectively. The spatial-wideband effect and frequency-wideband effect in the THz beamforming are analyzed and the corresponding solutions are provided.  Further, we present the corresponding fabrication techniques and illuminate the emerging applications benefiting from THz beamforming. Open challenges and future research directions on THz UM-MIMO systems are finally highlighted.
\end{abstract}

\section{Introduction}
Since the beginning of the 21st century, the evolution of online applications on social networks has led to an unprecedented growing number of wireless subscribers, who require real-time connectivity and tremendous data consumption \cite{5g,5g2}. In this backdrop, various organizations and institutions have published their standards to support the wireless traffic explosion, such as the 3rd generation partnership project (3GPP) long term evolution (LTE) \cite{lte}, the wireless personal area network (WPAN) \cite{wpan}, the wireless high definition (WiHD) \cite{wihd}, IEEE 802.15 \cite{8015}, and IEEE 802.11 wireless local area network (WLAN) \cite{wlan}. During the evolution of these standards, one common feature is that the consumed frequency spectrum increased steadily to satisfy the dramatic demands for instantaneous information. According to Edholm's Law of Bandwidth \cite{edh}, the telecommunications data rate doubles every 18 months. The wireless data rate requirement is expected to meet $100$ Gbps to fulfill different growing service requirements before 2030. 

The emerging millimeter-wave (mmWave)-band communications in the fifth-generation (5G) standard are able to achieve considerable improvements in the network capacity, with the vision of meeting demands beyond the capacity of previous-generation systems. Although some advanced means like the multiple-antenna technologies \cite{mimo}, the coordinated multi-point (CoMP) technologies \cite{comp}, and the carrier aggregation (CA) technologies \cite{caf} may boost the data rate reaching several Gbps in mmWave communications, the achievable data rate on each link will be significantly suppressed when the number of connected users and/or devices is growing. To further raise the throughput, enhance the spectral efficiencies, and increase the connections, academic and industrial research proposed to deploy a large number of low-power small cells giving rise to the heterogeneous networks (HetNets) \cite{het,het2}. Meanwhile, novel multiple access schemes, such as orthogonal time-frequency space (OTFS) \cite{otfs1,otfs2}, rate-splitting \cite{rates1,rates2}, and non-orthogonal multiple access (NOMA) \cite{noma1,noma2} technologies have been investigated for meeting the above requirements by sharing the resource block, e.g., a time slot, a frequency channel, a spreading code, or a spatial beam. However, due to the frequency regulations, current mmWave communications have limited available bandwidth and at present, there is even no frequency block wider than $10$ GHz left unoccupied in the bands below $100$ GHz.

\begin{figure}[t]
\centering
\includegraphics[width=3.5in]{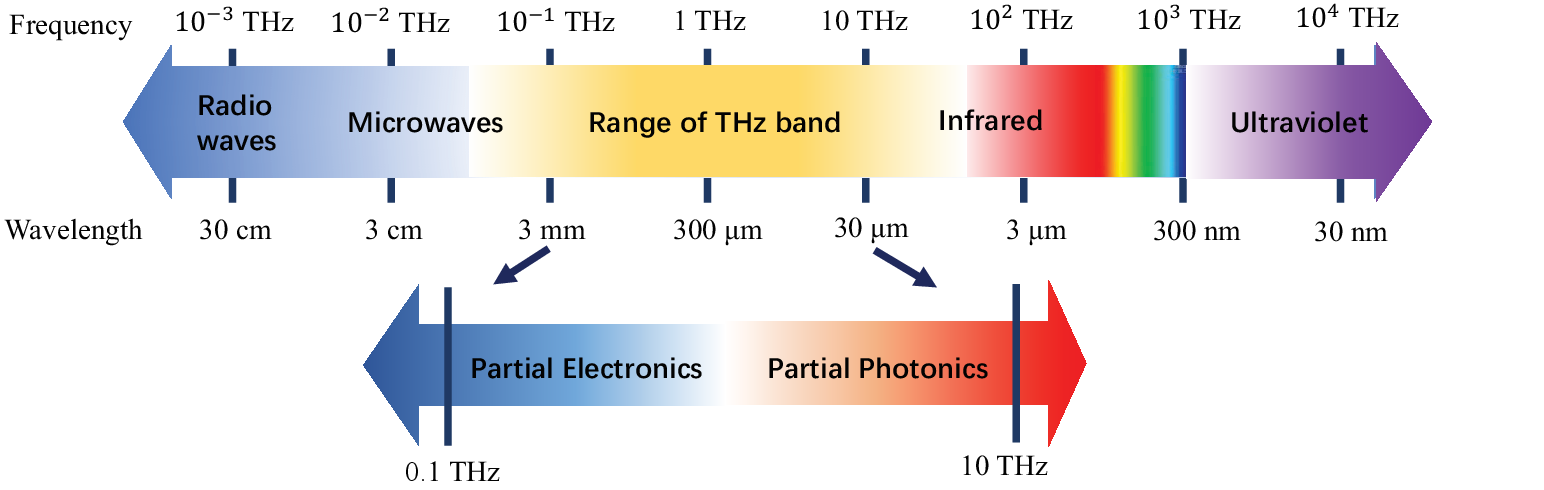}
\caption{The range of THz band in the electromagnetic spectrum.}\label{tba}\vspace{-10pt}
\end{figure}

To alleviate the spectrum bottleneck and realize at least 100 Gbps communications in the future, new spectral bands should be explored to support the data-hungry applications, e.g., virtual reality (VR) and augmented reality (AR), which require microsecond latency and ultra-fast download. To this end, the terahertz (THz) band ($0.1$-$10$ THz) has received noticeable attention in the research community as a promising candidate for various scenarios with high-speed transmission. As shown in Fig. \ref{tba}, the bands below and above THz band have been extensively explored, including radio waves, microwave/mmWave, and free-space optical (FSO). In terms of signal generation, the THz band is exactly between the frequency regions generated by oscillator-based electronic and emitter-based photonic approaches, which incurs difficulty of electromagnetic generation, known as ``the last piece of radio frequency (RF) spectrum puzzle for communication systems''\cite{sur320}. In terms of wireless transmission, the most urgent challenges lie in the physical-layer hindrance, i.e.,  high spreading loss and severe molecular absorption loss in THz electromagnetic propagation.

Multiple-input-multiple-output (MIMO) systems, which can generate high-gain directional beams via beamforming technologies, are considered a pragmatic solution to compensate for the propagation loss.  For metallic-antenna arrays, the element spacing is commonly half wavelength. With the increase of frequency adopted in wireless systems, more elements can be packed in the same size of the antenna array. Thus, compared to the MIMO in microwave systems, the concept of \emph{massive MIMO} with tens to hundreds of elements in mmWave systems has been raised in\cite{mmimo0}. In 2016, Ian F. Akyildiza and Josep Miquel Jornet introduced the concept of \emph{ultra-massive MIMO} (UM-MIMO) in THz systems\cite{gramimo2}, which showed that the number of elements of the metamaterial-based or graphene-based nano-antenna THz array can be much larger than that of the mmWave massive MIMO. This is because the antenna spacing of nano-antenna arrays can be much smaller than half wavelength, meanwhile, the wavelength of the THz band is smaller than its mmWave counterpart. 

In this paper, we reuse the term ``UM-MIMO'' to describe \emph{the THz antenna array with hundreds to thousands of elements}, which is not specific to nano-antenna array (as in \cite{gramimo2}) but compatible with conventional half-wavelength metallic-antenna array.
We start with the fundamental concepts related to the THz UM-MIMO and precisely illuminate the principle of beamforming via affluent graphs, which are significant and instructive for communication researchers or engineers. Moreover, this paper covers some on-trend research points and classifies the existing THz MIMO arrays, which sheds light on the current research status and the progress of THz UM-MIMO. Finally, the emerging applications and open challenges are elaborated. This paper serves as a review to provide useful technical guidance and inspiration for future research and implementation of beamforming in THz UM-MIMO systems.

\subsection{Related Tutorials, Magazines, and Surveys}
Despite that the THz technology is not as mature as microwave/mmWave and FSO, the gap is progressively being closed thanks to its rapid development in recent years. Table I summarizes the representative works in this field including tutorials, magazines, and surveys published in the last two decades. 

The first survey was conducted by P. H. Siegel, in which existing THz applications, sensors, and sources are presented \cite{sur1}. In 2004, M. J. Fitch and R. Osiander further discussed the sources, detectors, and modulators for practical THz systems \cite{sur2}. In 2007, several magazines came up with reports of THz technology progress status and applications of THz systems \cite{sur3,sur4,sur5}. In 2010, J. Federici and L. Moeller provided the first overview focusing on the THz communications including channel coding, generation methods, detection, antennas, and link measurements \cite{sur6}. After that, K. C. Huang and Z. Wang provided a tutorial on constructing robust, low-cost THz wireless systems, in 2011 \cite{sur7}. In the same year, T. Nagatsuma \emph{et al.} discussed the current progress of THz communications applications and highlighted some issues that need to be considered for the future of THz systems \cite{sur8,sur9,sur10}. In 2012, K. Wu \emph{et al.} provided a tutorial on THz antenna technologies, and T. Kurner \emph{et al.} reported the standardization at IEEE 802.15 IG THz \cite{sur11,sur12}. In 2014, T. Kurner and S. Priebe reported the current research projects, spectrum regulations, and ongoing standardization activities in THz communication systems \cite{sur15}. I. F. Akyildiz \emph{et al.} reported the state-of-the-art THz technologies and highlighted the challenges from the communication and networking perspective as well as in terms of experimental testbeds \cite{sur16}. In 2015, A. Hirata and M. Yaita gave a brief overview of the THz technologies and standardization of wireless communications \cite{sur17}. In 2016, C. Lin and G. Y. Li reported an array-of-subarrays structure for THz wireless systems and discussed the benefits in terms of circuit and communication \cite{sur18}. M. Hasan \emph{et al.} provided an overview of the progress on graphene-based devices and Nagatsuma, T. \emph{et al.} gave a tutorial on the photonics technologies in THz communications \cite{sur19,sur20}. J. F. Federici \emph{et al.} gave a survey to illustrate the impact of weather on THz wireless links \cite{sur21}. 
In 2017, S. Mumtaz \emph{et al.} discussed the opportunities and challenges in THz communications for vehicular networks \cite{sur22}. In 2018, V. Petrov \emph{et al.} provided a tutorial on propagation modeling, antenna, and testbed designs \cite{sur23}. A. Boulogeorgos \emph{et al.} reported the basic system architecture for THz wireless links with bandwidths into optical networks \cite{sur24}. C. Han and Y. Chen reported three methods, e.g., deterministic, statistical, and hybrid methods, to model THz propagation channels \cite{sur25}. I. F. Akyildiz \emph{et al.} focused on the solution to the THz distance limitation \cite{sur251}. N. Khalid \emph{et al.} provided a tutorial on performing THz modulation schemes \cite{sur26}. D. Headland\cite{sur261} provided a tutorial on the basic principles of beam control in the THz band.

Since 2019, the number of papers on THz communications has increased notably. There were two magazines and five surveys that contain the development progress, unresolved problems, latest solutions, standardization works, and opportunities 
\onecolumn
\begingroup
\renewcommand{\arraystretch}{1.5} 

 {\fontsize{9}{9}\selectfont 
\begin{center} 
\begin{longtable}[]{|m{0.11\textwidth}<{\centering}|m{0.03\textwidth}<{\centering}|m{0.24\textwidth}|m{0.06\textwidth}<{\centering}|m{0.45\textwidth}|}
\caption{Tutorials, magazines, and surveys on THz communications} \label{table} 
\\ \hline
Authors   & Year & Title & Type  & Brief Description 
\\ \hline
\endfirsthead
    \hline
    Authors   & Year & Title & Type  & Brief Description \\
    \hline
    \endhead 
    \hline
    \endfoot 

P. H. Siegel \cite{sur1}& 2002 & Terahertz technology & Survey   & This paper gives an overview of THz technology applications, sensors, and sources, with some discussion on science drivers, historical background, and future trends.
\\ \hline
M. J. Fitch and R. Osiander \cite{sur2}& 2004 & Terahertz waves for communications and sensing & Magazine  & This article reports the THz technology for communications and sensing applications. Sources, detectors, and modulators are also discussed for practical systems.
\\ \hline
R. Piesiewicz \emph{et al.} \cite{sur3} & 2007 & Short-range ultra-broadband terahertz communications: concepts and perspectives & Magazine  & This article reports the concept of ultra-broadband THz communication and gives the potential applications of such a system supporting multi-gigabit data rates.
\\ \hline
I. Hosako \emph{et al.} \cite{sur4} & 2007 & At the dawn of a new era in terahertz technology & Magazine  & This article reports the developments in these fields such as THz quantum cascade lasers, THz quantum well photodetectors, an ultra-wideband THz time domain spectroscopy system, an example of a database for materials of fine art, and results from measuring atmospheric propagation.
\\ \hline
M. Tonouchi \cite{sur5}& 2007 & Cutting-edge terahertz technology & Magazine  & This article reports the THz technology progress status and expected usages in wireless communication,   agriculture, and medical applications.
\\ \hline
J. Federici and L. Moeller \cite{sur6}& 2010 & Review of terahertz and subterahertz wireless communications & Survey   & This paper gives an overview of THz communication systems, which demonstrate basic channel coding, generation methods, detection, antennas, and link measurements. 
\\ \hline
K. c. Huang and Z. Wang  \cite{sur7} & 2011 & Terahertz terabit wireless communication & Tutorial & This paper provides a tutorial to construct robust, low-cost wireless systems for THz terabit communications.  
\\ \hline
T. Kleine-Ostmann and T. Nagatsuma \cite{sur8} & 2011 & A review on terahertz communications research & Magazine  & This article reports the emerging technologies and system researches that might lead to ubiquitous THz communication systems in the future.
\\ \hline
T. Nagatsuma \cite{sur9} & 2011 & Terahertz technologies: present and future & Magazine  & This article reports the latest progress in THz technologies in sources, detectors, and system applications. Future challenges toward market development are also discussed.
\\ \hline
H. J. Song and T. Nagatsuma  \cite{sur10}      & 2011 & Present and future of terahertz communications & Survey & This paper gives an overview of the current progress of THz communications applications and discusses some issues that need to be considered for the future of THz systems. 
\\ \hline
K. Wu \emph{et al.} \cite{sur11} & 2012 & Substrate-integrated millimeter-wave and terahertz antenna technology & Tutorial & This paper provides a tutorial on mmWave and THz antenna technologies including the planar/nonplanar antenna structures and provides a promising technological platform for mmWave and THz wireless systems.
\\ \hline
T. Kurner \emph{et al.} \cite{sur12} & 2012 & Towards future terahertz communications systems  & Magazine  & This article reports the technology development, demonstrations of data transmission, ongoing activities in standardization at IEEE 802.15 IG THz, and the regulation of the spectrum beyond $300$ GHz.
\\ \hline
T. Nagatsuma \emph{et al.} \cite{sur13} & 2013 & Terahertz wireless communications based on photonics technologies & Tutorial  & This paper provides a tutorial on recent works on THz wireless communications systems based on photonic signal generation at carrier frequencies of over $100$ GHz.
\\ \hline

K.Jha, G.Singh \cite{sur131} & 2013 & 
Terahertz planar antennas for future wireless communication: A technical review & Magazine & This paper gives an overview of high directivity antennas, high-power sources, and efficient detectors for compact THz communication systems, with a special focus on improving planar antenna gain by reducing conductor and substrate losses.
\\ \hline

I. F. Akyildiz \emph{et al.} \cite{sur14} & 2014 & Terahertz band: Next frontier for wireless communications   & Magazine  & This article reports the THz applications and challenges in the generation, channel modeling, and communication systems, along with a brief discussion on experimental and simulation testbeds. 
\\ \hline
T. Kurner and S. Priebe \cite{sur15}& 2014 & Towards THz communications - Status in Research, standardization and regulation & Magazine  & This article reports the current research projects, spectrum regulations, and ongoing standardization activities in THz communication systems.
\\ \hline
I. F. Akyildiz \emph{et al.} \cite{sur16} & 2014 & TeraNets: ultra-broadband communication networks in the terahertz band   & Magazine  & This article reports the state of the art in THz Band device technologies and highlights the challenges and potential solutions from the communication and networking perspective as well as in terms of experimental testbeds.
\\ \hline
A. Hirata and M. Yaita \cite{sur17} & 2015 & Ultrafast terahertz wireless communications technologies & Survey & This paper gives an overview of the development of the THz technologies and standardization of wireless communications.
\\ \hline
C. Lin and G. Y. Li \cite{sur18}& 2016 & Terahertz communications: An array-of-subarrays solution  & Magazine  & This article reports the indoor multi-user THz communication systems with antenna arrays and discusses how the array-of-subarrays structure benefits THz communications from both the circuit and communication perspectives.
\\ \hline
M. Hasan \emph{et al.} \cite{sur19} & 2016 & Graphene terahertz devices for communications applications & Survey  & This paper gives an overview of recent progress on graphene-based devices for modulation, detection, and generation of THz waves, which are among the key components for future THz band communications systems. 
\\ \hline

T. Nagatsuma \emph{et al.} \cite{sur20} & 2016 & Advances in terahertz communications accelerated by photonics  & Tutorial &  This paper provides a tutorial on the latest trends in THz communications research, focusing on how photonics technologies have played a key role in the development of first-age THz communication systems. 
\\ \hline

J. F. Federici \emph{et al.} \cite{sur21} & 2016 & Review of weather impact on outdoor terahertz wireless communication links & Survey   & This paper gives an overview of the impact of weather on THz wireless links and emphasizes THz attenuation and channel impairments caused by atmospheric gases, airborne particulates, refractive index inhomogeneities, and their associated scintillations.
\\ \hline

S. Mumtaz \emph{et al.} \cite{sur22} & 2017 & Terahertz communication for vehicular networks & Survey & This paper gives an overview of the opportunities and challenges in THz communications for vehicular networks. 
\\ \hline

V. Petrov \emph{et al.} \cite{sur23} & 2018 & Last meter indoor terahertz wireless access: Performance insights and implementation roadmap  & Magazine  & This article reports the propagation modeling, antenna, and testbed designs, along with a step-by-step roadmap for THz Ethernet extension for indoor environments. 
\\ \hline

A. A. A. Boulogeorgos \emph{et al.} \cite{sur24} & 2018 & Terahertz technologies to deliver optical network quality of experience in wireless systems beyond 5G & Magazine  & This article reports the basic system architecture for THz wireless links with bandwidths of more than 50 GHz into optical networks.
\\ \hline

C. Han and Y. Chen \cite{sur25}  & 2018 & Propagation modeling for wireless communications in the terahertz band & Magazine  & This article reports the channel modeling in the THz band, based on the deterministic, statistical, and hybrid methods. The state-of-the-art THz channel models in single-antenna and UM-MIMO systems are extensively reviewed, respectively.
\\ \hline

I. F. Akyildiz \emph{et al.}  \cite{sur251}& 2018 & Combating the distance problem in the millimeter wave and terahertz frequency bands  & Magazine   &  This article reports the research advances on physical layer distance adaptive design, UM-MIMO, reflectarrays, and hyper-surfaces to show the direction to solve the problem of THz limited transmission distance.
\\ \hline

N. Khalid \emph{et al.} \cite{sur26}  & 2018 & Energy-efficient modulation and physical layer design for low terahertz band communication channel in 5G femtocell Internet of Things & Tutorial & This paper provides a tutorial on the modulation schemes, the hardware parameters, and the circuit blocks in the THz band which are suitable for mass market production.
\\ \hline

D. Headland \emph{et al.} \cite{sur261}  & 2018 & Tutorial: Terahertz beamforming, from concepts to realizations & Tutorial & This paper provides a tutorial on the basic principles of beam control in the THz range from the perspective of array antenna theory and diffraction optics and demonstrates significant demonstrations, including conventional optics, phased array antennas, leaky-wave antennas, and passive arrays.
\\ \hline

Z. Chen \emph{et al.} \cite{sur27}  & 2019 & A survey on terahertz communications & Survey   & This paper gives an overview of the development towards THz communications and presents some key technologies faced in THz wireless communication systems.    
\\ \hline

K. Tekbıyık \emph{et al.} \cite{sur28}  & 2019 & Terahertz band communication systems: Challenges, novelties and standardization efforts   & Survey   & This paper gives an overview of the unresolved problems, the latest solutions, and the standardization works in the THz communication systems. 
\\ \hline

T. S. Rappaport  \emph{et al.} \cite{sur29} & 2019 & Wireless communications and applications above 100 GHz: Opportunities and challenges for 6G and beyond & Survey   & This paper gives an overview of the technical challenges and opportunities for wireless communication and sensing applications above 100 GHz and presents several promising discoveries, novel approaches, and recent results.  
\\ \hline

K. K. O \emph{et al.} \cite{sur31} & 2019 & Opening terahertz for everyday applications  & Magazine  & This article reports the devices in CMOS, the challenges in implementing THz circuits, the performance of CMOS THz circuits, and their applications and expected advances.  
\\ \hline

K. M. S. Huq  \emph{et al.} \cite{sur32}  & 2019 & Terahertz-enabled wireless system for beyond-5G ultra-fast networks: A brief survey & Magazine  & This article reports the applications utilizing THz bands and hints at future research directions in this rapidly developing area.  
\\ \hline

X. Fu \emph{et al.} \cite{sur321} & 2020 & Terahertz beam steering technologies: From phased arrays to field-programmable metasurfaces  & Survey   &
This article gives a comprehensive summary of the principles and characteristics of THz beam steering technology from two aspects, the conventional technologies, and the reconfigurable metasurface-based technologies.
\\ \hline

H. Elayan \emph{et al.} \cite{sur320} & 2020 & Terahertz Band: The last piece of RF spectrum puzzle for communication systems  & Survey   & This paper gives an overview of the recent activities on the development,  standardization, and applications in THz communications.    
\\ \hline

H. Sarieddeen \emph{et al.} \cite{sur33} & 2020 & Next generation terahertz communications: A rendezvous of sensing, imaging, and localization & Magazine  & This article reports the THz technologies that bring significant advances to the areas of wireless communications, imaging, sensing, and localization. 
\\ \hline

L. Zhang \emph{et al.} \cite{sur34} & 2020 & Beyond 100 Gb/s optoelectronic terahertz communications: Key technologies and directions  & Magazine  & This article reports the key technologies of optoelectronic THz communications in the physical layer, including approaches of broadband devices, baseband signal processing technologies, and the design of advanced transmission system architectures.  
\\ \hline

M. A. Jamshed \emph{et al.} \cite{sur35}  & 2020 & Antenna selection and designing for THz applications: Suitability and performance evaluation: A survey & Survey   & This paper gives an overview of the characteristics of THz band, THz-enabled applications, materials of THz antenna, design parameters, and approaches to measure the performance of a THz-enabled antenna. 
\\ \hline

S. Ghafoor \emph{et al.} \cite{sur36} & 2020 & MAC Protocols for Terahertz Communication: A Comprehensive Survey & Survey   & This paper gives an overview of THz MAC protocols with classifications, band features, design issues, and future challenges. 
\\ \hline

C. X. Wang \emph{et al.} \cite{sur37} & 2020 & 6G Wireless Channel Measurements and Models: Trends and Challenges & Magazine  & This article reports the application scenarios, performance metrics, potential key technologies, and future research challenges of 6G wireless communication networks.
\\ \hline

A. Faisal \emph{et al.} \cite{sur38}  & 2020 & Ultramassive MIMO systems at terahertz bands: Prospects and challenges  & Magazine  & This article reports recent advances in transceiver design and channel modeling and discusses the major challenges and shortcomings by deriving the relationships among communication range, array dimensions, and system performance.  
\\ \hline

J. Tan \emph{et al.} \cite{sur381}  & 2020 & THz precoding for 6G: Applications, challenges, solutions, and opportunities  & Magazine  & This article reports three THz precoding architectures to cope with the challenges of severe path loss and limited coverage in THz communications. 
\\ \hline

F. Lemic \emph{et al.} \cite{sur39} & 2021 & Survey on Terahertz Nanocommunication and Networking: A Top-Down Perspective  & Survey   & This paper gives an overview of the current THz applications, different layers of the protocol stack, as well as the available channel models and experimentation tools. 
\\ \hline

C. Han \emph{et al.} \cite{sur40} & 2021 & Hybrid Beamforming for Terahertz Wireless Communications: Challenges, Architectures, and Open Problems  & Magazine & This article reports the challenges and characteristics of THz hybrid beamforming design and compares different hybrid beamforming architectures for THz communications.
\\ \hline

H. Sarieddeen \emph{et al.} \cite{sur41} & 2021 & An Overview of Signal Processing Techniques for Terahertz Communications  & Survey & This paper gives an overview of the recent developments in signal processing techniques for THz communications, focusing on waveform modulation, channel estimation, beamforming, and precoding in the THz UM-MIMO systems.
\\ \hline


H. J. Song \cite{sur44} & 2021 & Terahertz Wireless Communications: Recent Developments Including a Prototype System for Short-Range Data Downloading  & Magazine & 
This article reports device technologies that can be employed for THz communication prototype systems, including photonic devices, Si-CMOS, SiGe HBTs, and III-V HEMTs/HBTs. 
\\ \hline

H. Do \emph{et al.}  \cite{sur45} & 2021 & Terahertz line-of-sight MIMO communication: Theory and practical challenges  & Magazine & This paper reports that MIMO spatial multiplexing in the THz band is feasible even under LoS conditions with reconfigurable array architectures, and provides insights from the information-theoretic perspective.
\\ \hline

Z. Chen \emph{et al.} \cite{sur46} & 2021 &  Terahertz wireless communications for 2030 and beyond: A cutting-edge frontier & Magazine & This paper reports four promising directions for future THz communications, including integrated sensing and communications, ultra-massive MIMO, and dynamic hybrid beamforming, intelligent surfaces.
\\ \hline

Z. Chen \emph{et al.} \cite{sur47} & 2021 & Intelligent reflecting surface assisted terahertz communications toward 6G & Magazine & This paper reports the expectation and realization progress of incorporating the transformative features of the intelligent reflecting surface into THz wireless communications.
\\ \hline

C. Chaccour  \emph{et al.} \cite{sur42} & 2022 & Seven Defining Features of Terahertz (THz) Wireless Systems: A Fellowship of Communication and Sensing  & Survey & This paper gives an overview of seven features of THz wireless systems, including THz band characteristics, tailored network architectures, synergy with lower frequencies, communication, and sensing integration, physical layer signal processing, multiple access, and network optimization.
\\ \hline

C. Han \emph{et al.}\cite{holis} & 2022 & Terahertz Wireless Channels: A Holistic Survey on Measurement, Modeling, and Analysis  & Survey & This paper provides a comprehensive overview of the study of THz wireless channels, and categorizes existing channel modeling methodologies into deterministic, stochastic, and hybrid approaches.
\\ \hline

\end{longtable}
\end{center}
}
\endgroup

\twocolumn
\noindent  for THz communications \cite{sur320,sur27,sur28,sur29, sur31, sur32, sur321}. In 2020, H. Sarieddeen \emph{et al.} reported the current THz technologies in wireless communications, imaging, sensing, and localization \cite{sur33}. L. Zhang  \emph{et al.} reported the key technologies of optoelectronic THz communications in the physical layer \cite{sur34}. M. A. Jamshed   \emph{et al.} conducted a survey on the antenna selection design for THz applications \cite{sur35}. S. Ghafoor \emph{et al.} gave an overview of THz media access control (MAC) protocols with classifications, band features, design issues, and future challenges \cite{sur36}. C. X. Wang  \emph{et al.} reported the channel measurements, characteristics, and models for THz frequency band \cite{sur37}. A. Faisal \emph{et al.} reported the advantages of UM-MIMO systems in THz communications and discussed the challenges and shortcomings \cite{sur38}. In 2021, F. Lemic \emph{et al.} conducted a  survey on THz nano-communication and network from a top-down perspective \cite{sur39}. C. Han \emph{et al.} investigated the architectures and challenges of hybrid beamforming in THz communications \cite{sur40}. H. Sarieddeen \cite{sur41} gave an overview of signal processing techniques and propagation characteristics  of THz wireless systems, respectively. H. J. Song \emph{et al.} reported the device technologies employed for THz prototype systems\cite{sur44}. H. Do  \emph{et al.} reported that spatial multiplexing is feasible even the under line-of-sight  (LoS) conditions with reconfigurable array architectures\cite{sur45}. Z. Chen \emph{et al.} reported promising directions for future THz communications\cite{sur46,sur47}. C. Chaccour  \emph{et al.} \cite{sur42} gave an overview of seven features of THz wireless systems. C. Han \emph{et al.}\cite{holis} provided a comprehensive overview of the study of THz wireless channels.

With the rapidly advancing of THz technologies in terms of new manufacturing materials, transceiver architectures, and antenna designs, THz UM-MIMO has been envisioned as a key paradigm for future wireless systems\cite{umm}. In the above-mentioned works, the authors in \cite{sur38} provided a brief overview of THz UM-MIMO systems and highlighted a few challenges. The authors in \cite{sur25} reviewed several channel models of THz UM-MIMO systems. However, to the best of our knowledge, there is still lacking a detailed tutorial overview tailored for THz UM-MIMO beamforming technologies, to cover their theoretic breakthroughs, novel technological developments, engineering fabricating issues, and practical deployment considerations. This tutorial thus aims to fill this gap by providing a holistic view of the THz UM-MIMO beamforming, including the basic principles, wideband effects, beamforming scenarios, existing MIMO arrays, emerging applications, and future challenges.
\begin{figure}[t]
\centering
\includegraphics[width= 3.5in]{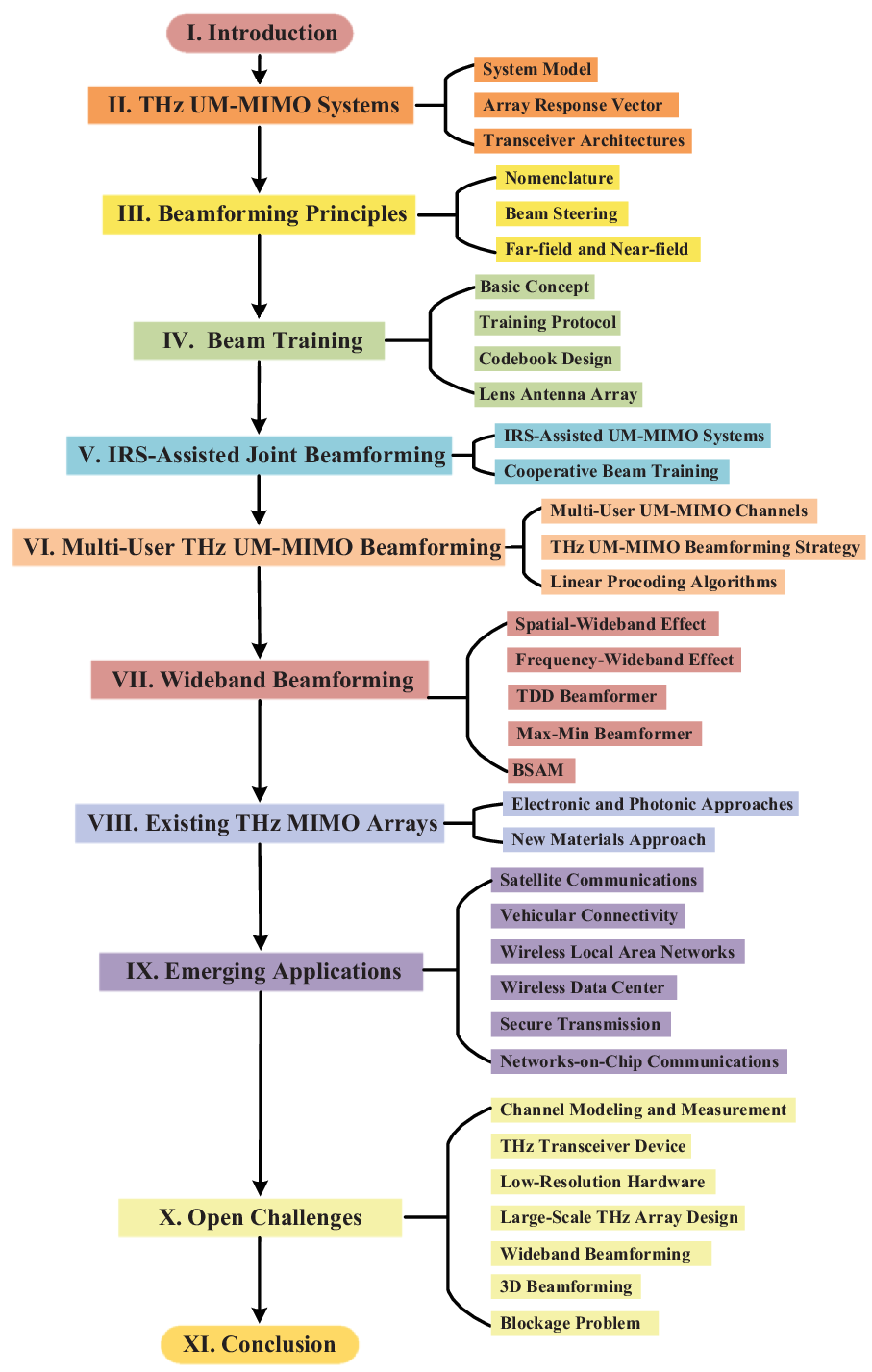}
\caption{The structure of this paper.}\label{struct}\vspace{-10pt}
\end{figure}
\subsection{Contributions of this Paper}
 The main contributions of this paper are summarized below.
\begin{itemize}
\item We present a basic system model for THz UM-MIMO and propose a way to determine the channel parameters. The effects of the antenna geometry and transceiver architecture on the THz UM-MIMO systems are further discussed with precise definitions.

\item We illustrate the basic principles of beamforming by visualizing the electromagnetic field distribution in the physical layer. We characterize the features of beam patterns generated by large-scale array and discuss the near-field and far-field assumptions in THz UM-MIMO systems.

\item We introduce a beam training scheme that can realize the beamforming in THz UM-MIMO systems without using the channel state information (CSI). The basic concept of the beam training is elaborated and two essential components, i.e., training protocol and codebook design, are discussed. The lens antenna array is introduced to realize beam training as a special manner.

\item We study an intelligent reflecting surface (IRS)-assisted system for THz communication, which helps enhance the signal coverage and improve spectral efficiency with very low energy consumption. The system model of the IRS-assisted THz UM-MIMO is presented and the joint active and passive beamforming strategies for THz communication are discussed. 

\item We introduce the multi-user UM-MIMO channels and provide an effective beamforming strategy for multi-user THz UM-MIMO systems. Based on this strategy, we revisit the popular linear beamforming algorithms that can eliminate the inter-user interference and evaluate their performance in THz UM-MIMO systems.

\item We present two main issues, i.e., spatial-wideband effect and frequency-wideband effect, which are notable and need to be considered in the wideband THz beamforming. We explain why these effects occur and provide some available solutions to addressing these issues.

\item We provide an overview of the existing THz MIMO arrays based on different fabrication techniques, including electronic-based, photonics-based, and new materials-based. We pay special attention to those THz antenna arrays with dynamic beam steering capabilities.

\item We identify the transformative functions of the THz beamforming technologies in emerging applications, including satellite communications, vehicle connectivity, indoor wireless networks, wireless data centers, secure transmission, and inter-chip communications. These applications demonstrate the great potential of THz UM-MIMO systems in shaping future wireless networks.

\item We point out various open challenges faced by the THz UM-MIMO systems, in terms of channel modeling and measurement, THz transceiver device, low-resolution hardware, large-scale THz array design, wideband beamforming, 3D beamforming, and blockage problem, to light up new horizons and stimulate enthusiasm for future research.
\end{itemize}

\subsection{Organization and Notations}
As shown in Fig. \ref{struct}, the rest of this paper is organized as follows. Section II introduces the THz UM-MIMO systems. In Section III, we illuminate the basic beamforming principles. Sections IV and V present respectively the beam training and the IRS-assisted joint beamforming in THz UM-MIMO systems. Section VI and VII respectively discuss the multi-user beamforming schemes wideband beamforming solutions  in THz UM-MIMO systems. Section VIII provides an overview of the existing THz MIMO arrays. Section IX identifies the emerging applications with THz beamforming technologies. Open challenges and future research directions are discussed in Section X. Finally, we conclude the paper in Section XI.

\indent \emph{Notation:} We use small normal face for scalars, small bold face for vectors, and capital bold face for matrices. The superscript ${{\rm{\{ }} \cdot \}^T}$, ${{\rm{\{ }} \cdot \}^{\dag}}$, and ${{\rm{\{ }} \cdot \}^H}$ denote the transpose, conjugate, and Hermitian transpose, respectively. $|\cdot|$ and $||\cdot||_{F}$ represent the modulus operator and Frobenius norm, respectively. ${\rm{diag}}( \cdot )$ denotes a diagonal matrix whose diagonal elements are given by its argument. $\mathbb{R}$, $\mathbb{C}$ stand for the set of real numbers and the set of complex numbers, respectively.

\section{THz UM-MIMO Systems}
In this section, we start with describing the system model and specifying the parameters of THz UM-MIMO systems. Then, some important system characteristics are elaborated, i.e., array response vectors and transceiver architectures.

\begin{table}[t]
\centering
\caption{Comparision of channel characteristics in indoor office scenarios.} \label{tablej}
\includegraphics[width=3.5in]{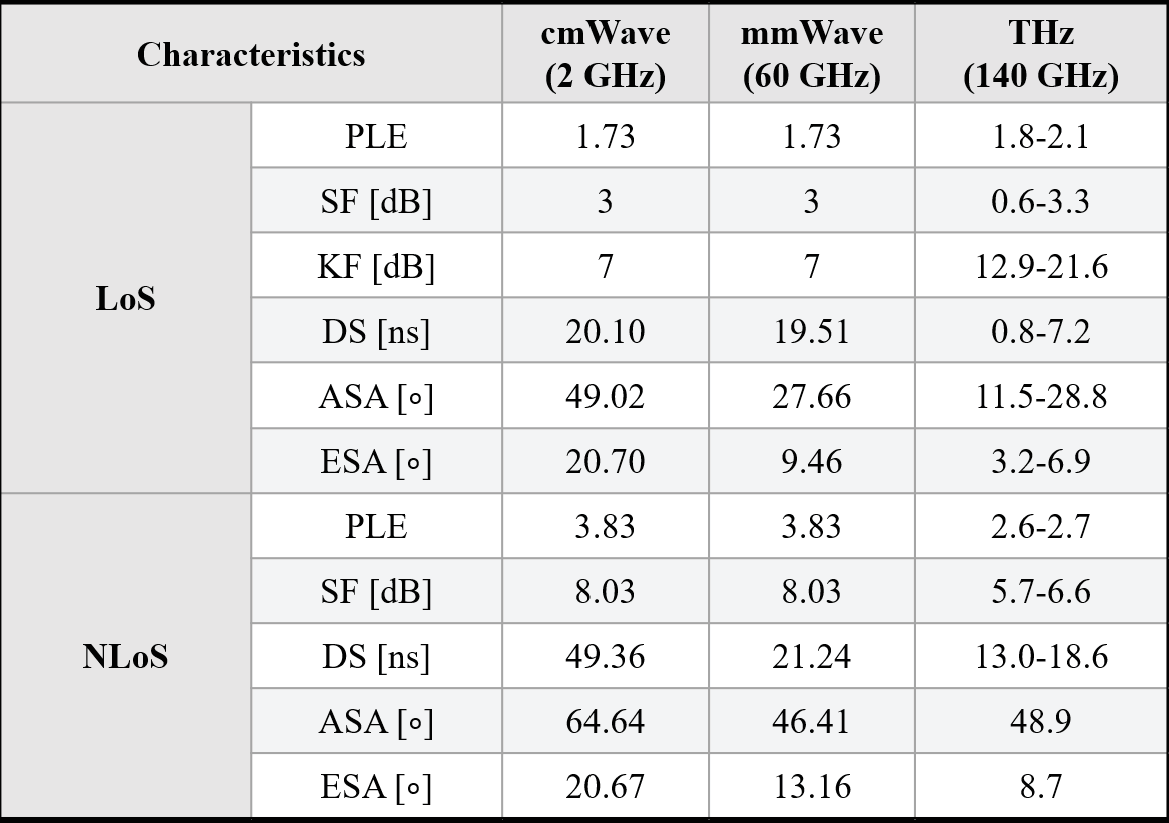}
\end{table}

{\fontsize{9}{11}\selectfont 
\begin{table*}[t]
\centering
\caption{Three attenuation models of electromagnetic wave between $0.1$ and $1$ THz.} \label{table2}
\begin{tabular}{|c|l|l|l|}
\hline
\multicolumn{4}{|c|}{Three Evaluation Models for the Attenuation of Electromagnetic Wave between $0.1$ and $1$ THz} \\ \hline
MPM model \cite{mpm}  & \multicolumn{3}{l|}{\tabincell{l}{MPM is a broadband model for complex refractivity to predict the propagation effects of loss \\and delay for the neutral atmosphere. Input variables are barometric pressure, temperature,\\ relative humidity, suspended water droplet concentration, and rainfall rate.}}                                      \\ \hline
AM model \cite{am} & \multicolumn{3}{l|}{\tabincell{l}{AM is a tool for radiative transfer computations at the microwave to submillimeter wavelengths. \\The program can also be applied to a variety of other radiative transfer problems and has been \\used for applications ranging from laboratory receiver testing to radio spectrum management.}}                                       \\ \hline
ITU-R model \cite{itu} & \multicolumn{3}{l|}{\tabincell{l}{ITU-R P.676-11 provides methods to estimate the attenuation of atmospheric gases on terrestrial\\ and slant paths by using a summation of individual gaseous absorption lines that is valid for the \\ frequency range
$1$ GHz to $1$ THz, based on atmospheric pressure, temperature, and water vapor.}}                                      \\ \hline
\end{tabular}
\end{table*}
}
\subsection{System Model}\label{sm}
Consider a point-to-point THz UM-MIMO system with the quasi-static block fading channel. Let $N_t$ and $N_r$ denote the number of transmit and receive antenna elements, respectively. The received signal vector ${\mathbf{y}}\in \mathbb{C}^{N_r}$  can be expressed as 
\begin{equation}\label{system}
{\mathbf{y}} = {\mathbf{Hx}} + {\mathbf{n}},
\end{equation}
where ${\mathbf{x}} \in \mathbb{C}^{N_t}$ is the transmitted signal vector. ${\mathbf{H}} \in \mathbb{C}^{N_r \times N_t}$ is the channel matrix and ${\mathbf{n}} \in \mathbb{C}^{N_r}$ is the noise vector. 

\subsubsection{Channel Model}
The channel matrix  ${\mathbf{H}}$ can be modelled based on the channel measurements in THz band\cite{mod2,mod3,mpm,am,itu}. Important channel parameters include path loss exponent (PLE), shadow fading (SF), $K$ factor (KF), delay spread (DS), azimuth spread of arrival (ASA), and elevation spread of arrival (ESA).  These parameters in THz band are quite different from those in cmWave and mmWave scenarios, as shown in Table \ref{tablej} \cite{holis}.  Here, we present a popular statistical channel model, i.e.,  Saleh-Valenzuela (S-V) channel model \cite{sv}, which is widely used in UM-MIMO systems.\footnote{In particular,  THz channel model can be classified as deterministic \cite{umcm1, umcm2}, statistical \cite{umcm3}, and hybrid model \cite{umcm4, umcm5}.  Deterministic model requires detailed geometric knowledge of the propagation environment, which is site-specific and subject to high complexity.  In comparison, statistical channel model allows fast channel construction based on key channel statistics. } With the employment of ultra-massive antenna elements, the THz channel generally shows sparsity and strong directivity, which is composed of a strong LoS cluster and several weak NLoS cluster\cite{mod1,mod4,mod5}. As such, the channel matrix in the time-domain can be  written as 
\begin{equation}\label{chan}
{\bf{H}}(t) = \sum\limits_{i = 1}^{{N_{{\rm{cl}}}}} {\sum\limits_{j = 1}^{N_{{\rm{ray}}}^i} {\delta (t - {\tau _{ij}})\sqrt {{\alpha _{ij}}{G_{t,ij}}{G_{r,ij}}} {{\bf{a}}_{r,ij}}} } {\bf{a}}_{t,ij}^H,
\end{equation}
where ${N_{{\mathrm{cl}}}}$ and ${N_{{\mathrm{ray}}}^i}$ are the number of clusters and the number of rays in the $i$-th cluster, respectively.  $\delta (\cdot)$ is the impulse response function.  ${\alpha _{ij}}$ denotes the path gain of the $j$-th ray in the $i$-th cluster. ${\tau _{ij}} = {T_i} + {T_{ij}}$, in which ${T_i}$ (with ${T_1}=0$) and ${T_{ij}}$ (with ${T_{i1}}=0$) represent the arrival time of the $i$-th cluster and that of the $j$-th ray in it, respectively. ${G_{t,ij}}$ and ${G_{r,ij}}$ (resp. ${{\mathbf{a}}_{t,ij}}$ and ${\mathbf{a}}_{r,ij}$) represent the transmit and receive antenna gains (resp. array response vectors at transmitter and receiver) of the $j$-th ray in the $i$-th cluster, respectively.   In particular, the number of clusters ${N_{{\mathrm{cl}}}}$ is commonly assumed to follow Poisson processes, while the arrival time difference between two clusters ${T_i}-{T_{i-1}}$ is assumed to be exponentially distributed\cite{ang1}.  The angles of departure/arrival (AoDs/AoAs) in ${{\mathbf{a}}_{t,ij}}$ and ${\mathbf{a}}_{r,ij}$ can be assumed to be uniformly distributed within $[0,2\pi)$, and the ray angle within a cluster can be assumed to follow zero-mean Laplacian distribution \cite{ang1} or a zero-mean second-order Gaussian mixture model (GMM) \cite{ang2}. According Table \ref{tablej}, the $K$ factor in THz band is $12.9$-$21.6$, which indicates that  $90\%$-$95\%$ of path gain comes from the LoS cluster. Moreover, as the ASA and the ESA in THz are much smaller than their counterparts in cmWave and mmWave, we can assume only one ray in each cluster. Then, the THz channel can be modelled as a LoS rank-one matrix, i.e.,
\begin{equation}\label{losa}
{{\bf{H}}_{{\rm{LoS}}}} = \sqrt G  \cdot {{\bf{a}}_r}{\bf{a}}_t^H,
\end{equation}
where $G = \alpha {G_t}{G_r}$ is the channel gain.
Due to the LoS dominant property, the beamforming technologies in the THz band show some unique aspects compared to those in the cmWave and mmWave band. First, the conventional beamforming is designed based on the estimated CSI, which accounts for the propagation on many (or even infinite) paths. While the THz beamforming merely takes into account of the LoS path, which can be realized without CSI by beam training (as will be detailed in Sec. \ref{BT}). Second,  LoS blockage is destructive for THz beamforming. Thus, the cooperative beamforming, e.g., IRS-assisted joint beamforming (as will be detailed in Sec. \ref{IRSJB}), is worthy of consideration to reduce the LoS blockage.

\begin{figure*}[t]
\centering
\includegraphics[width=7.2in]{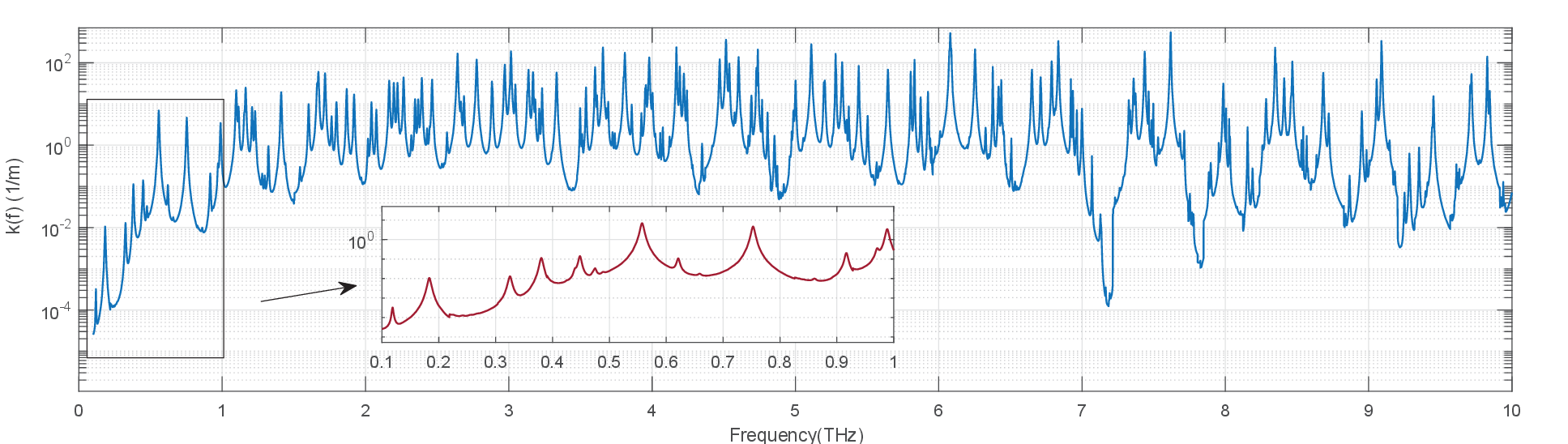}
\caption{$k(f)$ at frequencies from $0.1$ THz to $10$ THz in HITRAN database (temperature ${\mathcal{T}}$=$296$ K and pressure ${\mathcal{P}}$=$1$ atm in sunny day).}\label{kk}\vspace{-10pt}
\end{figure*}

\subsubsection{Path Gain}\label{PG}
The path gain of THz channels suffers from severe free spreading loss due to the extremely high frequency. According to the Friis’ formulation \cite{af}, the free spreading loss is given by 
\begin{equation}\label{p4}
{L_{{\mathrm{spread}}}}(f,d) = {\left( {\frac{{4\pi fd}}{c}} \right)^2},
\end{equation}
where $c$ is the speed of light in free space, $f$ is the carrier frequency, and $d$ is the path distance. Hence, the spreading loss increases with the frequency squared. In addition, the molecular absorption also causes severe attenuation of THz radial signals, which is not negligible. Hence, the path gain ${\alpha _{ij}}$ in (\ref{chan}) can be explicitly written as 
\begin{equation}\label{44}
{\alpha _{ij}}(f,{d_{ij}}) = \frac{1}{{{L_{{\mathrm{spread}}}}(f,{d_{ij}}){L_{{\mathrm{abs}}}}(f,{d_{ij}})}}{e^{ - \frac{{{T_i}}}{{{\Gamma _c}}}}}{e^{ - \frac{{{T_{ij}}}}{{{\Gamma _r}}}}},
\end{equation}
where $d_{ij}$ is the path distance of the $j$-th ray in the $i$-th cluster. ${L_{{\mathrm{abs}}}}(f,{d_{ij}})$ is the molecular absorption which mainly comes from water vapor \cite{mod3}. ${\Gamma _c}$ and ${\Gamma _r}$ are the exponential attenuation factors of the arrive cluster and ray, respectively, which are frequency and wall-material dependent in general \cite{sv}.

\begingroup
\renewcommand{\arraystretch}{1.3} 
\begin{table}[t]
\centering
\caption{Total molecular absorption at some frequencies.} \label{table3}
\begin{tabular}{|c|c|c|c|c|c|c|}
\hline
\multicolumn{7}{|c|}{Total Molecular Absorption at some Frequencies} \\ \hline
$f$ (THz)    & 0.14    & 0.26    & 0.35    & 0.41    & 0.67     & 0.85    \\ \hline
$k(f)$ (dB) &  -42.2 &  -38.5 &  -27.8 &  -22.4 &  -18.5 &  -20.9 \\ \hline
\end{tabular}
\vspace{-2pt}
\end{table}
\endgroup

The molecular absorption at the frequency below $1$ THz can be evaluated by the atmospheric mmWave propagation
model (MPM) \cite{mpm}, atmospheric model (AM) \cite{am}, and ITU-R P.676-10 model \cite{itu} for various environments. A brief introduction of these models is relegated to Table \ref{table2}. In general, the molecular absorption can be expressed as 
\begin{equation}\label{55}
{L_{{\mathrm{abs}}}}(f,{d_{ij}}) = {e^{k(f){d_{ij}}}},
\end{equation}
where $k(f)$ is the total absorption coefficient that is comprised of a weighted sum of different molecular absorption in the medium, i.e., $k(f) = \sum\nolimits_i {{m_i}{k_i}(f)}$, in which ${m_i}$ is the weight and ${{k_i}(f)}$ is the molecular absorption coefficient of the $i$-th species. The exact ${{k_i}(f)}$ on condition of any temperature and pressure can be obtained from the high resolution transmission (HITRAN) database \cite{hitran}.  Based on the details therein, we plot $k(f)$ at frequencies from $1$ to $10$ THz in Fig. \ref{kk}. As can be seen, the total absorption coefficient is relatively large at frequencies above $1$ THz. Despite that two obvious drops can be witnessed in the range within $7$-$8$ THz, the near-future THz systems are generally considered below $1$ THz since the conventional metallic-antenna array requires extremely sophisticated process technology in the band above $1$ THz. This also motivates the works on the nano-antenna array, e.g., meta-material, liquid crystal, and graphene arrays, for higher THz bands. It is observed from the red line in Fig. \ref{kk} that there are some frequency bands with smaller absorption, which hopefully will be first used for wireless communication in the future. For ease of access, we specify $k(f)$ at the center points of these candidates in Table \ref{table3}. These results are expected to provide rational and equitable validation, evaluation, and simulation in future works.
\begin{figure}[t]
\centering
\includegraphics[width=3.3in]{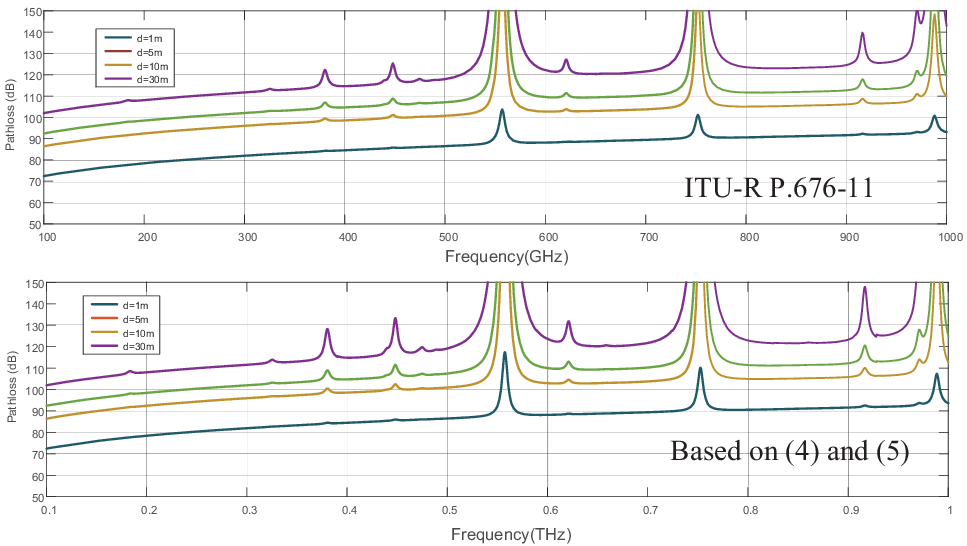}
\caption{The path loss computed by two methods.}\label{tt}\vspace{-2pt}
\end{figure}

Let us consider the total path loss of the  channel in (\ref{chan}). By using the molecular absorption coefficient shown in Fig. {\ref{kk}}, based on (\ref{55}), we can obtain the exact path gain of ${{{\alpha _{11}}(f,{d_{11}})}}$ in (\ref{44}).
By this means, we select $k(f)$ within $0.1$-$1$ THz and compute the corresponding path loss based on (\ref{44}) and (\ref{55}). In the meanwhile, we plot the counterpart by using the ITU-R model tool for comparison. As shown in Fig. {\ref{tt}}, both figures almost share the same trend of the loss on different transmit distances. The striking jump points in both figures stand at the same frequencies, albeit with the difference of the order of magnitude. Except for these jump points, the path gain calculated by our provided model is accurate and reliable.

\subsubsection{Antenna Gain}\label{AG}
\begin{figure}[t]
\centering
\includegraphics[width=2.6in]{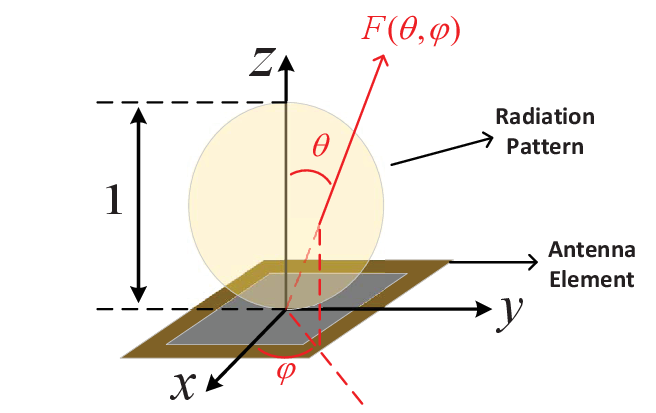}
\caption{Normalized radiation pattern of antenna element.}\label{ele}\vspace{-2pt}
\end{figure}

The transmit/receive antenna gain is the product of \emph{element gain} and \emph{array gain}. The element gain describes how much power is transmitted/received compared to an isotropic element. Let $G_e(\varphi,\theta )$ denote the element gain for a propagation path with the azimuth angle $\varphi$ and the elevation angle $\theta$, we have 
\begin{equation}
{G_e}(\varphi ,\theta ) = \varsigma F(\varphi ,\theta )\mathcal{D},
\end{equation}
where
\begin{equation}
\mathcal{D}  = \frac{{4\pi }}{{\int\limits_{\varphi  = 0}^{2\pi } {\int\limits_{\theta  = 0}^\pi  {F(\varphi ,\theta )\sin \theta d\theta d\varphi } } }}\label{dire}
\end{equation}
is the directivity, $\varsigma$ is the antenna efficiency, and $F(\varphi,\theta)$ is the normalized radiation pattern. The radiation pattern generally focus more energy in the direction perpendicular to the element, as shown in Fig. {\ref{ele}}.  An exemplary $F\left( {\varphi,\theta } \right) $ is given by \cite{pat} 

\begin{equation}\label{erp}
F\left( {\varphi,\theta } \right) = \left\{ {\begin{aligned}
\cos \theta,&{}&{\varphi \in \left[ {0,2\pi} \right]{\mathrm{and}}\;\theta \in \left[ {0,\frac{\pi}{2} } \right]}\\
0,&{}&{\mathrm{otherwise}\quad}
\end{aligned}} \right..
\end{equation}
If the radiation pattern has only one major lobe,  the directivity in (\ref{dire}) can be empirically calculated as  \cite{rs} 
\begin{equation}
\mathcal{D} \simeq  \frac{{4\pi }}{{\theta _{3{\mathrm{dB}}}^x\theta _{3{\mathrm{dB}}}^y({\mathrm{radians}})}} = \frac{{41253}}{{\theta _{3{\mathrm{dB}}}^x\theta _{3{\mathrm{dB}}}^y({\mathrm{degrees}})}},\;
\end{equation}
where $\theta _{3{\mathrm{dB}}}^x$ and $\theta _{3{\mathrm{dB}}}^y$ is the half-power beamwidth (HPBW) of the major lobe on $x$- and $y$-axis, respectively.

  The array gain achieved if the signal is coherently added from $N$ array elements. Let $P$ represent the transmit power.  By using $N$ antenna elements (each equally assigned a power of $P/N$), the maximum amplitude of the signal in the desired direction is $N \times \sqrt{P/N} = \sqrt{PN}$, and the corresponding signal power is $PN$. Thus, compared to using only one antenna element with power $P$, using $N$ antenna elements achieves an array gain of $N$. As such, the antenna gain ${G_{t,ij}}$ and ${G_{r,ij}}$ in (\ref{chan}) are respectively given by \cite{sun}
\begin{equation}\label{arra}
\begin{aligned}
{G_{t,ij}} = {N_t}{G_e}(\theta _{ij}^t,\varphi _{ij}^t),\;{G_{r,ij}} = {N_r}{G_e}(\theta _{ij}^r,\varphi _{ij}^r).
\end{aligned}
\end{equation}



\subsubsection{Noise}
The noise term in (\ref{system}) generally comes from two parts, i.e., the electronic thermal noise and the re-radiation noise. 
The first part, also called background noise, is caused by the thermal motion of molecules, which is related to the frequency but independent of the transmitted signal power. Its power spectral density (PSD) can be expressed as \cite{bcnoise}
\begin{equation}\label{bn}
S_{bn}(f)= \frac{hf}{\exp(hf/k_{B}\mathcal{T})-1},\;
\end{equation}
where $h$ stands for Planck’s constant, $k_{B}$ is the Boltzmann constant, and $\mathcal{T}$ is the reference temperature. The second part, re-radiation noise, is highly correlated with the transmission signal power.  As described in \cite{abnoise}, atmospheric molecules will return to stability after being excited by the electromagnetic waves, thus the absorbed energy will be re-radiated with random phases. The PSD of the re-radiation noise comes from the lost power caused by molecular absorption, i.e., 
\begin{equation}\label{abn}
{S_{an}}(f,d) = \frac{{{S_t}(f)}}{{{L_{spread}}(f,d)}}\left[ {1 - \frac{1}{{{L_{abs}}(f,d)}}} \right].
\end{equation}
Consequently, the PSD of the total noise can be written as
\begin{equation}
S_{n}(f) = {{S_{bn}}\left( f \right) + \sum\limits_{i,j}  \eta _{ij} {{S_{an}}\left( {f,{d_{ij}}} \right)} },
\end{equation}
where $\sum\nolimits_{i,j} {S_{an}} ({f,{d_{ij}}})$ represents the power density of re-radiation noise from all paths, $\eta _{ij}$ is a loss factor that indicates how much power can be detected by the receiver. It is worth pointing out that some works \cite{abnoise,abnoise2,abnoise3} assumed that $\eta _{ij}=1,\forall ij$.  However, the molecular absorption happens everywhere along with the propagation, and the spread of the re-radiated power is omnidirectional, which will not be all captured by the receiver. Regarding this, $\eta _{ij}$ should be rather small and need to be properly modeled.

Based on the analyse of path/antenna gain above, we observe that the path gain decrease with frequency. However, if the antenna footprint is kept, the number of antenna elements increases with frequency, which improves the antenna gain. Regarding this, an interesting problem is how the received power changes with the frequency, if the antenna footprint $\mathcal{S}$ and transceiver distance $d$ are kept. Let $P_t$ denote the transmit power.  The received power can be written as 
\begin{equation}
{P_r} = {P_t}\frac{{{G_t}{G_r}}}{{{L_{{\rm{spread}}}}(f,d){L_{{\rm{abs}}}}(f,d)}}.
\end{equation}
Considering the isotropic antenna elements, i.e., $G_e=1$, we have ${G_t}=G_r=N$. The footprint  of a uniform square array is approximately given by 
\begin{equation}
{\mathcal{S}} = {\left( {N\frac{\lambda }{2}} \right)^2} \;\Rightarrow \;N = \frac{{2f}}{c}\sqrt {\mathcal{S}},
\end{equation}
where $\lambda$ is wavelength. For multi-input single-output (MISO) systems, there is no antenna gain at the receiver, and thus $G_r=1$. In this case,  based on (\ref{p4}) and (\ref{55}), the received power can be rewritten as 
\begin{equation}
{P_r} = {P_t}\cdot\frac{1}{f} \cdot {e^{ - k(f)d}} \cdot \frac{{c\sqrt {{\mathcal{S}}} }}{{8{{\left( {\pi d} \right)}^2}}},
\end{equation}
where $1/f$ is the remaining component of the spreading loss. That is, the increased antenna gain cannot compensate for the spreading loss in MISO systems, and the received power decreases with the large-scale increase in frequency (without considering the small-scale fluctuation). In MIMO systems, the received power can be written as 
\begin{equation}
{P_r} = {P_t} \cdot {e^{ - k(f)d}} \cdot \frac{S}{{4{{\left( {\pi d} \right)}^2}}},
\end{equation}
where the antenna gain can compensate for the spreading loss and only the loss of molecular absorption remains. As can be seen from Fig. \ref{kk}, the absorption factor does not show monotonicity with frequency. Thus, unlike MISO systems, the received power in MIMO systems might be higher with the large-scale increase in frequency, e.g., $P_r$ at $7.2$ THz is larger than  $P_r$ at $3$ THz.

\begin{figure*}[t]
\centering
\includegraphics[width=7in]{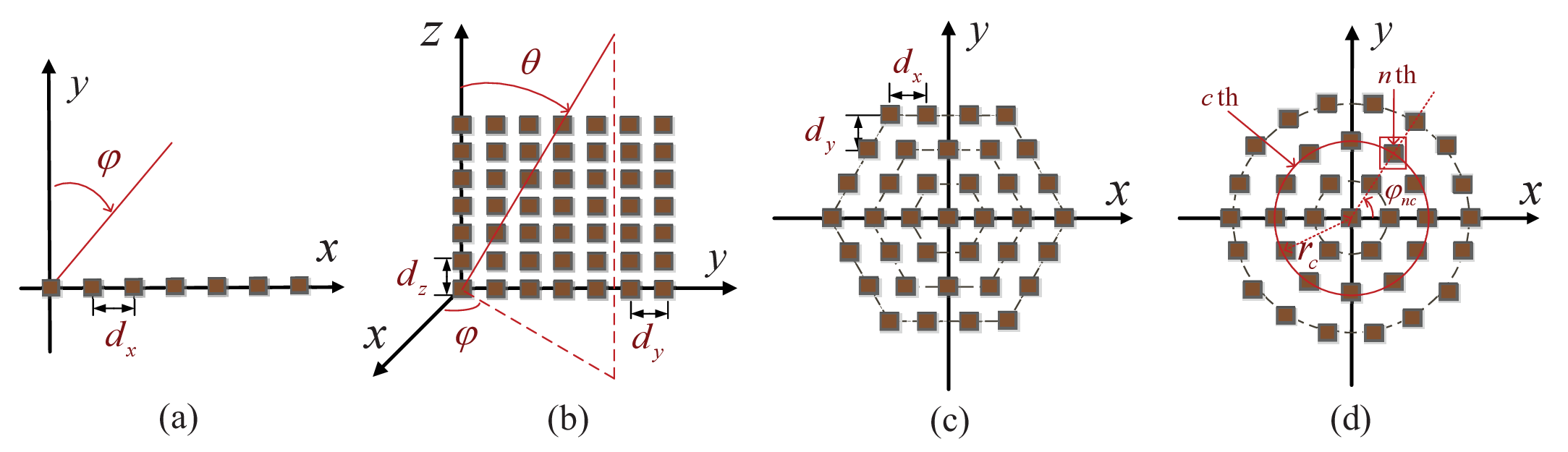}
\caption{Illustration of array geometries: (a) ULA; (b) URPA; (c) UHPA; (d) UCPA.}\label{ulafig}
\vspace{-2pt}
\end{figure*}
\subsection{Array Response Vector}\label{ed}
Each array response vector ${{\mathbf{a}}_{r,ij}}$ in (\ref{chan}) presents a propagation path on a specific AoD or AoA, which could be one or two dimensional, depending on the the antenna geometry. Commonly, there are four typical array geometries: uniform linear array (ULA), uniform rectangular planar array (URPA), uniform hexagonal planar array (UHPA), and uniform circular planar array (UCPA).  Next, we present the expression of the array response vectors in the above geometries and the corresponding path angles are plotted in Fig. \ref{ulafig}. 

\subsubsection{ULA (see Fig. \ref{ulafig} (a))}
Considering an $N_a$-element ULA. The array response vector can be written as \cite{array2}
\begin{equation}\label{ula0}
{\mathbf{a}}{\left( \varphi  \right)_{\rm{ULA}}} = \frac{1}{{\sqrt N_a}}{\left[ {1,{e^{jk{d_x}\sin \left( \varphi  \right)}},...,{e^{jk(N_a - 1){d_x}\sin \left( \varphi  \right)}}} \right]^T},
\end{equation}
where $d_x$ is the inter-element spacing, $k = \frac{{2\pi }}{\lambda }$ is the phase constant. 
\subsubsection{URPA (see Fig. \ref{ulafig} (b))}
Consider an URPA with $N_y$ times $N_z$ elements lying on the $yz$-plane. The array response vector is given by \cite{array1}
\begin{equation}
\begin{aligned}
&{\mathbf{a}}{\left( {\varphi ,\theta } \right)_{\rm{URPA}}}\\
&= \frac{1}{{\sqrt {N_a} }}\left[ {1,...,{e^{jk\left( {m{d_y}\sin \left( \theta  \right)\sin \left( \varphi  \right) + n{d_z}\cos \left( \theta  \right)} \right)}},} \right.\\
&{\left. {{\kern 1pt} {\kern 1pt} {\kern 1pt} {\kern 1pt} {\kern 1pt} {\kern 1pt} {\kern 1pt} {\kern 1pt} {\kern 1pt} {\kern 1pt} {\kern 1pt}  \ldots ,{e^{jk\left( {\left( {{N_y} - 1} \right){d_y}\sin \left( \theta  \right)\sin \left( \varphi  \right) + \left( {{N_z} - 1} \right){d_z}\cos \left( \theta  \right)} \right)}}} \right]^T},
\end{aligned}
\end{equation}
where $d_y$ and $d_z$ are the inter-element spacing on the $y$-axis and $z$-axis, $0 \le m \le {N_{y}-1}$ and $0 \le n \le {N_{z}-1}$ are the indices of antenna element, $N_a = {N_y}{N_z}$ is the total number of elements, and $\varphi$ and $\theta$ are the azimuth and elevation angles of arrival, respectively. 
\subsubsection{UHPA (see Fig. \ref{ulafig} (c))}
Consider a UHPA consisting of $V$ hexagon rings on the $xy$-plane. The inter-element spacing on the horizontal and vertical direction are $d_x$ and ${d_y} = {{\sqrt 3 {d_x}} \mathord{\left/
 {\vphantom {{\sqrt 3 {d_x}} 2}} \right.
 \kern-\nulldelimiterspace} 2}$, respectively, and the array response vector is given by \cite{array2}
\begin{equation}
{\mathbf{a}}{\left( {\varphi ,\theta } \right)_{\rm{UHPA}}} = \frac{1}{{\sqrt {N_a} }}{\left[ {{\mathbf{f}}_V},...,{{\mathbf{f}}_1},{{\mathbf{f}}_0},{{\mathbf{f}}_{-1}},...,{{\mathbf{f}}_{-V}} \right]^T},
\end{equation}
where $\{\mathbf{f}_v\}_{v = -V}^{V}$ denote the ULA vectors at different rows, $N_a =1+ \sum\nolimits_{v=1}^{V}6v$, is the total number of elements. For different parities of the subscript $v$, $\mathbf{f}_v$ can be expressed as \cite{array2}
\begin{align}\label{arra_uhpa}
\begin{aligned}
&{{\mathbf{f}}_v}\left| {_{v = 2n}} \right.\\
&= {e^{ - jvk{d_y}\sin \left( \theta  \right)\cos \left( \varphi  \right)}}\left[ {{e^{ - j\left( {V - \frac{v}{2}} \right)k{d_x}\sin \left( \theta  \right)\cos \left( \varphi  \right)}},} \right.\\
&\qquad \ldots ,{e^{ - jk{d_x}\sin \left( \theta  \right)\cos \left( \varphi  \right)}},1{\kern 1pt} {\kern 1pt} {\kern 1pt} ,{e^{jk{d_x}\sin \left( \theta  \right)\cos \left( \varphi  \right)}}, \\
&\qquad\qquad\qquad \ldots ,{\left. {{e^{j\left( {V - \frac{v}{2}} \right)k{d_x}\sin \left( \theta  \right)\cos \left( \varphi  \right)}}} \right]},
\end{aligned} \\
\begin{aligned}
&{{\mathbf{f}}_v}\left| {_{v = 2n - 1}} \right.\\
&= {e^{ - jvk{d_y}\sin \left( \theta  \right)\cos \left( \varphi  \right)}}\left[ {{e^{ - j\left( {V - \frac{{v - 1}}{2}} \right)k{d_x}\sin \left( \theta  \right)\cos \left( \varphi  \right)}}} \right.,\\
&\qquad   \ldots ,{e^{ - jk{d_x}\sin \left( \theta  \right)\cos \left( \varphi  \right)}},{\kern 1pt} {e^{jk{d_x}\sin \left( \theta  \right)\cos \left( \varphi  \right)}},\\
&\qquad\qquad\qquad{\left. { \ldots ,{e^{j\left( {V - \frac{{v - 1}}{2}} \right)k{d_x}\sin \left( \theta  \right)\cos \left( \varphi  \right)}}} \right]}.
\end{aligned}
\end{align}
\subsubsection{UCPA (see Fig. \ref{ulafig} (d))}
Consider a UCPA consisting of $C$ circles on the $xy$-plane, each element uniformly distributed over the circle. The array response vector is given by \cite{array2}

\begin{equation}
\begin{aligned}
&{\mathbf{a}}{\left( {\varphi ,\theta } \right)_{\rm{UCPA}}}= \frac{1}{{\sqrt N_a}}\left[ {1, \ldots ,{e^{jk{r_c}\sin \left( \theta  \right)\cos \left( {\varphi  - {{\overline \varphi  }_{nc}}} \right)}},} \right.\\
&\qquad\qquad\qquad \qquad\ldots {\left. {,{e^{jk{r_C}\sin \left( \theta  \right)\cos \left( {\varphi  - {{\overline \varphi  }_{nC}}} \right)}},...} \right]^T},
\end{aligned}
\end{equation}
where $N_a =1+ \sum\nolimits_{c=1}^{C}6c$ is the total number of elements, $r_c$ is the radius of the $c$-th circle, $\overline {{\varphi}} _{nc}$ is the angle of the $n$-th element in the $c$-th circle to the $x$-axis.

\begin{figure}[t]
\centering
\includegraphics[width=3.4in]{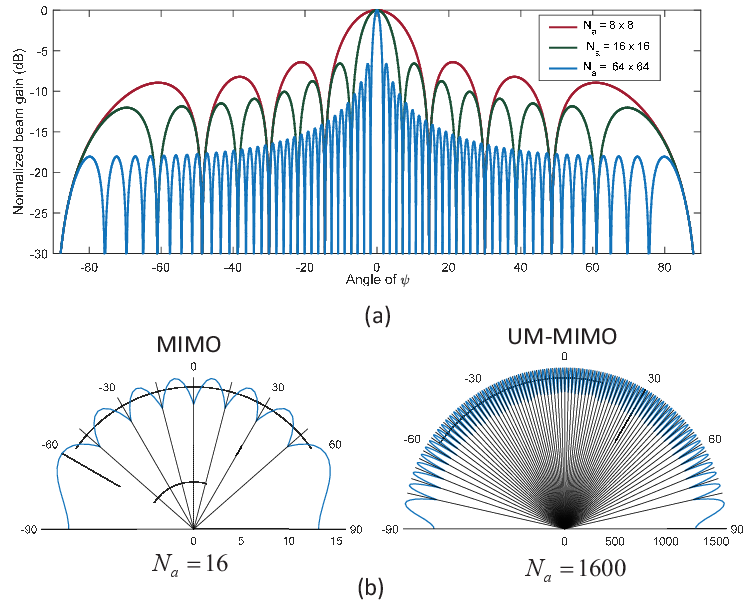}
\caption{(a) The normalized beam pattern of array response vector. (b) Comparison of URPA beam patterns between MIMO and UM-MIMO.}\label{UMM}
\vspace{-12pt}
\end{figure}

It should be mentioned that the array response vector can be set as a beamforming codeword to transmit a beam that covers a dedicated zone, as shown in  Fig. \ref{UMM}(a). With increasing $N_a$, the beam gain is improved, while the beam becomes narrower and thus more beams are needed to cover the whole space, which incurs difficulty in beam alignment and management. For example,  Fig. \ref{UMM}(b) shows the URPA beam patterns at $\theta=0$, with $N_a=16\; (4 \times 4)$ in MIMO systems and $N_a=1600\; (40 \times 40)$ in UM-MIMO systems. The maximum beam gain raises from $16$ to $1600$, while the number of required beams raises from $64 \;(8\times 8)$ to $6400\;(80\times 80)$.  Table \ref{table_link} shows the link budget for some THz scenarios with different antenna geometry.


\begingroup
\renewcommand{\arraystretch}{1.1} 
\begin{table}[]
\caption{Link budget for THz UM-MIMO scenarios.} 
\label{table_link}
\begin{tabular}{|>{\columncolor[HTML]{C0C0C0}}m{2.7cm}|m{1.5cm}<{\centering}|m{1.5cm}<{\centering}|m{1.5cm}<{\centering}|}
\hline
\rowcolor[HTML]{DAE8FC} 
\makecell[c]{\textbf{Cases}}  & \textbf{Kiosk download} & \textbf{Data center} & \textbf{Fronthaul /backhaul} 
\\ \hline
\textbf{Frequency (THz)} 
& $0.14$ & $0.22$ & $0.30$
\\ \hline
\textbf{Bandwidth (GHz)} 
& $5.00$ & $20.00$ & $50.00$
\\ \hline
\textbf{Distance (m)}   
&  $10.00$  & $50.00$  & $200.00$
\\ \hline
\textbf{Propagation loss (dB)}  
& $-95.36$ & $-113.27$ & $-128.00$
\\ \hline
\textbf{Molecules absorption (dB)} 
& $-0.0026$  & $-0.024$ & $-0.44$
\\ \hline
\textbf{\shortstack[l]{Thermal PSD \\ (dBm/Hz)}} 
& $-174$  & $-174$  & $-174$
\\ \hline
\textbf{\shortstack[l]{Thermal noise \\ (dBm)}}  
& $-77.01$ & $-70.99$ & $-67.01$ 
\\ \hline
\textbf{\shortstack[l]{Molecular noise \\ (dBm)}} 
& $-92.36$  & $-103.27$ & $-105.00$
\\ \hline
\textbf{\shortstack[l]{Atmospheric\\ attenuation (dB)}} 
& $0.00$ & $0.00$ & $2.00$
\\ \hline
\beizhu{在较强降水下，天气导致的大气衰减仅为o.1dB/m}
\textbf{Tx antenna size}  
& $16\times16$  & $32\times 32$ & $32\times32$
\beizhu{仅考虑阵列增益即可满足需求}
\\ \hline
\textbf{Tx power (dBm)} 
&  $3.00$  &  $10.00$  & $25.00$
\\ \hline
\textbf{\shortstack[l]{Tx antenna gain\\ (dBi)}} 
&  $24.00$  & $30.00$ & $30.00$
\\ \hline
\textbf{Rx antenna size} 
& $4\times4$ & $32\times32$ & $32\times 32$
\\ \hline
\textbf{Rx antenna gain (dBi)} 
& $12.00$ & $30.00$ & $30.00$
\\ \hline
\textbf{Noise figure (dB)} 
& $8.00$ & $8.00$  & $8.00$
\\ \hline
\textbf{Other losses (dB)}  
& $10.00$  & $10.00$ & $10.00$
\\ \hline
\textbf{SNR (dB)}
& $2.52$  & $9.70$  & $3.57$ 
\\ \hline
\textbf{Data rate (Gbps)}  
& $ 7.39$  &  $67.36$   & $ 85.59$ 
\\ \hline
\end{tabular}
\end{table}
\endgroup

\begin{figure*}[t]
\centering
\includegraphics[width=5.2in]{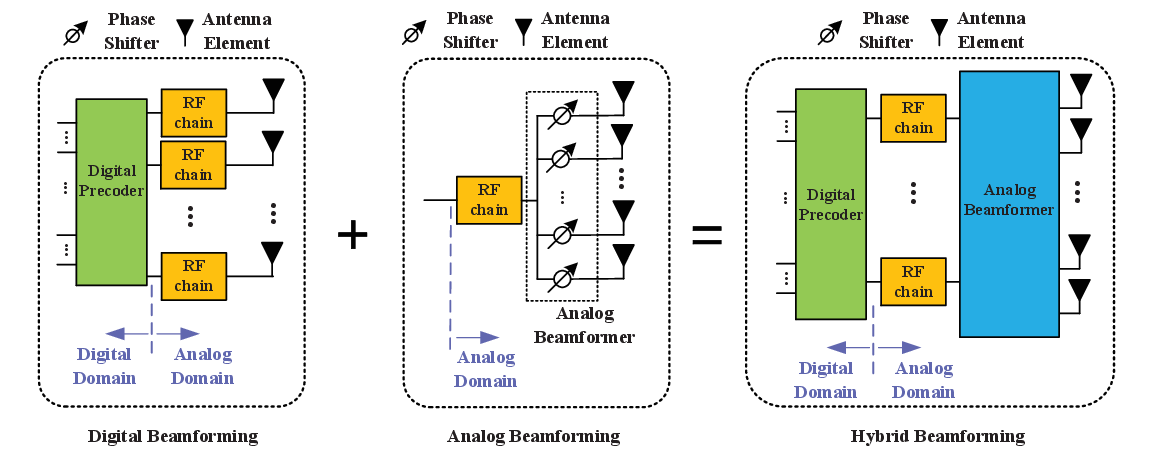}
\caption{Illustration of the hybrid beamforming that combines the digital and analog beamforming.}\label{hba}
\vspace{-6pt}
\end{figure*}

\subsection{Transceiver Architectures}\label{mmar}
In this subsection, we review the progress of transceiver architecture in THz UM-MIMO systems. The earliest beamforming architecture can be traced back to the attempt to construct high directional gain beams using phase shifters in antenna arrays, which is now known as analog beamforming. In analog beamforming architecture, all antennas adjust the phase of the same symbol through the phase shifters in the analog domain \cite{archi1}. Since the phase shifters can not change the magnitude of the symbol, the analog beamforming vector is subject to constant modulus constraint, which limits the flexibility of control and impairs the beamforming performance and capacity improvement. 
In contrast, as a high-cost architecture, digital beamforming or precoding can realize any linear transformation of multiple signal streams from the digital baseband to the antenna elements, which provides more degree of freedom (DoF) \cite{archi2}. Despite the ease of beam control, it is unaffordable to be applied in UM-MIMO systems due to the high power consumption and high system cost \cite{archi3}. 

As a cost-performance trade-off, hybrid beamforming (HB) architecture has emerged as an attractive solution for UM-MIMO systems \cite{archi4, archi6}. The HB can be expressed as a combination of analog beamforming and digital precoder, as presented in Fig. \ref{hba}. HB architecture operates in both the baseband and analog domains, which has been shown to achieve the performance of the digital beamforming in some special cases but with much lower hardware cost and power consumption \cite{archi61, archi7}. There are mainly three types of HB architectures that have been reported extensively: the fully-connected architecture, the partially-connected (or array-of-subarray (AoSA) ), and the dynamically-connected architecture.

\subsubsection{Fully-Connected Architecture}
\begin{figure*}[t]
\centering
\includegraphics[width=7in]{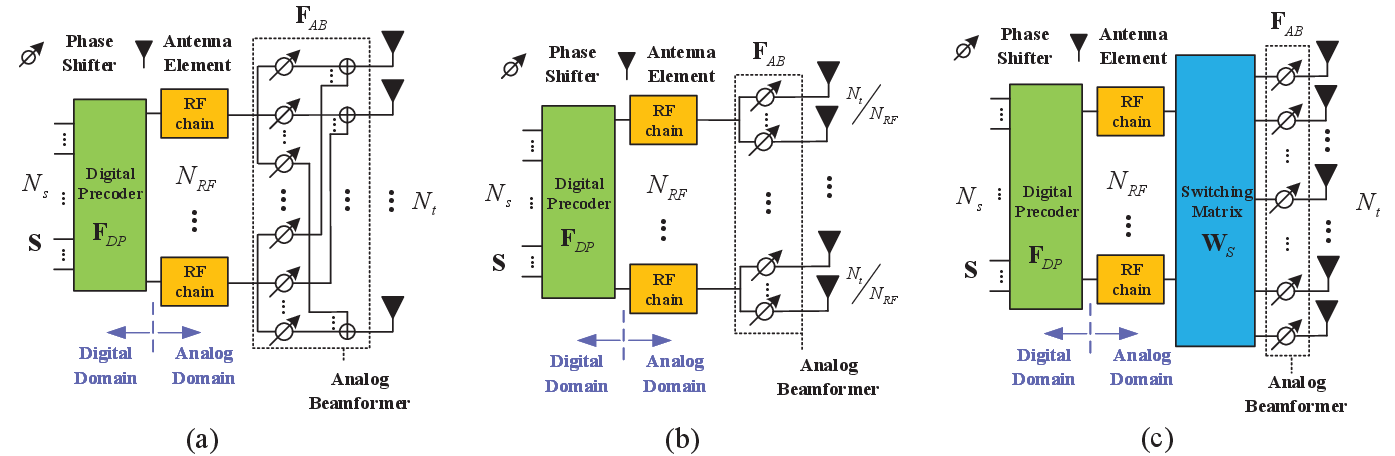}
\caption{Three types of hybrid beamforming architecture: (a) fully-connected (FC); (b) partially-connected (PC); (c) dynamically-connected (DC).}\label{HBfcfig}
\vspace{-6pt}
\end{figure*}

As shown in Fig. \ref{HBfcfig} (a), the fully-connected architecture equipped with $N_{RF}$ RF chains and $N_t$ antenna elements.  The output of each antenna element is the overlapped signals from all RF chains. The beamforming process of fully-connected HB architecture can be expressed as
\begin{equation} \label{hbfc}
{\mathbf{x}} = {{\mathbf{F}}_{\rm{AB}}}{{\mathbf{F}}_{\rm{DP}}}{\kern 1pt}{\kern 1pt}{\mathbf{s}},
\end{equation}
where ${{\mathbf{x}}}\in {\mathbb{C}}^{N_t}$, ${{\mathbf{F}}_{\rm{AB}}}\in {\mathbb{C}}^{N_t \times N_{RF}}$, ${{\mathbf{F}}_{\rm{DP}}} \in {\mathbb{C}}^{N_{RF}\times N_s}$, and ${\mathbf{s}}\in {\mathbb{C}}^{N_s}$ denote the signal emitted by the antenna arrays, analog beamformer, digital precoder, and the transmitted symbol vector, respectively, and $N_s$ is the number of data stream. The elements of each column in matrix ${{\mathbf{F}}_{\rm{AB}}}$ are phase shifters connected by one RF chain, while the elements in each row are phase shifters connected to one antenna port. In the fully-connected architecture, the antenna elements are fully used for every RF chain to provide high beamforming gain with a high complexity of ${N_{t}} \times {N_{RF}}$ RF links (mixer, power amplifier, phase shifter, etc.) \cite{archi8}.

\begingroup
\renewcommand{\arraystretch}{1.2} 

\begin{table*}[]
\caption{Advantages and limitations of different beamforming architectures in the THz UM-MIMO.} 
\label{table_archi}
\begin{tabular}{|
>{\columncolor[HTML]{C0C0C0}}m{0.75cm} >{\columncolor[HTML]{C0C0C0}}m{0.75cm}<{\centering}|m{4.7cm}|m{4.7cm}|m{5.5cm}|}
\hline
\multicolumn{2}{|m{2cm}|}{\cellcolor[HTML]{C0C0C0}\textbf{Beamforming architectures}}
&  {\textbf{Advantages}}
&  {\textbf{Limitations}}
&  {\textbf{Feasibility to THz UM-MIMO}}
\\ \hline

\multicolumn{2}{|c|}{\cellcolor[HTML]{C0C0C0}\textbf{Digital}}
& Optimal performance, extremely high beam flexibility, full DoF.
& Lots of RF chains, extremely high hardware costs, high power consumption.
& Highly feasible but high overhead.
\\ \hline

\multicolumn{2}{|c|}{\cellcolor[HTML]{C0C0C0}\textbf{Analog}}
& The simplest circuits, the lowest power cost, easy to implement.
& Magnitude constraint or quantized bits, only support single stream, poor  beam flexibility.
& Highly feasible but limited DoF.
\\ \hline

\multicolumn{1}{|c|} {\multirow{3}*{\cellcolor[HTML]{C0C0C0}\textbf{HB}}} 
& \multicolumn{1}{c|} {\cellcolor[HTML]{C0C0C0}\textbf{FC}}          
&  Optimal or near-optimal  performance with fewer RF chains, low power consumption.
& Lots of phase shifters, difficult layout, difficult integration.
& Medially feasible.
\\  \cline{2-5} 

\multicolumn{1}{|c|}{\cellcolor[HTML]{C0C0C0}} 
& \multicolumn{1}{c|} {\cellcolor[HTML]{C0C0C0}\textbf{PC}} 
& Fewer phase shifters and fewer circuit lines compared to FC.
& Each stream cannot obtain the array gain from all antennas.
& Highly feasible but low array gain.
\\  \cline{2-5} 
\multicolumn{1}{|c|}{\cellcolor[HTML]{C0C0C0}}
& \multicolumn{1}{c|} {\cellcolor[HTML]{C0C0C0}\textbf{DC}}     
&  Higher beam flexibility than PC.
& Too many switches, difficult layout, difficult integration.
& Medially feasible.
\\ \hline
%
\end{tabular}
\end{table*}
\endgroup

\subsubsection{Partially-Connected Architectures}
The partially-connected architecture is also known as AoSA. As shown in Fig. \ref{HBfcfig} (b), each RF chain is connected to a unique subset of antenna elements \cite{archi9}, i.e., a subarray.
With this architecture, ${{\mathbf{F}}_{AB}}$ in (\ref{hbfc}) has the form of a block diagonal matrix as ${{\mathbf{F}}_{AB}} = \mathrm{diag}\left[ {{{\mathbf{f}}_1}, \ldots ,{{\mathbf{f}}_{{N_{RF}}}}} \right]$, where $\{{\mathbf{f}}_i\}_{i=1}^{N_{RF}}$ are $\frac{{{N_t}}}{{{N_{RF}}}}\times 1$ phase shifter vectors connected to the RF chains. ${{\mathbf{F}}_{AB}}$ can also be expressed as
\begin{equation} \label{hbpc}
\begin{aligned}
&{{\mathbf{F}}_{AB}} = {{\mathbf{\Phi }}_{AB}}{\mathbf{M}}\\
&{\kern 1pt} {\kern 1pt} {\kern 1pt} {\kern 1pt} {\kern 1pt} {\kern 1pt} {\kern 1pt} {\kern 1pt} {\kern 1pt} {\kern 1pt} {\kern 1pt} {\kern 1pt} {\kern 1pt} {\kern 1pt} {\kern 1pt} {\kern 1pt} {\kern 1pt} {\kern 1pt} {\kern 1pt}  = \left[ {\begin{array}{*{20}{c}}
{{\phi _1}}& \cdots &0\\
 \vdots & \ddots & \vdots \\
0& \cdots &{{\phi _{{N_t}}}}
\end{array}} \right]
\left[ {\begin{array}{*{20}{c}}
{\mathbf{m}_1}& \cdots &{\mathbf{0}}\\
 \vdots  & \ddots & \vdots \\
{\mathbf{0}}& \cdots &{\mathbf{m}_{N_{RF}}}
\end{array}} \right],
\end{aligned}
\end{equation}
where $\mathbf{M}$ can be regarded as a switch matrix of dimension $N_t \times N_{RF}$, ${\mathbf{m}_i}$ is an all-one column vector, which means the state of corresponding phase shifters are on. The diagonal matrix ${{\mathbf{\Phi }}_{AB}}$ represents the phase shifters connected to the antenna elements, where ${\phi _i} = {e^{j{\theta _i}}}$ is the phase shift factor of the $i$-th phase shifter.
Compared with the fully-connected architecture, the partially-connected architecture further reduces the hardware cost to $N_{t}$ links \cite{archi8,archi10}, which is much appealing in THz communications\cite{archadd1, archadd2}.  In particular, the authors in \cite{archadd3} proposed a novel partially-connected HB architecture with two digital beamformers, wherein the additional one is developed to compensate for the performance loss caused by frequency selective fading.

\subsubsection{Dynamically-Connected Architecture}
One disadvantage of the partially-connected architectures is the fixed circuit connection, which prevents adaptive and dynamic control \cite{archi11}. To improve the beam flexibility, a dynamically-connected HB architecture was reported in \cite{archi11, archi12, archi13, archi14}. As shown in Fig. \ref{HBfcfig} (c),  a switching network $W_s$ is added between the RF chains and the antenna elements \cite{archi11}, and the transmit signal can be written as
\begin{equation} \label{hbdc}
{\mathbf{x}} = {{\mathbf{\Phi }}_{\rm{AB}}}{{\mathbf{W}}_{\rm{S}}}{{\mathbf{F}}_{\rm{DP}}}{\kern 1pt}{\kern 1pt}{\mathbf{s}},
\end{equation}
where ${{\mathbf{\Phi }}_{\rm{AB}}}$ follows the definition in (\ref{hbpc}), ${{\mathbf{W}}_{\rm{S}}} = \left\{ {{w_{i,j}}} \right\},i = 1, \ldots ,{N_t},j = 1, \ldots ,{N_{RF}}$, is a Boolean matrix, and $w_{i,j}$ represents the switch from the $i$-th antenna element to the $j$-th RF chain.  It is worth noting that since each antenna element is only allowed to connect to one RF signal at a time, ${{\mathbf{W}}_{\rm{S}}}$ should satisfy constraint $\sum\nolimits_j {{w_{i,j}}} = 1$. While maintaining the low-cost advantage of partial connection architecture, dynamic connection architecture improves processing freedom through the flexible switch network, which can be regarded as the transition architecture from full connection to partial connection.  In particular, the authors in \cite{archi15} proposed a dynamic AoSA (DAoSA), where each antenna is connected to $N_{RF}$ phase shifters, claiming to save power compared to the fully-connected HB architecture. From the perspective of hardware structure, the DAoSA is more sophisticated than the fully-connected HB architecture, which may not be feasible for UM-MIMO systems.  In addition, if we expect to balance the energy cost and performance between fully connected architecture and AoSA, a straightforward way is to turn off some phase shifters, rather than adding switches (which may cost additional power). We summarize the advantages and limitations of above architectures for THz UM-MIMO in Table \ref{table_archi}.

\begin{table*}[]
\caption{Term interpretation of beamforming/combining, precoding/decoding, and beam steering.} \label{clas}
\begin{tabular}{|>{\columncolor[HTML]{C0C0C0}}m{3cm}|m{4.5cm}<{\centering}|m{4.5cm}<{\centering}|m{4.5cm}<{\centering}|}
\hline
\rowcolor[HTML]{C0C0C0} 
\makecell[c]{\textbf{Term}}  & \textbf{Beamforming/combining} & \textbf{Precoding/decoding} & \textbf{Beam steering} 
\\ \hline
\textbf{Processor name} 
& Beamformer/combiner & Precoder/decoder & Beamformer/combiner
\\ \hline
\textbf{Digital domain} 
& \checkmark &  \checkmark &  $\times$
\\ \hline
\textbf{Analog domain} 
& \checkmark &  $\times$ &  \checkmark
\\ \hline
\textbf{Data stream} 
& $N_s \ge 1$  & $N_s \ge 1$ & $N_s = 1$
\\ \hline
\textbf{Codebook} 
& Not necessary & Not necessary & Necessary 
\\ \hline
\textbf{Applications}   
&  MIMO and radar system  & MIMO system  & Radar system
\\ \hline
\textbf{Main advantages} 
& Pattern diversity $\&$ Spatial multiplexing  & Pattern diversity $\&$ Spatial multiplexing   & Pattern diversity 
\\ \hline

\end{tabular}
\end{table*}


\section{Beamforming Principles}
In this section, we introduce the beamforming principles for THz UM-MIMO systems. To begin with, we specify some terms that are usually used in beamforming literature. Next, we endeavor to visualize how to steer a beam to desired directions via multiple antenna elements and unveil some important ideas behind it.  Finally, we discuss the far-field and near-field assumptions in THz UM-MIMO systems.

\subsection{Nomenclature}
Considering a point-to-point system in (\ref{system}), the received signal after spatial post-processing can be expressed as $\widetilde {\bf{y}} = {{\bf{W}}^H}{\bf{HFs}} + {{\bf{W}}^H}{\bf{n}}$,
in which ${{\mathbf{F}}}\in {\mathbb{C}}^{N_t \times N_s}$ is the \emph{beamformer} and ${{\mathbf{W}}}\in {\mathbb{C}}^{N_r \times N_s}$ is the \emph{combiner}. The processing from the data stream $\bf{s}$ to the transmit signals $\bf{x=Fs}$ is called \emph{beamforming}. On the contrary, the processing from the received signals ${\bf{y}}$ to the data stream $\widetilde {\bf{y}}={{\bf{W}}^H\bf{y}}$ is called \emph{combining}. This process can be applied in the digital domain (e.g. by FPGA), or analog domain (e.g., by phase shifters), or in both domains. Thus, the beamforming technologies can be specified as digital beamforming, analog beamforming, and hybrid beamforming.

Rigorously, \emph{precoding} is referred to as the digital case of beamforming, wherein the combining of this case is named \emph{decoding}. The processor for precoding and combining is called \emph{precoder} and \emph{decoder}, respectively. For ease of presentation, most papers do not strictly distinguish the terms beamforming and precoding in the digital domain, i.e., digital beamforming is equivalent to precoding. \emph{beam steering} is considered as a special case of analog beamforming with $N_s\!=\!1$. In particular, in beam steering schemes,  each beam is realized by an array response vector and the signal energy is steered to a specific direction\cite{steer1}. The features of precoding/decoding, beamforming/combining, and beam steering  are summarized in Table \ref{clas}.





\subsection{Beam Steering}\label{bpaa}
As a classic and basic technology to realize beamforming, the phased array has been well developed over decades. A phased array can steer the beam electronically in different directions, without moving the antennas \cite{vhj}. Specifically, the power from the RF chain is fed to the antenna elements via a phase shifter on each. By this means, the radio waves from the separate antennas add together to increase the radiation in the desired direction, while suppress radiation in undesired directions. If phased arrays are employed at both transmitter and receiver, the resulting signal can be written as 
\begin{equation}\label{y19}
y = {{\mathbf{w}}^H}{\mathbf{Hf}}s + {{\mathbf{w}}^H}{\mathbf{n}},
\end{equation}
where ${\mathbf{f}}\in \mathbb{C}^{N_t\times1}$ is the beamformer with $N_t$ phase shifters, ${\mathbf{w}}\in \mathbb{C}^{N_r\times1}$ is the combiner with $N_r$ phase shifters, and $s$ is a data symbol. The phase shifter only adjusts the phase without changing the amplitude of signal, which is subjected to a constraint, i.e., $\left| {{\mathbf{f}}(i)} \right| =\sqrt{ \frac{P}{{{N_t}}}} $ and $P$ is the total transmit power.  Since there is only one RF chain, the beamforming via phased array is commonly known as \emph{analog beamforming}. 

\begin{figure}[t]
\centering
\includegraphics[width=2.5in]{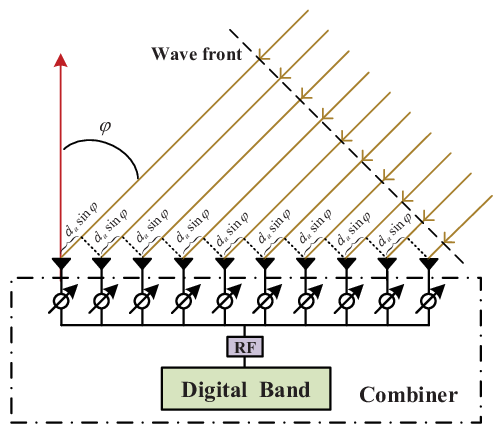}
\caption{Incoming signals with arrival angle $\varphi$.}\label{ula}\vspace{-2pt}
\end{figure}
\indent Consider a LoS THz channel given in (\ref{losa}). It can be verified that the received signal vector has a form of array response vector, i.e., 
\begin{equation}\label{y20}
{\mathbf{y}} ={\mathbf{Hf}}s= a \cdot {{\mathbf{a}}_r},
\end{equation}
where $a$ is a complex constant. In Section \ref{ed}, we straightforwardly present the expressions of ${{\mathbf{a}}_r}$ for different types of arrays. Here, we focus on a simple example, i.e., ULA, to illustrate the relation between the mathematical expression (\ref{ula0}) and its physical mechanism.

Fig. \ref{ula} plots the incoming signal wave to the receiver, in which $d_a$ is the antenna space and $\varphi$ is the arrival angle. It is obvious that for the same wavefront, the rightmost element receives it first, and the leftmost element receives it last. The wave-path difference between adjacent elements is ${d_a}\sin \varphi$. As the distance increases $\lambda$, the phase increases $2\pi$. Thus, the phase difference between  adjacent elements is 
\begin{equation}\label{ph}
\frac{{2\pi {d_a}\sin \varphi }}{\lambda }.
\end{equation}
As such, if the received signal at the first element is $c_r \in \mathbb{C}$, the received signal at the $n$-th element is 
$c_r \cdot {e^{-j\frac{{(n - 1)2\pi {d_a}\sin \varphi }}{\lambda }}}.$
As a result, the normalized array response vector for the receive ULA can be written as 
\begin{equation}\label{ar}
{{\mathbf{a}}_r}(\varphi ) = \frac{1}{{\sqrt {{N_r}} }}{\left[ {1,{e^{ - jk{d_a}\sin \varphi }},...,{e^{ - jk{d_a}(N_r - 1)\sin \varphi }}} \right]^T},
\end{equation}
where $k = 2\pi /\lambda $. Next, we study how to use the phased array to combine the received signals. Substituting (\ref{y20}) into (\ref{y19}), we aim to find the optimal combiner to maximize the resulting power, which is equivalent to 
\begin{equation*}\label{y34}
\begin{aligned}
{\rm{P}}(1):\; &\mathop {\max }\limits_{\mathbf{w}} {\left| {{{\mathbf{w}}^H}{{\mathbf{a}}_r}(\varphi )} \right|^2}\\
&{\mathrm{s}}{\mathrm{.t}}{\mathrm{.}}\;\;\;\left| {{\mathbf{w}}(i)} \right| = \frac{1}{{\sqrt {{N_r}} }},\;\;i = 1,2,...,{N_r}.
\end{aligned}
\end{equation*}
It is easy to verify that an optimal solution is given by
\begin{equation}\label{y35}
{\mathbf{w}} = {{\mathbf{a}}_r}(\varphi ),
\end{equation}
The combining process can be regarded as an inverse beamforming. As ${{\mathbf{a}}_r}(\varphi )$ can maximumly combine the signal from the direction of  $\varphi$, we can also refer to this process as receiving a narrow beam in the direction of $\varphi$\cite{mrcode}.
\begin{figure}[t]
\centering
\includegraphics[width=2.5in]{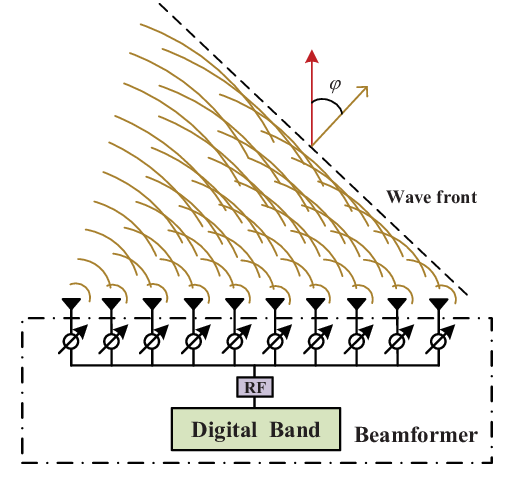}
\caption{Steering beam with departure angle $\varphi$.}\label{ula2}\vspace{-2pt}
\end{figure}

The array response vector in (\ref{ar}) represents the signal coming from the direction $\varphi$. Vise versa, if the array transmits signal by using the array response vector in (\ref{ar}) as beamformer, the wavefront moves in the direction of $\varphi$. To be exact, as shown in Fig. \ref{ula2}, the signal waves in the direction of $\varphi$ have the same phase and will add together to increase the radiation power. Generally, the antenna space is considered to be half-wavelength, i.e., ${d_a} = \lambda /2$, to reduce the self-interference. For ease of expression, we assume that ${N_t} = {N_r} = {N_a}$, then the array response vector at both the transmitter and the receiver can be unified as
\begin{equation}\label{arr}
{\mathbf{a}}(\varphi ) = \frac{1}{{\sqrt {{N_a}} }}{\left[ {1,{-e^{j\pi \sin \varphi }},...,{e^{-j\pi (N_a - 1)\sin \varphi }}} \right]^T}.
\end{equation}

\subsection{Far-field and Near-field}
\begin{figure}[t]
\centering
\includegraphics[width=2.5in]{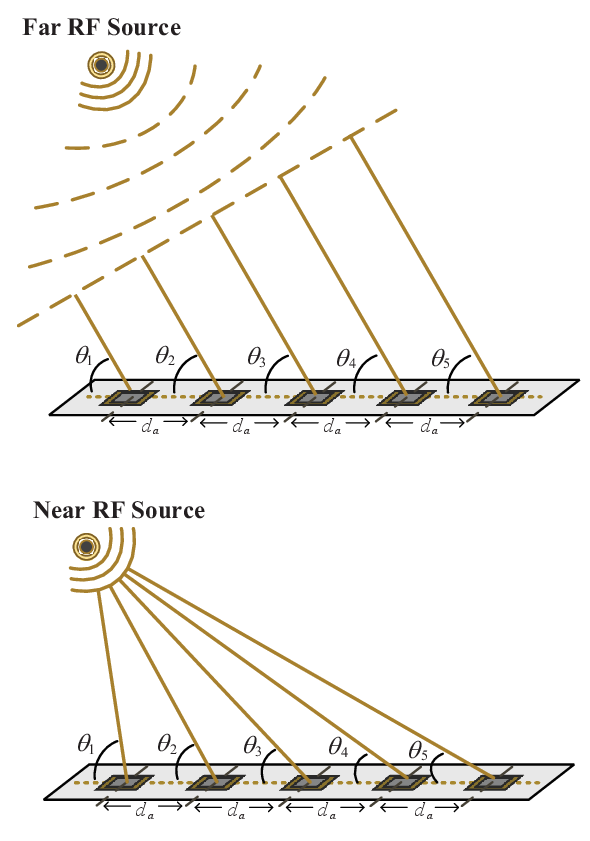}
\caption{Illustration of the near field and the far field.}\label{far}\vspace{-2pt}
\end{figure}

We should mention that all the array response vectors discussed above are based on an essential assumption, that is, the wavefront is the flat plane and all the elements have the same AoAs and AoDs. This assumption holds when the RF source is far away from the receiver. To illustrate this point intuitively, Fig. {\ref{far}} plots the radiation cases of near-field and far-field. As can be seen, when the RF source is far away, the large radius of the spherical wavefront results in wave propagation paths that are approximately parallel. As such, we have ${\theta _1}={\theta _2}=... = {\theta _5}$. With the near RF source, the AoA varies for each element, i.e., ${\theta _1}>{\theta _2}>...>{\theta _5}$.  In other words, the far-field array response vector is inaccurate if applied to the near-field scenarios, or cannot maximally combine energy from a near-field source. Thus, an interesting question is when can we make the far-field assumption? In general, the boundary between near-field and far-field is given by the Rayleigh distance (RD) \cite{ca}, i.e.,
\begin{equation}\label{df}
{\rm{RD}}= \frac{{2{D^2}}}{\lambda },
\end{equation}
where $D = ({N_a} -1)d_a$ is the side length of the antenna array. That is to say, with the \emph{same aperture of antenna array}, the RD for THz communication could be quite large due to its extremely small wavelength $\lambda$. As the THz frequency is thousands of orders higher than microwave frequency, it seems that the RD could be many kilometers. Is it true?

In fact, from another perspective, the THz antenna elements are commonly packed in a small footprint as the element spacing holds $d_a=\lambda/2$. As such, by substituting $D = (N_a - 1)\lambda /2$, the RD can be rewritten as
\begin{equation}\label{de}
{\rm{RD}} = D\left( {N_a} - 1 \right) = {\left( {N_a} - 1 \right)^2}\frac{\lambda }{2},
\end{equation}
which indicates that with the \emph{same number of antenna elements}, the RD for THz communication could be quite small.  Compared to the mmWave MIMO, THz UM-MIMO has more antennas but a smaller aperture of the array. Thus, there are two opposite effects on RD. Moreover, although the far-field array response model is inaccurate when the distance of the source is slightly less than RD, the array gain may not lose much if  the far-field beam is still used. In sight of this, the authors in \cite{ERD} derived the effective Rayleigh distance (ERD), to show the boundary at which the array gain loss of the far-field beam is $5\%$.  The ERD is smaller than conventional RD and related to the AoA $\varphi$, which is given by
\begin{equation}
{\rm{ERD}} = 0.367{\cos ^2}\varphi \frac{{2{D^2}}}{\lambda }.
\end{equation}

\begin{table}[t]
  \begin{center}
    \caption{Some representative RD and ERD in THz UM-MIMO systems}\label{erd}
    \begin{tabular}{c|ccc}
    \toprule
     Array size & Antenna number & RD & ERD \\
    \midrule
     $10\times 10$ & 100 & 0.04 m & $\leq 0.15$ m\\
     $50\times 50$ & 2500 & 1.20 m & $\leq 0.44$ m\\
     $100\times 100$ & 10000 & 4.9 m & $\leq 1.80$ m \\
    \bottomrule
  \end{tabular}
  \end{center}
 \vspace{-2pt}
\end{table}
Table \ref{erd} presents some representative values in terms of RD and ERD in THz UM-MIMO systems, where the operation frequency is $0.3$ THz. It can be seen that even using $10,000$ antenna elements, the far-field beamforming is effective (with performance loss less than $5 \%$) when the communication distance is more than $1.8$ m. Thus, the far-field beamforming is still suitable for most scenarios while the near-field beamforming can be a supplement for a few scenarios.

\section{Beam Training}\label{BT}
In this section, we introduce a beam training scheme that can realize the beamforming in THz UM-MIMO systems without using the CSI. To start with, the basic concept of beam training is introduced. Then, two essential components in the beam training, i.e., training protocol and codebook design, are discussed, subsequently. Finally, we briefly introduce the lens antenna array to show a special manner to realize beam training.

\subsection{Basic Concept}
To enable a beam-connected wireless communication, the beamformer ${\mathbf{f}}$ and the combiner ${\mathbf{w}}$ need to be optimized to maximize the decoding SNR, which is equivalent to 
\begin{equation*}
\begin{aligned}
{\rm{P}}(2):\;\;\{ {\mathbf{w}}^{{\mathrm{opt}}},{\mathbf{f}}^{{\mathrm{opt}}}\}  = \arg &\max {\left| {{\mathbf{w}}^H{{\mathbf{H}}}{{\mathbf{f}}}} \right|^2}\\
&{\mathrm{s}}{\mathrm{.t}}{\mathrm{.}}\;\;{\left\| {{{\mathbf{f}}}} \right\|^2} = 1,\;\;{\left\| {{{\mathbf{w}}}} \right\|^2} =1.
\end{aligned}
\end{equation*}
Provided that ${{\mathbf{H}}}$ is  known at the transmitter and receiver by channel estimation, the optimal beamformer ${\mathbf{f}}^{{\mathrm{opt}}}$ and the combiner ${\mathbf{w}}^{{\mathrm{opt}}}$ can be obtained by applying the singular value decomposition (SVD) on ${{\mathbf{H}}}$. However, the conventional channel estimation methods may not apply to THz UM-MIMO systems due to the unaffordable operation complexity on large-scale array. Besides, the beam pattern of conventional pilot signals is almost omni-directional, which suffers from severe path loss over THz channels and may not be effectively detected by the receiver. Regarding the above issues, an interesting question is whether we can find ${\mathbf{f}}^{{\mathrm{opt}}}$ and ${\mathbf{w}}^{{\mathrm{opt}}}$ without the need of  CSI?

The beam training scheme provides a promising solution \cite{tran1,tran2,tran3}. Since the LoS path is dominant in THz channel, we can replace ${{\mathbf{H}}}$ in ${\rm{P}}(2)$ with the LoS channel in (\ref{losa}), i.e.,
\begin{equation*}
{\rm{P}}(3):\;\;\{ {\mathbf{w}}^{{\mathrm{opt}}},{\mathbf{f}}^{{\mathrm{opt}}}\}  = \arg \max {\left| {{{\mathbf{w}}^H} {{\mathbf{a}}_r}{\mathbf{a}}_t^H{\mathbf{f}}} \right|^2},
\end{equation*}
where the optimal solution is given by the array response vectors, e.g., ${\mathbf{w}}^{{\mathrm{opt}}}\!=\!{{\mathbf{a}}_r}$ and ${\mathbf{f}}^{{\mathrm{opt}}}\!=\!{{\mathbf{a}}_t}$.  Due to this fact, the feasible region of ${\mathbf{f}}$ and ${\mathbf{w}}$ can be reduced from a vector space to an AoA/AoD space. Thus, we can predefine a codebook ${\mathcal{F}}$  that includes many antenna response vectors representing the narrow beams corresponding to different AoAs/AoDs \cite{steer1,steer2,steer3}.  For ULA, the beamforming codebook can be written as 
\begin{equation}
{\mathcal{F}} = \left\{ {{{\mathbf{a}}}(\varphi _1),{{\mathbf{a}}}(\varphi _2),...,{{\mathbf{a}}}(\varphi _N)} \right\},
\end{equation}
where $N$ is the number of beams.  Despite ${\varphi } \in [0,2\pi )$, it suffices to consider the $N$ beams only within $[ - \frac{\pi }{2},\frac{\pi }{2}]$ (equivalent to $[0,\frac{\pi }{2}] \cup [\frac{{3\pi }}{2},2\pi )$) due to the following lemma \cite{tran4}.
\begin{lemma}
Beams within $[ - \frac{\pi }{2},\frac{\pi }{2}]$ and beams within $[ \frac{\pi }{2},\frac{3\pi }{2}]$ are isomorphic for ULA. In particular, the narrow beam in direction of $\varphi$ is equivalent to that in direction of $\pi-\varphi$, i.e.,
\begin{equation}
{{\mathbf{a}}}(\varphi)= {{\mathbf{a}}}(\pi-\varphi ).\label{lem1}
\end{equation}
\end{lemma}

\begin{figure}[t]
\centering
\includegraphics[width=3.5in]{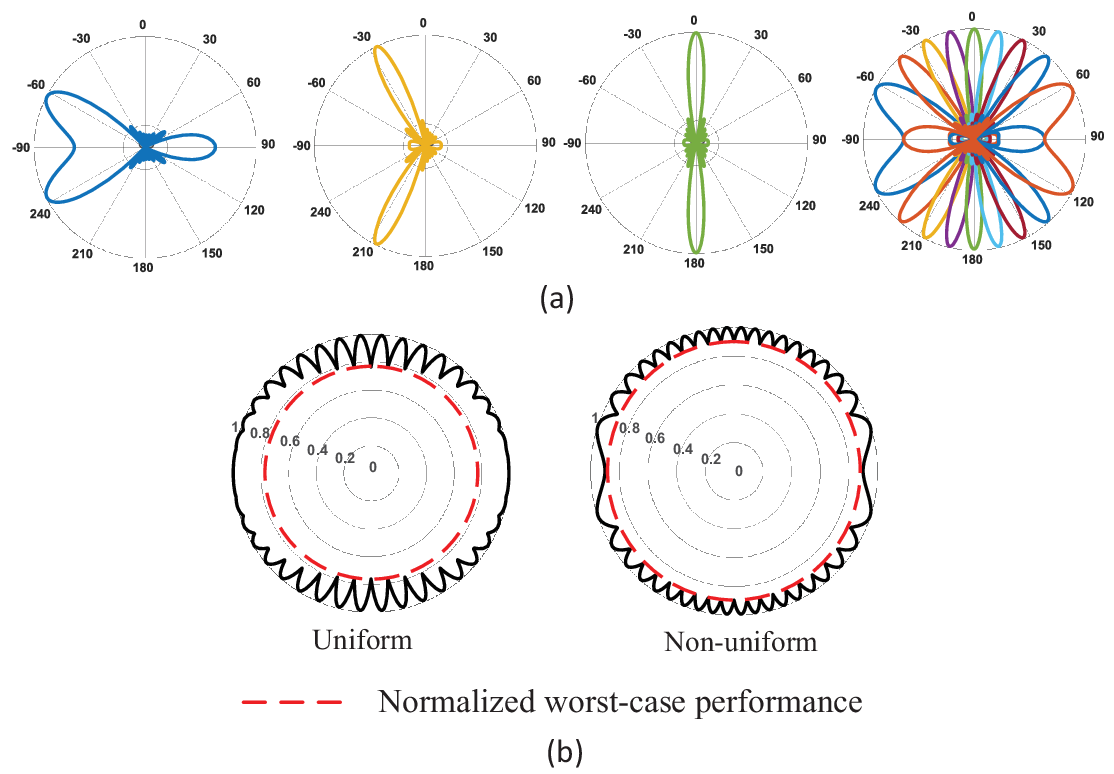}
\caption{ (a) The narrow beams' pattern (with different color representing different beams) when $N_a=10$. (b) Comparison between uniformly and non-uniformly distributed beams.} \label{b4}
\vspace{-2pt}
\end{figure}

As can be seen from Fig. \ref{b4}(a), each beam shows the symmetrical radiation patterns within the range $[ - \frac{\pi }{2},\frac{\pi }{2}]$ and $[\frac{\pi }{2},\frac{3\pi }{2}]$, which validates Lemma 1. Moreover, the beam is narrower with the AoA/AoD around $0$  and is wider with the AoA/AoD around $ \pm \frac{\pi }{2}$.  This implies that with the same number of beams, the codebook with non-uniformly distributed beams may achieve a higher worst-case performance than that with uniformly distributed beams (see Fig. \ref{b4}(b)). The optimal distribution of the beams' AoA/AoD that maximize the worst-case performance is uniform distribution in sine space \cite{3DB2}.

\begin{figure}[t]
\centering
\includegraphics[width=3.3in]{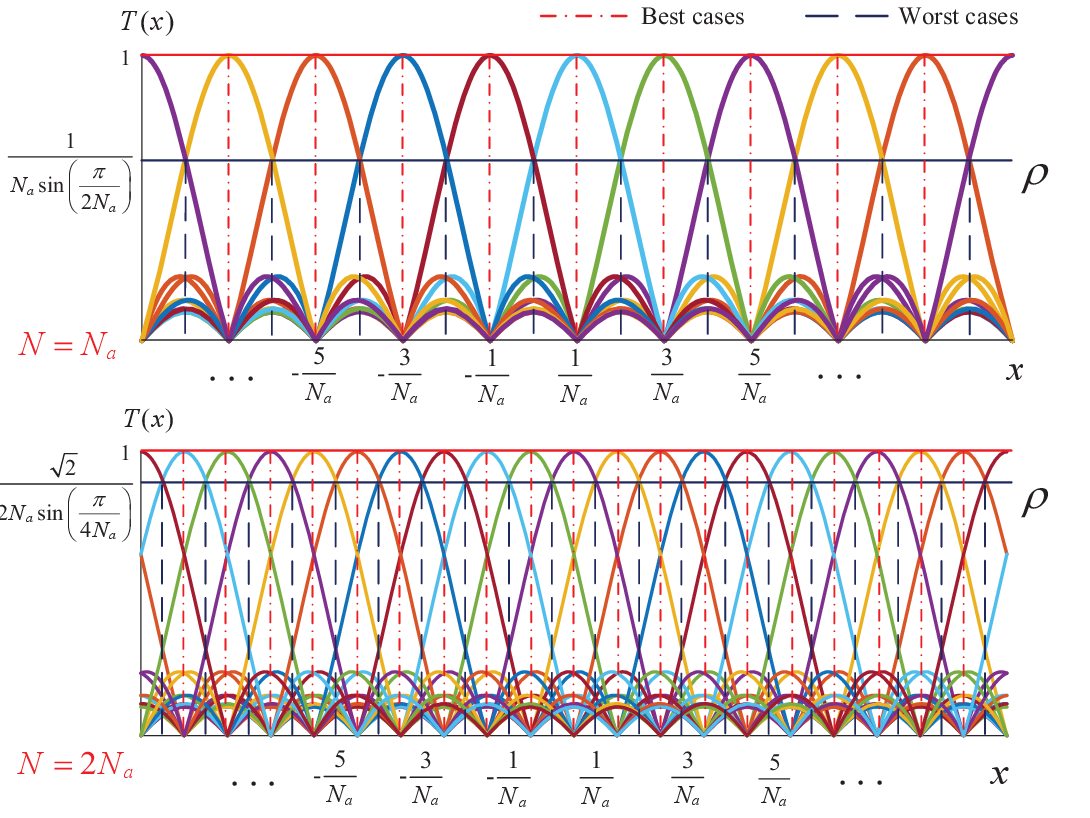}
\caption{$N=N_a$ beams and $N=2N_a$ beams covering the whole space.}\label{NNN}\vspace{-2pt}
\end{figure}

The larger the codebook size,  the closer the performance of beam training is to that of the optimal beamforming.  Generally, the number of beams $N$ required in the codebook is proportional to the number of array antennas $N_a$. Fig. \ref{NNN} shows the cases of $N=N_a$ beams and $N=2N_a$ beams covering $[ - \frac{\pi }{2},\frac{\pi }{2}]$, where the AoA/AoD of the beams are uniformly distributed in sine space. The normalized worst-case performance of the above two cases is given by\footnote{It can be seen from Fig. \ref{NNN} that the best case of the beam training is that the path angle exactly coincides with the beam center, and the worst case is that the path angle lies on the intersection of the beams. } 
\begin{equation}
\rho {\mathrm{ = }}\left\{ {\begin{aligned}
&{\frac{1}{{{N_a}\sin \left( {\frac{\pi }{{2{N_a}}}} \right)}},\;\;\quad N = {N_a}\;}\\
&{\frac{{\sqrt 2 }}{{2{N_a}\sin \left( {\frac{\pi }{{4{N_a}}}} \right)}},\quad N = 2{N_a}\;\;}
\end{aligned}} \right. .
\end{equation}

The above process is referred to as \emph{exhaustive beam training}. Generally, the beam training strategy includes training protocol and codebook design.  The protocol of the exhaustive beam training is the simplest, that is, exhaustively testing the narrow beam pairs of the transmitter and the receiver. The codebook consists of array response vectors, where the AoAs or AoDs are uniformly distributed in the sine space. While other beam training strategies requires more complicated training protocol and codebook design, which are discussed in the next subsections.

\begin{remark}
Sometimes, the concept of beam training is confused with beam alignment. Commonly, the beam alignment is to find a wide-beam pair to initialize a reliable connection, while the beam might be further optimized before transmitting data \cite{al1,al2,al3}. Whereas, beam training focuses on seeking the strongest narrow-beam pair and then directly using this beam pair for data transmission.
\end{remark}

\subsection{Training Protocol}
The training protocol determines the training reliability and complexity. Even though the exhaustive search protocol yields high reliability, it incurs the training complexity of $N^2$ and thus is quite time-consuming in UM-MIMO systems.  Compared to the exhaustive search, some other training protocols are more appealing owing to their lower training complexity.

For examples,  IEEE 802.11ad proposed an one-side search protocol \cite{hbt2}, where each user exhaustively searches the beams in the codebook while the BS transmits the signal in an omnidirectional mode, which incurs the complexity of $2N$. The authors in \cite{hbt3} proposed a parallel search protocol which uses $N_{RF}$ RF chains at BS to transmit multiple beams simultaneously while all users exhaustively search the beams, which incurs the complexity of $\left. N^2 \middle/ N_{RF} \right.$. The authors in \cite{hbt1} proposed an adaptive search and the complexity is given by $4\log _2N$.  The authors in \cite{archadd2} proposed a two-stage training scheme that combines sector level sweeping and fine search, which results in the complexity of $\left. N^2  \middle/Q \right. + Q$, where $Q$ is the number of narrow beams covered in each sector level. To the best of our knowledge, the most efficient protocol is \emph{$M$-tree search}\cite{tran1}, which is widely adopted in the hierarchical beam training.

\begin{figure}[t]
\centering
\includegraphics[width=3.4in]{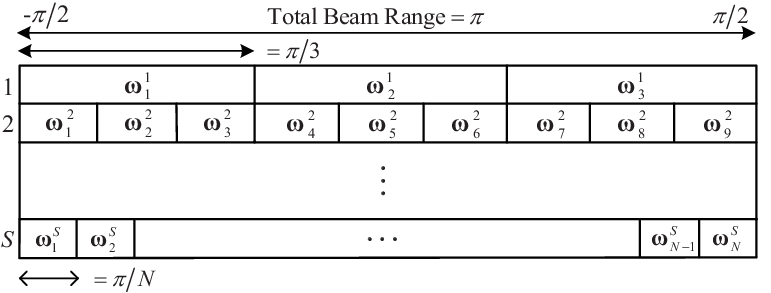}
\caption{A feasible zone division of ternary-tree search.}\label{fig3t}\vspace{-2pt}
\end{figure}

\begin{figure}[t]
\centering
\includegraphics[width=3.5in]{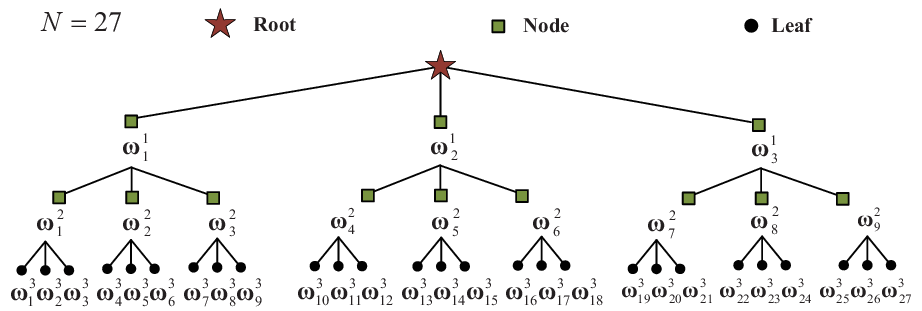}
\caption{Diagram of the ternary-tree search when $N=27$.}\label{t27}\vspace{-5pt}
\end{figure}  

\begin{figure}[t]
\centering
\includegraphics[width=3.3in]{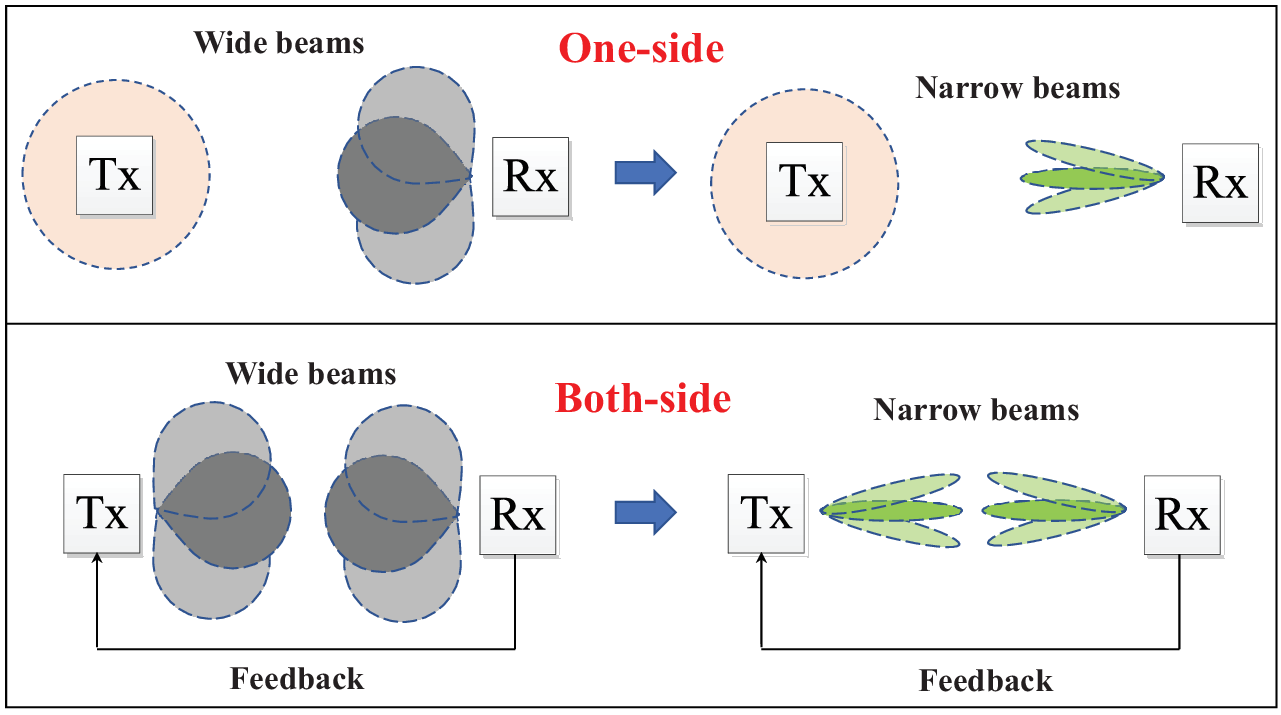}
\caption{Illustration of the one-side and both-side hierarchical search.}\label{obo}\vspace{-2pt}
\end{figure}

In the $M$-tree search, we search the beams stage by stage with decreasing beam width. Specifically, there are $S={\log _{M}}N$ stages and the $s$-th stage has $M^s$ beams. Fig. \ref{fig3t} shows a feasible zone division of the ternary-tree codebook,\footnote{It is worth mentioning that the zone division is a part of the protocol design and the uniform space division in Fig. \ref{fig3t} is not optimal.} i.e., $M=3$, where ${\bm{\omega }}_n^s$ denotes the $n$-th beam in the $s$-th stage. As can be seen, each wide beam exactly covers three narrower beams in the next stage.  Fig. \ref{t27} illustrates the diagram of the tree search. Specifically, we start with using an omnidirectional beam (root) for initial detection. Then, in each stage of the $M$-tree search, we find and follow the best beam (node) for the next stage search, until the best narrow beam (leaf) is found. It is worth mentioning that the $M$-tree search can be implemented on one side or both sides, as shown in Fig. \ref{obo}. In the one-side tree search \cite{tran1}, we fix the transmitter to be in an omnidirectional mode and run an $M$-tree search stage by stage to find the best receive narrow beam. And then we fix the receiver to be in a directional mode with the found best receive narrow beam, and then run the $M$-tree search stage by stage to find the best transmit narrow beam. Thus, the complexity of one-side $M$-tree search is given by
\begin{equation}\label{ones}
{T_{{\mathrm{one}}}} =M{\log _M}N+M{\log _M}N=2M{\log _M}N.
\end{equation}
In the both-side tree search \cite{hbt1}, we realize the beam training by selecting \emph{beam pairs} stage by stage with decreasing beamwidth, where the receiver determines the best pair in each stage and feedback to the transmitter for the search (within the last selected range) in the next stage. Thus, the complexity of both-side $M$-tree search is given by
\begin{equation}\label{boths}
{T_{{\mathrm{both}}}} =M^2{\log _M}N.
\end{equation}
Notice from (\ref{ones}) and (\ref{boths}) that the complexity of the  one-side $M$-tree search is less than that of the both-side $M$-tree search when $M\neq 2$, and the complexity is the same when $M=2$. Besides, the both-side $M$-tree search needs $S$ feedbacks while the one-side  $M$-tree search does not need any feedback.  In particular, the ternary-tree search has the lowest searching complexity among all $M$-tree  search schemes\cite{tran4}.  Table \ref{proto} summarizes the complexity of different training protocols. 

\begingroup
\renewcommand{\arraystretch}{1.2} 
\begin{table}[t]
\centering
\caption{Comparison of the beam training protocols.}\label{proto}
\vspace{-5pt}
\begin{tabular}{|c|c|}
\hline
Training Protocol &\tabincell{c}{Training complexity} \\ \hline
Exhaustive search   & ${N^2} $      \\ \hline
One-side search \cite{hbt2}   &  $2N$\\ \hline
Adaptive search \cite{hbt1}     & $4\log _2N$ \\ \hline
Parallel search \cite{hbt3}  & $\left. N^2 \middle/ N_{RF} \right.$ \\ \hline
Two-stage search \cite{archadd2}  &  $\left. N^2  \middle/Q \right. + Q$ \\ \hline
One-side $M$-Tree search   &  $2M\log _MN $ \\ \hline
Both-side $M$-Tree search    &  $M^2\log _MN $ \\ \hline
\end{tabular}
\end{table}
\endgroup

\subsection{Codebook Design}
Codebook design aims to optimize the weights of each codeword to make its beam pattern cover the zone specified by the protocol.  For example, in the $M$-tree search, the narrow beams in the bottom stage can be set as array response vectors whereas the wide-beam design is an open problem. 

Given a codeword $\bf{w}$, its beam pattern is characterized by the magnitude (or power) of the array factor, i.e.,
\begin{equation}\label{y48}
g(\phi ,{\bf{w}}) = \left| {{{\bf{w}}^H}{\bf{a}}(\phi )} \right|,\;\;\phi  \in \left[ { - \frac{\pi }{2},\frac{\pi }{2}} \right],
\end{equation}
where ${\bf{a}}$ is the array response vector. If we expect to design a wide beam covering the zone of $(\theta^{-},\theta^{+})$, the ideal beam pattern of ${\bm{\omega }}(\theta^{-},\theta^{+})$ should be such that the beam pattern within the coverage range is uniform, whereas the beam pattern outside the coverage range is $0$. However, this ideal pattern cannot be realized practically and an important problem is how to design ${\bm{\omega }}(\theta^{-},\theta^{+})$ which achieves a close beam pattern to the ideal one.

\begin{figure}[t]
\centering
\includegraphics[width=3.4in]{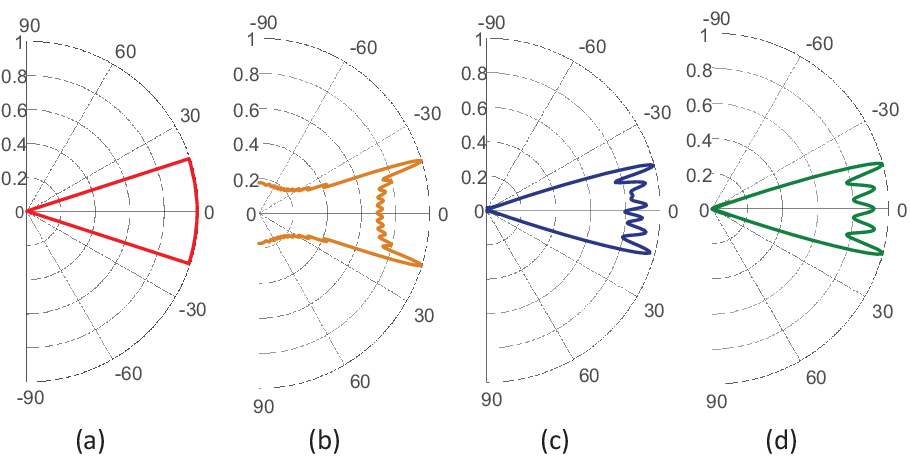}
\caption{Ideal beam pattern and practical beam patterns of ${\bm{\omega }}(-\pi/10,\pi/10)$ with $N_t=32$: (a) idea beam pattern. (b) beam pattern in \cite{hbt1}. (c) beam pattern in \cite{ps1}. (d) beam pattern in \cite{mrcode}.}\label{figwide}
\vspace{-12pt}
\end{figure}

There have been some works discussing the wide-beam designs for ULA. Fig. \ref{figwide} shows the ideal beam pattern of  ${\bm{\omega }}(-\pi/10,-\pi/10)$, as well as its practical beam pattern realized by different approaches.  From the perspective of design principles, these approaches can be referred to as: real-objective pursuit (ROP)\cite{hbt1}, complex-objective pursuit (COP)\cite{ps1}, sum narrow beams (SNB)\cite{SNB}, sum narrow beams with gradient phases (SNB-GP) \cite{mrcode}, successive convex approximation (SCA)-based auxiliary target pursuit (SCA-ATP)\cite{wbeam1}, and the sum of symmetrical array response vectors (S-SARV)\cite{wbeam2}. Next, we present the core idea and key steps of these approaches to obtain a wide beam with unit main lobe.\footnote{We mainly discuss the approaches of using all antenna elements in fully digital beamforming. Other approaches such as antenna  deactivation\cite{deact}, changing frequency\cite{hbt2a}, and analog beamforming\cite{tran1} are out of the scope. }

\subsubsection{ROP}
Uniformly sample $K>N_a$ discrete values $\{{\varphi _k}\}_{k=1}^K$ in $\left[ { - \frac{\pi }{2},\frac{\pi }{2}} \right]$. We expect that
\begin{equation}\label{td}
{{\bf{a }}_{{N_t}}}{({\varphi _k})^H}{\bf{w }}({\theta ^ - },{\theta ^ + }) = \left\{ {\begin{array}{*{20}{c}}
1&{{\rm{if}}\;{\varphi _k} \in \left[ {\theta ^ -,\theta ^ +} \right]}\\
0&{{\rm{otherwise}}}
\end{array}} \right.
\end{equation}
holds true for all ${\varphi _k}$. Let us define a lattice matrix as 
\begin{equation}
{\bf{A}} \buildrel \Delta \over = \Big[ {{{\bf{a }}_{{N_t}}}({\varphi _1}),{{\bf{a }}_{{N_t}}}({\varphi _2}),...,{{\bf{a }}_{{N_t}}}({\varphi _K})} \Big].
\end{equation}
Next, define an target vector ${{\bf{d}}_{{\theta ^ - },{\theta ^ + }}} \in \mathbb{R}^{K\times 1}$ in which
\begin{equation}
{{\bf{d}}_{{\theta ^ - },{\theta ^ + }}}(k) = \left\{ {\begin{array}{*{20}{c}}
1&{{\rm{if}}\;{\varphi _k} \in \left[{\theta ^ -,\theta ^ +}\right]},\\
0&{{\rm{otherwise}}}
\end{array}} \right., \; \forall k.
\end{equation}
As an approximate solution, one can minimize the Euclidean distance by solving
\begin{equation*}
{\rm{P}}(4):\;\;\mathop {\min }\limits_{\bf{w }} \left\| {{{\bf{A}}^H}{\bf{w }}({\theta ^ - },{\theta ^ + }) - {{\bf{d}}_{{\theta ^ - },{\theta ^ + }}}} \right\|.
\end{equation*}
Then, the optimal solution to ${\rm{P}}(4):$ can be written as
\begin{equation}
{\bf{w }}_{\rm{ROP}} = {({\bf{A}}{{\bf{A}}^H})^{ - 1}}{\bf{A}}{{\bf{d}}_{{\theta ^ - },{\theta ^ + }}}.
\end{equation}
\subsubsection{COP}
The COP is extended from ROP by adding a phase degree of freedom in the objective vector. As we expect the \emph{beam power} within the specified range is unit, (\ref{td}) should be corrected as 
\begin{equation}\label{td2}
|{{\bf{a }}_{{N_t}}}{({\varphi _k})^H}{\bf{w }}({\theta ^ - },{\theta ^ + })| = \left\{ {\begin{array}{*{20}{c}}
1&{{\rm{if}}\;{\varphi _k} \in \left[ {\theta ^ -,\theta ^ +} \right]}\\
0&{{\rm{otherwise}}}
\end{array}} \right., \;\forall \varphi _k.
\end{equation}
As such, a complex target vector ${{\bf{\hat d}}_{{\theta ^ - },{\theta ^ + }}} \in \mathbb{C}^{K\times 1}$ can be defined as
\begin{equation}
{{{\bf{\hat d}}}_{{\theta ^ - },{\theta ^ + }}}(k) = \left\{ {\begin{array}{*{20}{c}}
e^{j\lambda _k}&{{\rm{if}}\;{\varphi _k} \in \left[ {{\theta ^ - },{\theta ^ + }} \right]}\\
0&{{\rm{otherwise}}}
\end{array}} \right..
\end{equation}
where $\{\lambda _k\}_{k=1}^K$ are introduced phase variables and the Euclidian distance minimization problem can be expressed as
\begin{equation*}
{\rm{P}}(5): \;\;\mathop {\min }\limits_{{\bf{w }},\{ {\lambda _k}\} _{k = 1}^K} \left\| {{{\bf{A}}^H}{\bf{w }}({\theta ^ - },{\theta ^ + }) - {{{\bf{\hat d}}}_{{\theta ^ - },{\theta ^ + }}}} \right\|.
\end{equation*}
The authors in \cite{ps1} proposed an iterative coordinate decent method to solve ${\rm{P}}(5)$ and get the codeword ${\bf{w }}_{\rm{COP}}$.
\subsubsection{SNB}
 The two narrow beams ${{\bf{a}}_{{N_t}}}({\varphi _1})$ and ${{\bf{a}}_{{N_t}}}({\varphi _2})$ are orthogonal if the angles hold 
\begin{equation}
|\sin {\varphi _1} - \sin {\varphi _2}| = \frac{2}{N_t}.
\end{equation}
For ULA with $N_t$ antennas, there are $N_t$ orthogonal narrow beams within $[- \frac{\pi }{2},\frac{\pi }{2}]$, whose angles can be expressed as 
\begin{equation}
    {\varphi _n}\; = \arcsin \left[ {\frac{{2n - 1}}{N_t} - 1} \right],\;\;\;\;n = 1,2,...,N_t.
\end{equation}
To construct a wide beam, an intuitive idea is by adding the orthogonal narrow beams within  $(\theta^{-},\theta^{+})$. Let us define an angle set as
\begin{equation}
{\bf{\Psi }}= \left\{ {{\varphi _n}\;|\;{\theta ^ - } \le {\varphi _n} \le {\theta ^ + }} \right\}.
\end{equation}
Then, the SNB solution can be written as 
\begin{equation}
{{\bf{w}}_{{\rm{SNB}}}}({\theta ^ - },{\theta ^ + }) = \sum\limits_{{\varphi _n} \in {\bf{\Psi }}} {{{\bf{a}}_{{N_t}}}({\varphi _n})}.
\end{equation}
\subsubsection{SNB-GP}
The major defect of the SA method is that the main-lobe is notably  fluctuant. To circumvent this issue, a PSA method was proposed by introducing new variables $\{w_n\}_{n\in{\bf{\Psi }}}$ to minimize the variation in the main-lobe gain of the beam pattern. Specifically, the wide beam with coverage range $(\theta^{-},\theta^{+})$ can be expressed as 
\begin{equation}
{{\bf{w}}_{{\rm{PSA}}}}({\theta ^ - },{\theta ^ + }) = \sum\limits_{{\varphi _n} \in {\bf{\Psi }}} {{e^{j{\omega _n}}}{{\bf{a}}_{{N_t}}}({\varphi _n})} .
\end{equation}
Then, the optimization of $\{\omega_n\}_{n\in{\bf{\Psi }}}$ can be formulated as 
\begin{equation*}
\begin{aligned}
{\rm{P}}(6): \;\;&\mathop {\min }\limits_{{{\{ {\omega _n}\} }_{n \in {\bf{\Psi }}}}} {\mathop{\rm var}} \left[ {{{\bf{a}}_{{N_t}}}{{(\phi )}^H}{{\bm{\omega }}_{{\rm{PSA}}}}({\theta ^ - },{\theta ^ + })} \right]\\
&\;\quad{\rm{s}}{\rm{.t}}{\rm{.}}\;\;\;\;\phi  \in \left[ {{\theta ^ - },{\theta ^ + }} \right],
\end{aligned}
\end{equation*}
where ${\mathop{\rm var}}[\cdot]$ represent the variance operator.
${\rm{P}}(6)$ is a NP-hard problem. To reduce the solving complexity, the authors in \cite{mrcode} show that  $\{\omega_n\}_{n\in{\bf{\Psi }}}$ can be replaced by one variable $\omega$ with  ${\omega _n} = n \cdot \omega$. Moreover, the region of $\omega $ can be reduced as
\begin{equation}
\omega  \in \left[ { - \frac{\pi }{{{N_t}}},\pi \left( {1 - \frac{\pi }{{{N_t}}}} \right)} \right].
\end{equation}
Finally, the variables $\{\omega_n\}_{n\in{\bf{\Psi }}}$ in SNB-GP solution are obtained by exhaustive search on $\omega$.

\begin{figure}[t]
\centering
\includegraphics[width=3.4in]{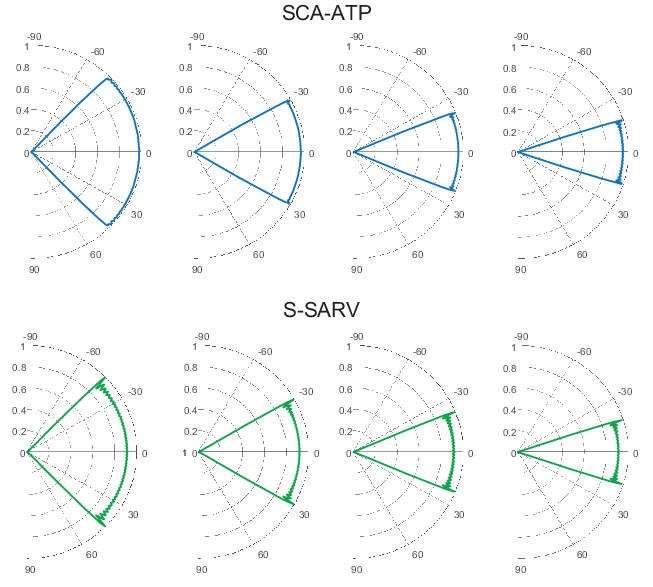}
\caption{Beam patterns of SCA-ATP\cite{wbeam1} and S-SARV\cite{wbeam2}.}\label{wbc}\vspace{-2pt}
\end{figure}
\subsubsection{SCA-ATP}
The above approaches are heuristic and lack a mathematical metric.
In sight of this, the authors in \cite{wbeam1} proposed a beam-pattern error (BPE) metric to characterize the gap between the practical beam pattern and the ideal one. The BPE can be expressed as
\begin{equation}\label{obj}
\begin{aligned}
\varepsilon &\left\{ {{\bf{w }}({\theta ^ - },{\theta ^ + })} \right\} \buildrel \Delta \over = \int\limits_{\varphi  \in [{\theta ^ - },{\theta ^ + }]} {{{\left| { 1 - \left| {{{\bf{a}}_{{N_t}}}{{(\varphi )}^H}{\bf{w }}({\theta ^ - },{\theta ^ + })} \right|} \right|}^2}d\varphi } \\
& + \int\limits_{\varphi  \in [\theta _*^ - ,{\theta ^ - }] \cup [{\theta ^ + },\theta _*^ + ]} {{{\left| {\mu (\varphi ) - \left| {{{\bf{a}}_{{N_t}}}{{(\varphi )}^H}{\bf{w }}({\theta ^ - },{\theta ^ + })} \right|} \right|}^2}d\varphi } \\
 &+ \int\limits_{\varphi  \in [ - \frac{\pi }{2},\theta _*^ - ] \cup [\theta _*^ + ,\frac{\pi }{2}]} {{{\left| {{{\bf{a}}_{{N_t}}}{{(\varphi )}^H}{\bf{w }}({\theta ^ - },{\theta ^ + })} \right|}^2}d\varphi } ,
\end{aligned}
\end{equation}
where $\mu (\varphi ) \in \mathbb{R}$ is an arbitrary variable with $0\leq \mu (\varphi )\leq 1$.  $\theta_*^{-}$ and $\theta_*^{+}$ are given by
\begin{equation}
\begin{aligned}
\theta _*^ -  &= \arcsin \left( {\sin ({\theta ^ - }) - \frac{2}{{{N_t}}}} \right),\\
\theta _*^ +  &= \arcsin \left( {\sin ({\theta ^ + }) + \frac{2}{{{N_t}}}} \right).
\end{aligned}
\end{equation}
 Then, the wide-beam design problem is formulated as
 \begin{equation*}
\begin{aligned}
&{\rm{P}}(7): \;\;\mathop {\min }\limits_{{\bm{\omega }}({\theta ^ - },{\theta ^ + }),\;\mu (\varphi )} \;\; \varepsilon \left\{ {{\bm{\omega }}({\theta ^ - },{\theta ^ + })} \right\}\\
&\quad {\rm{s}}.{\rm{t}}.\;\;0\leq \mu (\varphi )\leq \gamma,\;\;\varphi  \in [\theta _*^ - ,{\theta ^ - }] \cup [{\theta ^ + },\theta _*^ + ].  
\end{aligned}
\end{equation*}
Finally, the authors in \cite{wbeam1} proposed an SCA-ATP algorithm to solve ${\rm{P}}(7)$ and get the codeword ${\bf{w }}_{\rm{SCA-ATP}}$.
\subsubsection{S-SARV}
The conventional array response vector assumes that the signal phase at the first antenna is zero. Here, we define the symmetrical array response vector as 
\begin{equation}
\begin{aligned}
{{{\bf{\hat a}}}_N}(\varphi ) =& \frac{1}{{\sqrt N }}\Big[{e^{j\pi \left( { - \frac{{N - 1}}{2}} \right)\sin \varphi }},...,\\
&\qquad \qquad {e^{j\pi n\sin \varphi }},...,{e^{j\pi \left( {\frac{{N - 1}}{2}} \right)\sin \varphi }}\Big]^T,
\end{aligned}
\end{equation}
by assuming the signal phase at the center of ULA is zero, where ${n} =  - \frac{{({N} - 1)}}{2} + 1, - \frac{{({N} - 1)}}{2} + 2,...,\frac{{({N} - 1)}}{2} - 1$.  Define $M$ angles ($M\geq 2N_a$)  as 
\begin{align}
&{\psi _m} = \\
&\arcsin \left( {\sin {\theta ^ - } - \frac{1}{{{N_t}}} + \frac{{(m - 1)\left( {\sin {\theta ^ + } - \sin {\theta ^ - } + \frac{2}{{{N_t}}}} \right)}}{{M - 1}}} \right),\notag
\end{align}
for $m=1,2,...,M$. Then, the authors in \cite{wb2} proposed to construct the wide beam by directly adding the symmetrical array response vectors with above angles, i.e.,
\begin{equation}
{\bf{w }}_{\rm{S-SARV}}= \beta \sum\limits_{m = 1}^M {\hat{\bf{a}}_{{N_t}}}({\psi _m}),
\end{equation}
where 
\begin{equation}
\beta  = \frac{{\sqrt {{N_t}\left( {\sin {\theta ^ + } - \sin {\theta ^ - } + \frac{2}{{{N_t}}}} \right)} }}{{\sqrt 2  \cdot {\rm{norm}}\left[ {\sum\limits_{m = 1}^M {{{{\bf{\hat a}}}_{{N_t}}}({\psi _m})} } \right]}}.
\end{equation}

\begingroup
\renewcommand{\arraystretch}{1.2} 
\begin{table}[t]
\centering
\caption{Comparison of the wide-beam approaches. \\ More $\star$ represents lower complexity or higher performance}\label{wb}
\vspace{-2pt}
\begin{tabular}{|c|c|c|}
\hline
Wide-beam approaches & Complexity & Performance\\ \hline
ROP \cite{hbt1} & $\star\star\star\star\star$ &   $\star\star$  \\ \hline
COP\cite{ps1}   &   $\star\star$  & $\star\star\star \;\!\star$ \\ \hline
SNB \cite{SNB}    &  $\star\star\star\star\star$ & $\star$  \\ \hline
SNB-GP \cite{mrcode}&  $\star$  &  $\star\star\star \;\! \star$ \\ \hline
SCA-ATP  \cite{wbeam1}&   $\star\star\star$ & $\star\star\star\star \star$  \\ \hline
S-SARV  \cite{wbeam2}&   $\star\star\star\star\star$ & $\star\star\star\! \; \star$  \\ \hline
\end{tabular}
\end{table}
\endgroup

To the best of our knowledge, the beam pattern of the SCA-ATP is closest to the ideal one, and the S-SARV is the best approach to balance the performance and computational complexity. The beam patterns of the two approaches are presented in Fig. \ref{wbc}. Table \ref{wb} summarizes the performance and complexity of different wide-beam approaches.

\subsection{Lens Antenna Array}\label{tth}
\begin{figure}[t]
\centering
\includegraphics[width=2.8in]{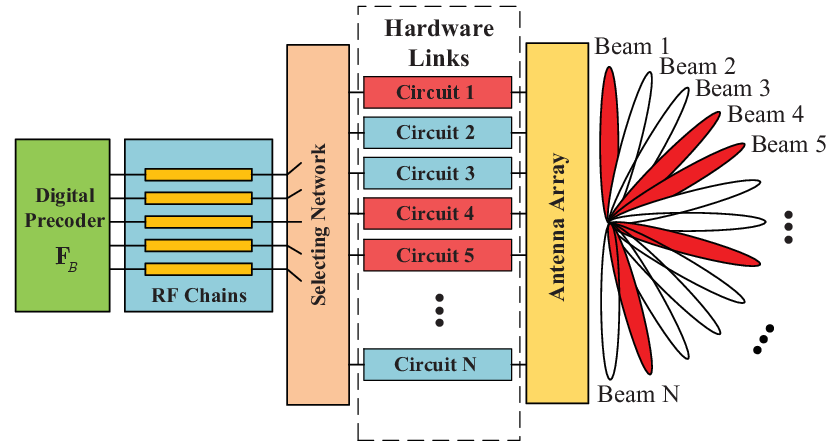}
\caption{Architecture of the training tailored hardware.}\label{hd}\vspace{-2pt}
\end{figure}

For implementing beam training with low cost, a concept of beamspace MIMO came up recently. This concept is particularly referred to as the beamforming realized by using the special hardware that makes the wireless communication system more like an optical one \cite{len1}. By modeling the optical lens as the approximate spatial Fourier transformer, beam training will be re-considered from the antenna space to the beam space that has much lower dimensions, to significantly reduce the number of required RF chains \cite{len2}. Fig. \ref{hd} shows the architecture of the training tailored hardware, in which each beam has a dedicated circuit and only the needed beams will be selected by limited RF chains via the selecting network. Compared to the conventional digital/analog hardware, the advantage of the training tailored hardware is that there is no need to control antenna weights. Instead, only a simple link selection is required, which makes the beamforming realized faster and more accurately.
\begin{figure}[t]
\centering
\includegraphics[width=3.5in]{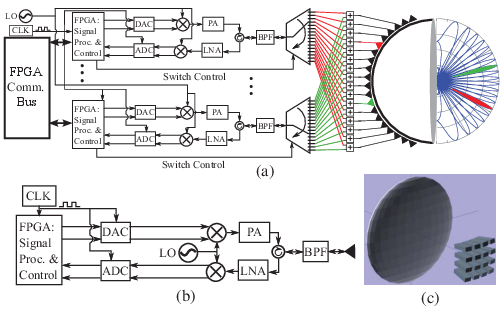}
\caption{An lens array prototype \cite{prot}: (a) block diagram of the transceiver; (b) block diagram of the RF chains; (c) structure of the lens array.}\label{28}\vspace{-2pt}
\end{figure}

One popular training tailored hardware is the \emph{lens antenna array}, which is composed of an electromagnetic lens and some antenna elements located in the focal region of the lens \cite{len3}. 
In general, there are three main fabrication technologies for the electromagnetic lens \cite{len4}: 1) by conventional antennas array connected with transmission lines with variable lengths \cite{l38,l39}; 2) by dielectric materials with carefully designed front/rear surfaces  \cite{l40,l41}; 3) by sub-wavelength periodic inductive and capacitor structures \cite{l42,l43}. As shown in Fig. \ref{28}, the authors in \cite{prot} presented a prototype of the lens array that consists of four main components: 1) field programmable gate array (FPGA)-based digital signal processor (DSP) back-end and analog-to-digital conversions (ADCs), 2) RF chains. 3) beam selector, and 4) front-end lens array. It can be observed that the fundamental principle of the lens array can focus the incident signals with sufficiently separated AoAs to different antenna elements (or a subset of elements), and vice versa. In works \cite{len4,l44,l45,l46,l47}, lens array has been shown to achieve significant performance gains as well as complexity reductions in UM-MIMO systems. Based on the lens array, the signal processing can be much simplified by treating the transmit/received signals on the virtual channels. Specifically, the conventional UM-MIMO system can be expressed as ${\mathbf{y}} = {\mathbf{Hx}} + {\mathbf{n}}$, where ${\mathbf{y}} \in {{\mathbb{C}}^{{N_r}}}$ and ${\mathbf{x}} \in {{\mathbb{C}}^{{N_t}}}$ are the transmit and receive signals on the antenna elements, which have large dimensions, respectively. By using the lens array, the UM-MIMO system can be expressed as 
\begin{equation}
\widetilde {\mathbf{y}} = \underbrace {{{\mathbf{W}}_{{\mathrm{lens}}}}{\mathbf{H}}{{\mathbf{F}}_{{\mathrm{lens}}}}}_{{\mathrm{virtual\;\; channel}}}\widetilde {\mathbf{x}} + \widetilde {\mathbf{n}} = \widetilde {\mathbf{H}}\widetilde {\mathbf{x}} + \widetilde {\mathbf{n}},
\end{equation}
where $\widetilde {\mathbf{y}} \in {{\mathbb{C}}^{{N_r}}}$ and $\widetilde {\mathbf{x}} \in {{\mathbb{C}}^{{N_t}}}$ are the transmit and receive signals on the lens array's elements, respectively. ${{\mathbf{W}}_{{\mathrm{lens}}}}\in {{\mathbb{C}}^{{N_r}\times N_r}}$ and ${{\mathbf{F}}_{{\mathrm{lens}}}}\in {{\mathbb{C}}^{{N_t}\times N_t}}$ are the fixed analog beamforming in the lens array architecture, which represents the signals' transformation from the lens to the antenna elements at the transmitter and receiver, respectively. $\widetilde {\mathbf{H}}={{\mathbf{W}}_{{\mathrm{lens}}}}{\mathbf{H}}{{\mathbf{F}}_{{\mathrm{lens}}}}$ is the virtual channel of the UM-MIMO systems based on lens array. Generally, the fixed analog beamforming matrices ${{\mathbf{W}}_{{\mathrm{lens}}}}$ and ${{\mathbf{F}}_{{\mathrm{lens}}}}$ can be expressed as the unitary discrete Fourier transform (DFT) matrices \cite{dft}, whose columns correspond to orthogonal beams with different spatial angles. An $N\times N$ DFT matrix ${\mathbf{U}}$ is given by 
\begin{figure}[t]
\centering
\includegraphics[width=3.2in]{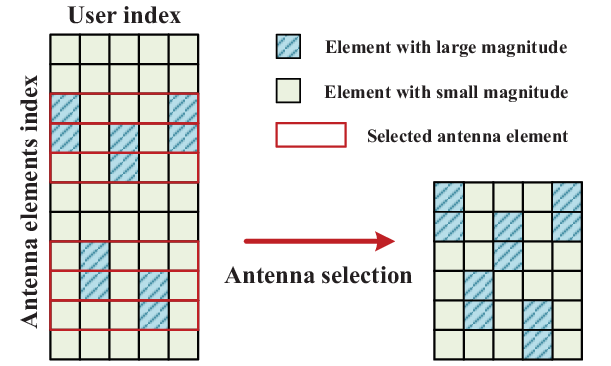}
\caption{Beam selection in beamspace MIMO systems with five users.}\label{slec}\vspace{0pt}
\end{figure}
\begin{align}
&{{\bf{U}} = \frac{1}{{\sqrt N }} \times }\\
&{\left[ {\begin{array}{*{20}{c}}
1&1& \cdots &1\\
{{e^{j\pi \sin {\varphi _1}}}}&{{e^{j\pi \sin {\varphi _2}}}}& \cdots &{{e^{j\pi \sin {\varphi _N}}}}\\
 \vdots & \vdots & \vdots & \vdots \\
{{e^{j\pi (N - 1) \sin {\varphi _1}}}}&{{e^{j\pi (N - 1) \sin {\varphi _2}}}}& \cdots &{{e^{j\pi (N - 1) \sin {\varphi _N}}}}
\end{array}} \right].}\notag
\end{align}

It is worth pointing out that the virtual channel $\widetilde {\mathbf{H}}$ has a sparse property in the THz UM-MIMO systems due to the limited number of spatial paths. Thus, the transmit vector $\widetilde {\mathbf{x}}$ and receive vector $\widetilde {\mathbf{y}}$ has limited elements with large magnitude. In view of this, we can merely process the signals from the elements with large magnitude and reduce the dimension from $N_t$/$N_r$ to $N_t^{RF}$/$N_r^{RF}$, where $N_t^{RF}$ and $N_r^{RF}$ represent the number of RF chains of the transmitter and receiver respectively. This is referred to as the antenna selection or beam selection \cite{len1,len4,sel}. For ease of understanding, an illustration of the antenna selection in beamspace multi-user MIMO system is presented in Fig. \ref{slec}. The less the number of users served, the less the number of antenna elements selected.

\section{IRS-Assisted Joint Beamforming}\label{IRSJB}
Despite UM-MIMO are able to offer great beamforming gains, its energy efficiency decreases with the increase of the number of antennas \cite{I1,I2}.  Besides, the LoS blockage problem significantly impairs the performance of THz beamforming. In this subsection, we introduce an emerging wireless technology, i.e., IRS, which can significantly improve the spectral efficiency with much-reduced energy consumption and reduce the impact of LoS blockage \cite{csa,cometa,basar,IRSAWN,IRSAWC}.

IRS is a metasurface consisting of a large set of tiny elements (i.e., controllable reflecting elements), each being able to passively reflect the incident wireless signal by adjusting its phase shifts. By judiciously optimizing the phase shifts at IRS, the reflected signals of different elements can be added/counteracted in intended/unintended directions\cite{shu}. Compared to the conventional MIMO systems whose performance is determined by their channels, the IRS-assisted UM-MIMO systems provide a programmable and controllable wireless environment \cite{act}. Given this advantage, the achievable data rates in IRS-assisted UM-MIMO systems can be enhanced by jointly optimizing the precoder/decoder (i.e., beamformer/combiner) at the transmitter/receiver and the phase shifts at the IRS. The earliest researches on IRS-assisted joint beamforming were aimed at the MISO IRS-assisted system \cite{qinte,ghy,huangchi,multicast,latency}. With different optimization targets, transmit power minimization\cite{qinte}, weighted sum-rate maximization\cite{ghy}, energy efficiency maximization\cite{huangchi}, multicast rate maximization \cite{multicast}, and latency minimization \cite{latency} have been investigated. 

\subsection{IRS-Assisted UM-MIMO Systems}

\begin{figure}[t]
\centering
\includegraphics[width=3.3in]{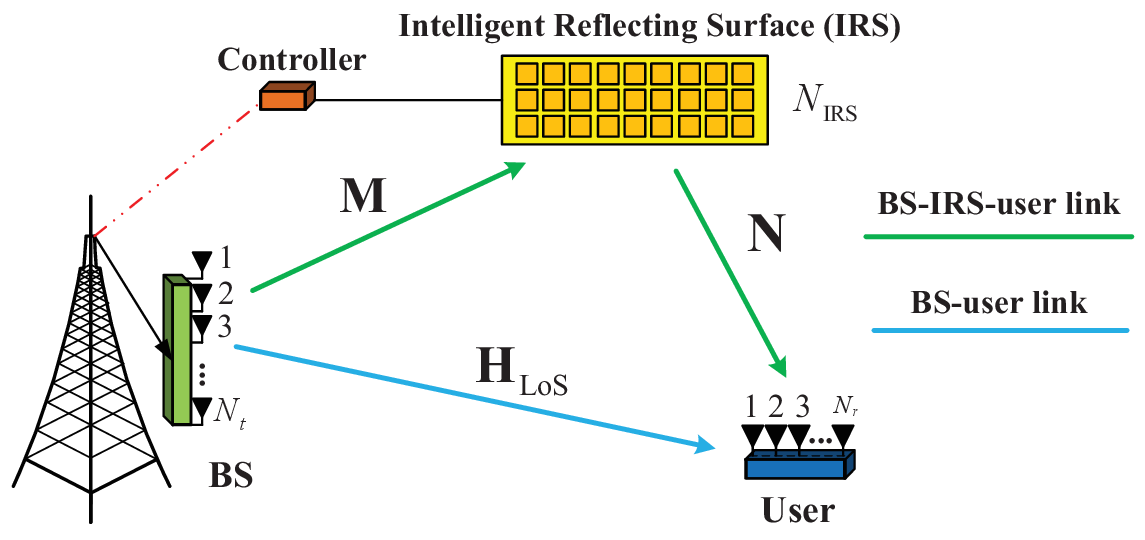}
\caption{A point-to-point IRS-assisted UM-MIMO system.}\label{irs}\vspace{-2pt}
\end{figure}
Consider a point-to-point IRS-assisted  UM-MIMO communication system as depicted in Fig. \ref{irs}, where the base station (BS), equipped with $N_t$ antennas, transmits $N_s \le N_t$ data streams to a user, equipped with $N_r$ antennas, with the help of an IRS equipped with $N_{\mathrm{IRS}}$ passive elements. In the communication, the BS sends its data message ${\mathbf{s}} \in {\mathbb{C}^{{N_s} \times 1}}, {\mathbf{s}} \sim \mathcal{CN}({\mathbf{0}},{\mathbf{I}}_{N_s})$ via a precoder ${\mathbf{F}} \in {\mathbb{C}^{{N_t} \times {N_s}}}$ to the user and the IRS simultaneously. Let ${\mathbf{H}}_{\mathrm{}}\in {\mathbb{C}^{{N_r} \times {N_t}}}$, ${\mathbf{M}} \in {\mathbb{C}^{{N_{\mathrm{IRS}}} \times {N_t}}}$, and ${\mathbf{N}} \in {\mathbb{C}^{{N_r} \times {N_{\mathrm{IRS}}}}}$ denote the channels from the BS to the user, from the BS to the IRS, and from the IRS to the user, respectively. The received signal at the IRS is first phase-shifted by a diagonal reflection matrix ${\mathbf{\Theta }} = {\mathrm{diag}}(\beta {e^{j{\theta _1}}},\beta {e^{j{\theta _2}}}, \cdots ,\beta {e^{j{\theta _{N_{\mathrm{IRS}}}}}}) \in {\mathbb{C}^{{N_{\mathrm{IRS}}} \times {N_{\mathrm{IRS}}}}}$ and then reflected to the user, where $j = \sqrt {-1}$ is the imaginary unit, $\{\theta _i \in [0,2\pi)\}_{i}^{N_{\mathrm{IRS}}}$ are the shifted phases, and $\beta \in [0,1]$ denote the amplitude of each reflection coefficient. As such, the overall received signal is expressed as
\begin{equation}
{{\mathbf{y}}} = \sqrt {\frac{P}{{{N_s}}}} ( {\underbrace {{\mathbf{N}}{\mathbf{\Theta MFs}}}_{{\text{BS-IRS-user}}} + \underbrace {{\mathbf{H}}_{\mathrm{LoS}}{\mathbf{Fs}}}_{{\text{BS-user}}}} ) + {{\mathbf{n}}},
\end{equation}
where $P$ is the total transmitted power and ${\|{\mathbf{F}}\|_F^2} = N_s$. In addition, ${\mathbf{n}} \sim \mathcal{CN}({\mathbf{0}},\sigma _n^2{{\mathbf{I}}_{{N_r}}})$ is zero-mean additive Gaussian noise. The aim is to maximize the spectral efficiency by jointly optimizing the precoding matrix $\mathbf{F}$ and the phase shifters $\{\theta _i\}_{i=1}^{N_{\mathrm{IRS}}}$, subject to the power constraint at the BS and the uni-modular constraints on the phase shifters. Let ${\mathbf{v}} = [{e^{j{\theta _1}}},{e^{j{\theta _2}}}, \cdots ,e^{j{\theta _{N_{\mathrm{IRS}}}}}]^H$ denote the phase shifter vector at the IRS, i.e., ${\mathbf{\Theta }} = \beta \cdot{\mathrm{diag}}({\mathbf{v}}^{\dag})$. Define the effective channel in IRS-assisted UM-MIMO systems as ${{\mathbf{H}}_{\text{eff}}} = {\mathbf{N}}{\mathbf{\Theta M}} + {\mathbf{H}}_{\mathrm{LoS}}.$ Thus, the IRS-assisted joint beamforming problem can be formulated as 
\begin{equation*}
\begin{aligned}
{\rm{P}}(8):\;\;&\mathop {\max }\limits_{{\mathbf{F}},{{\mathbf{v}}}} \;{\log _2}\det \left| {{{\mathbf{I}}_{{N_b}}} + \frac{P}{{\sigma _n^2}{N_s}}{{\mathbf{H}}_{\text{eff}}}{\mathbf{F}}{{\mathbf{F}}^H}{\mathbf{H}}_{\text{eff}}^H} \right|\\
&\;\;\;{\mathrm{s.t.}}\;\;\;{{\mathbf{H}}_{\text{eff}}} = {\mathbf{N}}{\mathbf{\Theta M}} + {\mathbf{H}}_{\mathrm{LoS}},\\
&\;\qquad\;\;{\left\| {\mathbf{F}} \right\|_F^2} = {N_s},\;{\mathbf{\Theta }} = \beta \cdot{\mathrm{diag}}({\mathbf{v}}^{\dag}),\\
&\;\qquad\;\;\left| {{\mathbf{v}}(i)} \right| = 1, \;\; i=1,2,\cdots,N_{\mathrm{IRS}},
\end{aligned}
\end{equation*}\label{ori}where ${\mathbf{v}}(i)$ denotes the $i$-th entry of ${\mathbf{v}}$. ${\rm{P}}(8)$ is a quite hard optimization problem as the non-convexity remains on both the objective function and the constraint imposed by IRS's phase shifts $\mathbf{v}$. To solve this problem, the authors in \cite{nby} proposed a sum-path-gain maximization approach to reach a suboptimal solution. Then, the authors in \cite{szhang} proposed an alternating optimization (AO)-based method to reach a high-performance near-optimal solution albeit compromised on the computational complexity. In sight of this, the authors in \cite{pwang} proposed a manifold optimization (MO)-based algorithm to achieve a better performance-complexity trade off. 
\begin{figure}[t]
\centering
\includegraphics[width=2.8in]{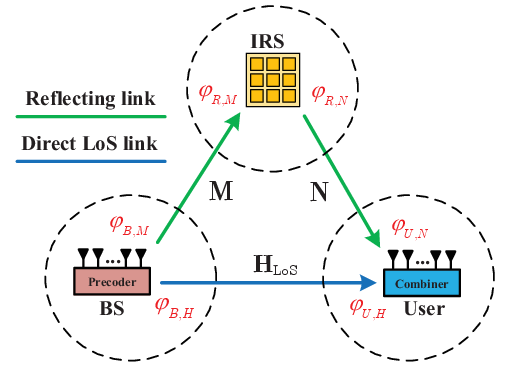}
\caption{Path angles in IRS-assisted UM-MIMO systems.}\label{3n}\vspace{-2pt}
\end{figure}

Hitherto, there have been many works focusing on the channel estimation solutions \cite{ch1,ch2,ch3,ch4,ch5} and the beamforming optimization problems \cite{op1,op2,op3,op4,op5,op6,op7} for various IRS-assisted UM-MIMO scenarios. However, it is practically inefficient to combine the channel estimation and the beamforming designs in UM-MIMO THz systems due to the in extremely high implementation complexity. Thus, it is of vital importance to consider the beam training strategies for IRS-assisted systems.

The main challenges of the beam training for IRS-assisted UM-MIMO systems are attributed to the passivity of IRS, i.e., unable to transmit and receive beams. Thus, the authors in \cite{irstran1} proposed to place the IRS relatively still to the BS and developed a fast beam training scheme by treating the BS and IRS as a whole. By skipping the signal processing, the authors in \cite{irstran2} proposed a machine learning empowered beam training framework for IRS-assisted UM-MIMO systems. Next, we introduce a cooperative beam training procedure for IRS-assisted systems proposed in \cite{irstran3} and \cite{tran4}, by using diagrams to illustrate its core idea. 

\begin{figure}[t]
\centering
\includegraphics[width=3.3in]{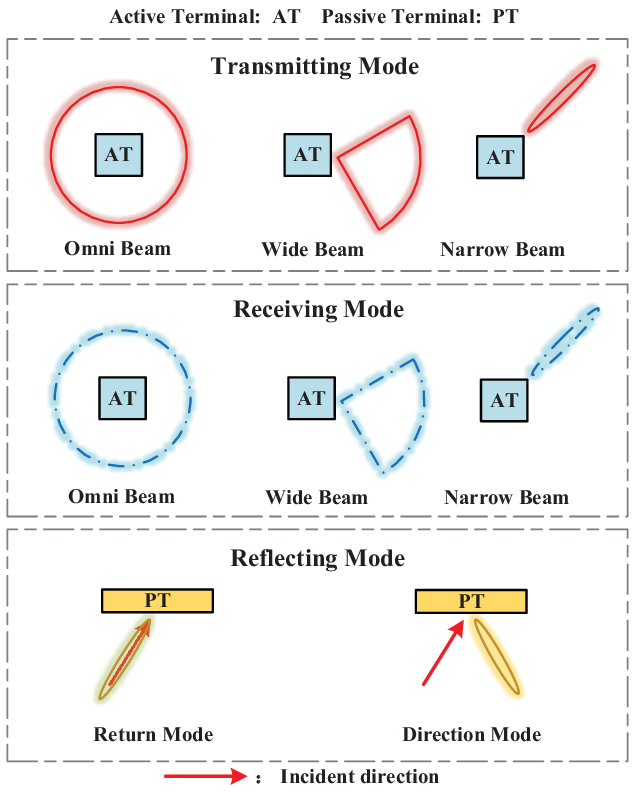}
\caption{Beam modes for active terminal and passive terminal.}\label{3mo}\vspace{-2pt}
\end{figure}

\subsection{Cooperative Beam Training}
Assume that the IRS is placed on the same horizontal level as the BS and the user, i.e., without loss of generality, we do not consider the elevation angle of IRS. As shown in Fig. \ref{3n}, the IRS-assisted system consists of six path angles  $\varphi _{B,H},\varphi _{U,H},\varphi _{B,M},\varphi _{R,M}, \varphi _{R,N},$ and $\varphi _{U,N}$. The cooperative beam training aims to find the narrow beams at these angles. For ease of exposition, we first define the training modes for the active terminal (BS and user) and the passive terminal (IRS), respectively. As shown in Fig. \ref{3mo}, we use a solid red line to represent the transmit beams, a solid-broken blue line to represent the receive beams, a solid red arrow and a solid yellow line to represent the incoming signal and the reflected signal, respectively. It is worth mentioning that in the return mode, the codewords of IRS are functions of the AoAs, i.e., $\{{{\bm{\Theta}} _{{\mathrm{ret}}}}(\varphi _n^{{\mathrm{in}}})\}_{n=1}^{N}$. If the AoA is $\varphi _k^{{\mathrm{in}}}$, the codeword ${{\bm{\Theta}} _{{\mathrm{ret}}}}(\varphi _k^{{\mathrm{in}}})$ ensures that the AoD of the reflected signals is in the back direction $\varphi _k^{{\mathrm{in}}}+\pi$, i.e., 
\begin{equation}
{{\mathbf{\Theta }}_{{\mathrm{ret}}}}(\varphi _k^{{\mathrm{in}}}){\mathbf{a}}(\varphi _k^{{\mathrm{in}}}) = {\mathbf{a}}(\varphi _k^{{\mathrm{in}}} + \pi).
\end{equation}
The codewords of the directions mode are functions of AoAs and AoDs, i.e., $\{{{\bm{\Theta}} _{{\mathrm{dir}}}}(\varphi _n^{{\mathrm{in}}},\varphi _p^{{\mathrm{out}}})\}_{n,p=1}^{N}$. If the AoA is  $\varphi _k^{{\mathrm{in}}}$, the codeword ${{\bm{\Theta}} _{{\mathrm{dir}}}}(\varphi _k^{{\mathrm{in}}},\varphi _m^{{\mathrm{out}}})$ ensures that the AoD of the reflected signals is   $\varphi _m^{{\mathrm{out}}}$, i.e.,
\begin{figure}[t]
\centering
\includegraphics[width=3.1in]{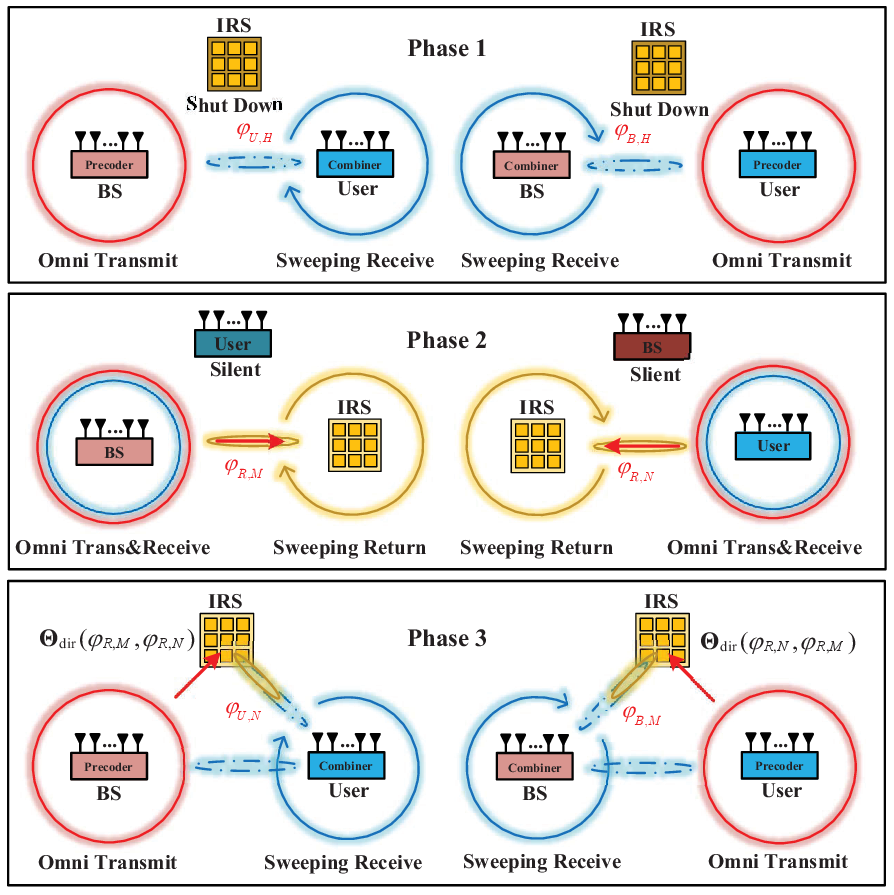}
\caption{Primary protocol of the IRS-assisted joint beam training.}\label{ideaofirs}\vspace{-2pt}
\end{figure}
\begin{equation}
{{\mathbf{\Theta }}_{{\mathrm{dir}}}}(\varphi _k^{{\mathrm{in}}},\varphi _m^{{\mathrm{out}}}){\mathbf{a}}(\varphi _k^{{\mathrm{in}}}) = {\mathbf{a}}(\varphi _k^{{\mathrm{out}}}).
\end{equation}
The design of the return-mode and direction-model codebooks are provided in \cite{irstran3}. In the next, we first present a primary protocol of the IRS-assisted joint beam training, albeit with some shortcomings, to draw some basic insights. We divide the overall protocol into three phases to achieve different groups of measurements as illustrated in Fig. \ref{ideaofirs}.  

\indent \emph{Phase 1}: Shut down the IRS. Fix the BS to be in an omni-beam transmitting mode and the user sweeps the beam to find the desired path direction. By switching the operation of the BS and the user, we obtain ${\varphi} _{U,H}$ and ${\varphi} _{B,H}$.

\indent \emph{Phase 2}: Keep the user silent and fix the BS to be concurrently in an omni-beam transmitting and receiving mode. Then, the IRS successively sweeps the codewords in return mode, i.e., ${\mathbf{\Theta }}_{\mathrm{ret}}({\varphi}_n^{\mathrm{in}})$, which are predefined with specific time slots, and is known to all terminals. The best codeword is informed to the BS by determining the time slot with the strongest receiving signal. By switching the operation of the BS and the user, we obtain ${\varphi} _{R,M}$ and ${\varphi} _{R,N}$.

\indent \emph{Phase 3}: With the obtained ${\varphi} _{R,M}$ and ${\varphi} _{R,N}$, we fix IRS to optimally bridge the BS-IRS-user link by direction mode, i.e., ${\mathbf{\Theta }}_{\mathrm{dir}}({\varphi} _{R,M},{\varphi} _{R,N})$. Fix the BS to be in an omni-beam transmitting mode and the user sweeps the beam to find the desired path direction. By switching the operation of the BS and the user, we obtain ${\varphi} _{U,N}$ and ${\varphi} _{B,M}$.

Based on the above three phases, we can find all the path angles in IRS-assisted systems\cite{irstran3}. However, this strategy suffers from the following main drawbacks.
\begin{itemize}
\item Omni-beam may not be effectively detected in THz communication.
\item Concurrently transmitting and receiving beams result in interference.
\item Sweeping the narrow beams incurs high complexity.
\end{itemize}
\begin{figure}[t]
\centering
\includegraphics[width=3.3in]{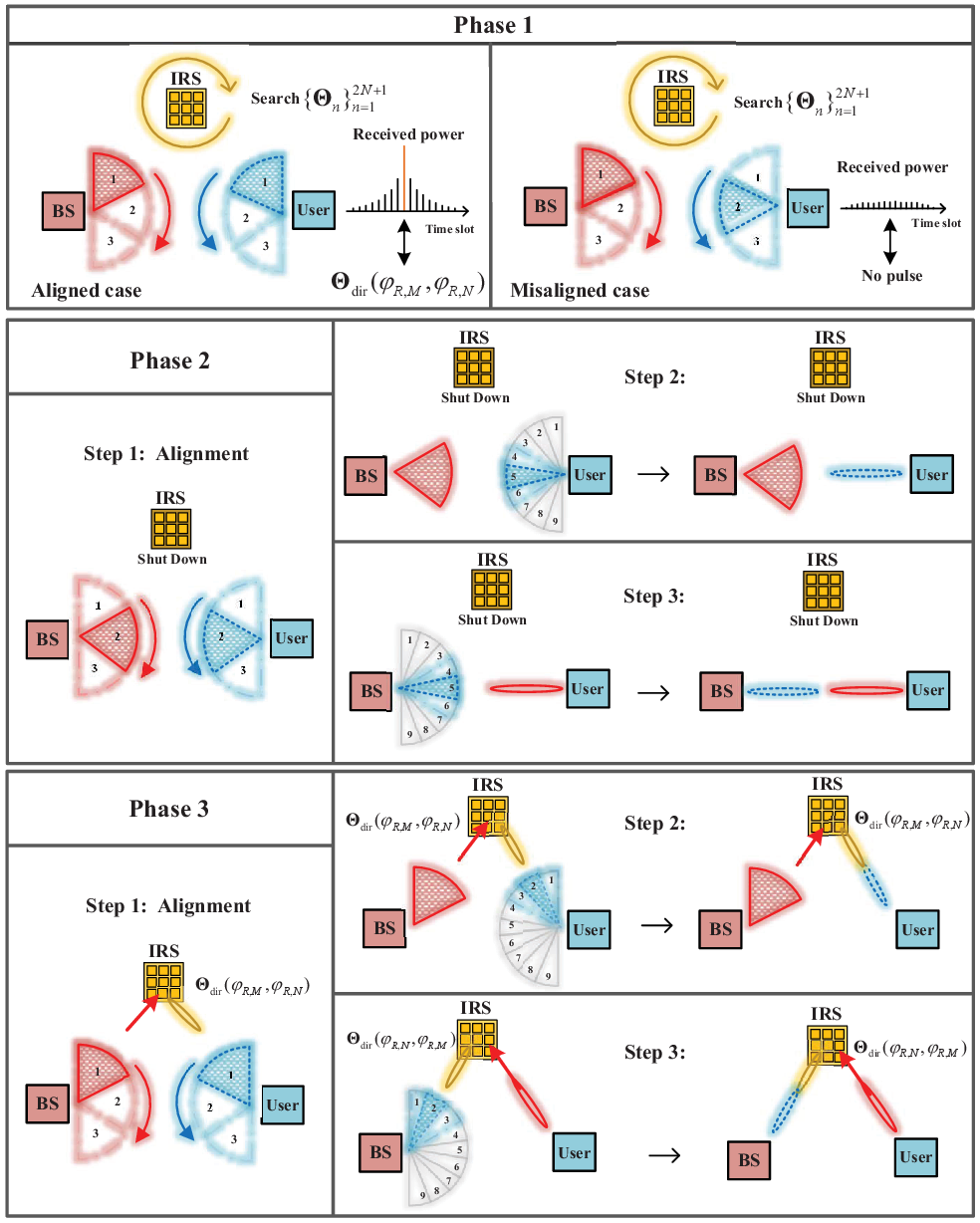}
\caption{Practical beam training protocol for IRS-assisted UM-MIMO
 systems.}\label{3tv}\vspace{-2pt}
\end{figure}
To circumvent these issues, we develop a more practical protocol, which combines the ternary-tree search (at BS and users) and the partial search (at IRS).  Let $N$ denotes the number of discretized AoAs/AoDs.  There are $(3N\!-\!2)\big/2$ codewords in the transit/receive codebook and $2N+1$  codewords in IRSs codebook. Details of the codebook design is provided in \cite{tran4}. The diagram of the training protocol is presented in Fig. \ref{3tv} and the process is described as below.  

\indent \emph{Phase 1}:
We aim to obtain the optimal codeword for IRS. To achieve the beam alignment, we first test $3\times3$ wide beams in $9$ successive intervals with BS in the transmitting mode and user $k$ in the receiving mode. In each interval, the IRS successively searches the codewords ${\rm{\{ }}{{\bf{\Theta }}_n}{\rm{\} }}_{n = 1}^{2N + 1}$. For the IRS, there is only one beam pair that covers both the BS-IRS link and the IRS-user link. During the interval when this beam pair (aligned case) is used, the user will detect an energy pulse in the time slot when IRS uses ${\mathbf{\Theta }}_{\mathrm{dir}}({\varphi} _{R,M},{\varphi} _{R,N})$. Thus, the user can utilize the pulse slots to identify this optimal codeword for IRS.

\indent \emph{Phase 2}:
We turn off the IRS and aim to obtain $\varphi _{B,H}$ and $\varphi _{U,H}$ via the following three steps. In step 1, $9$ wide-beam pairs are tested for alignment. The user compares the received energy in $9$ intervals and determines the aligned pair with the maximum power. The aligned pair is labeled by recording the beams chosen at both sides. In step 2, the BS transmits the labeled wide beam and the user uses a ternary-tree search to obtain $\varphi _{U,H}$. In step 3,  the user transmits the labeled narrow beam and BS uses a ternary-tree search to obtain $\varphi _{B,H}$.

\indent \emph{Phase 3}:
We aim to obtain $\varphi _{B,M}$ and $\varphi _{U,N}$ through three steps similar to Phase 2. We turn on IRS with the obtained optimal codeword, i.e., ${\mathbf{\Theta }}_{\mathrm{dir}}({\varphi} _{R,M},{\varphi} _{R,N})$. Note that there exist two propagation paths from the BS to the user, i.e., BS-IRS-user path and BS-user path. The BS-user path has been estimated in Phase 2. To estimate the AoA and AoD of the reflecting paths, we can use the ternary-tree search with a small modification. Specifically, in each stage of step 2, the receiver removes the signal component of the BS-user path, by signal processing, when determining the best beam. In step 3, the user transmits the labeled narrow beam and BS uses a ternary-tree search while removing the signal component of the BS-user path. By this means, we can obtain ${\varphi} _{B,M}$ and ${\varphi} _{U,N}$.

\begin{figure}[t]
\centering
\includegraphics[width=3.5in]{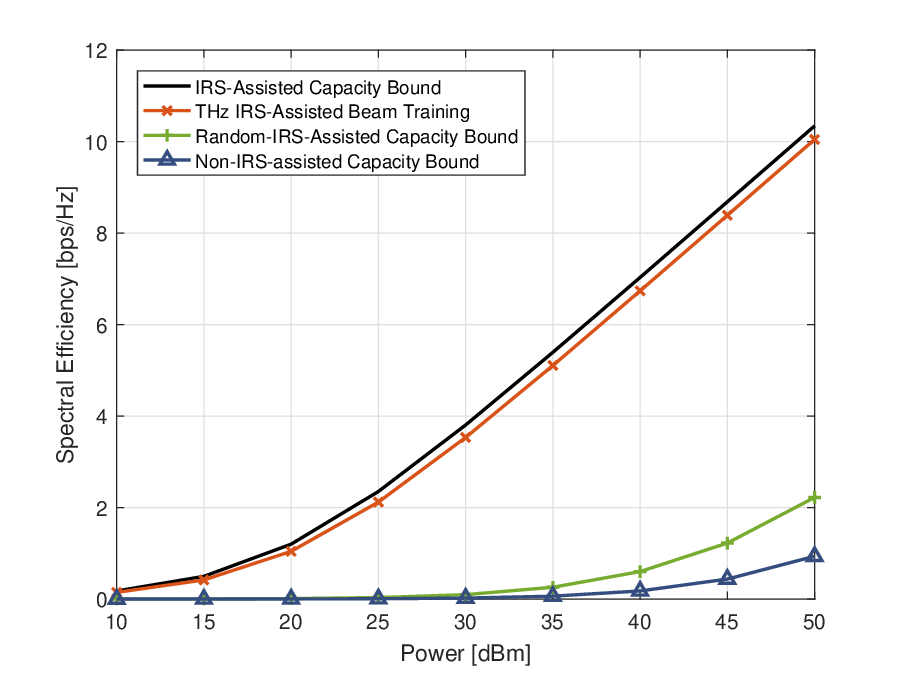}
\caption{The performance of different schemes in THz LoS-blockage case.}\label{irs11}\vspace{-12pt}
\end{figure}

\begin{figure}[t]
\centering
\includegraphics[width=3.5in]{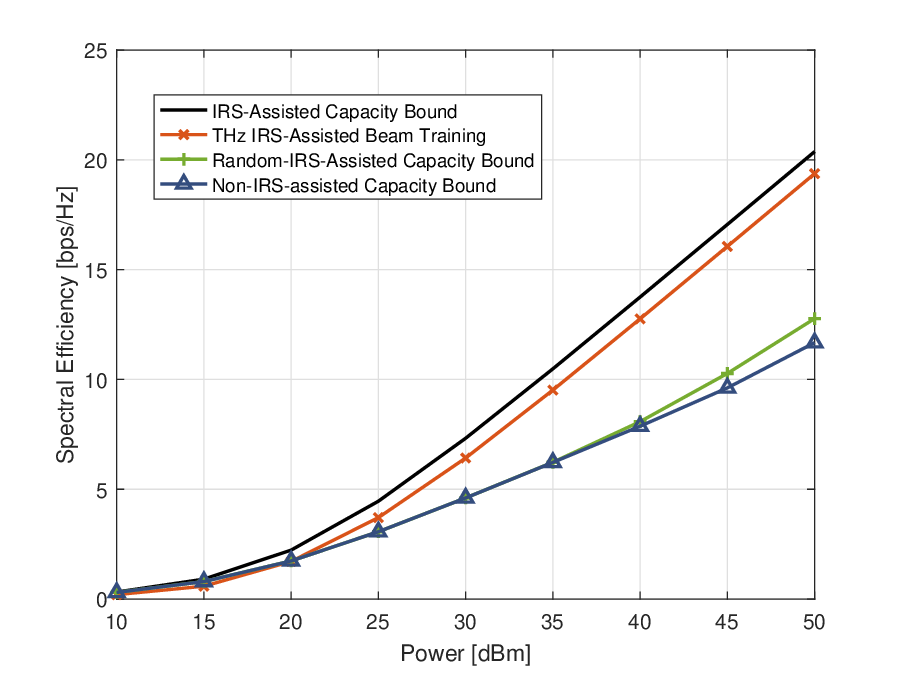}
\caption{The performance of different schemes in THz non-blockage case.}\label{irs22}\vspace{-12pt}
\end{figure}

\begin{figure*}[t]
\centering
\includegraphics[width=5.5in]{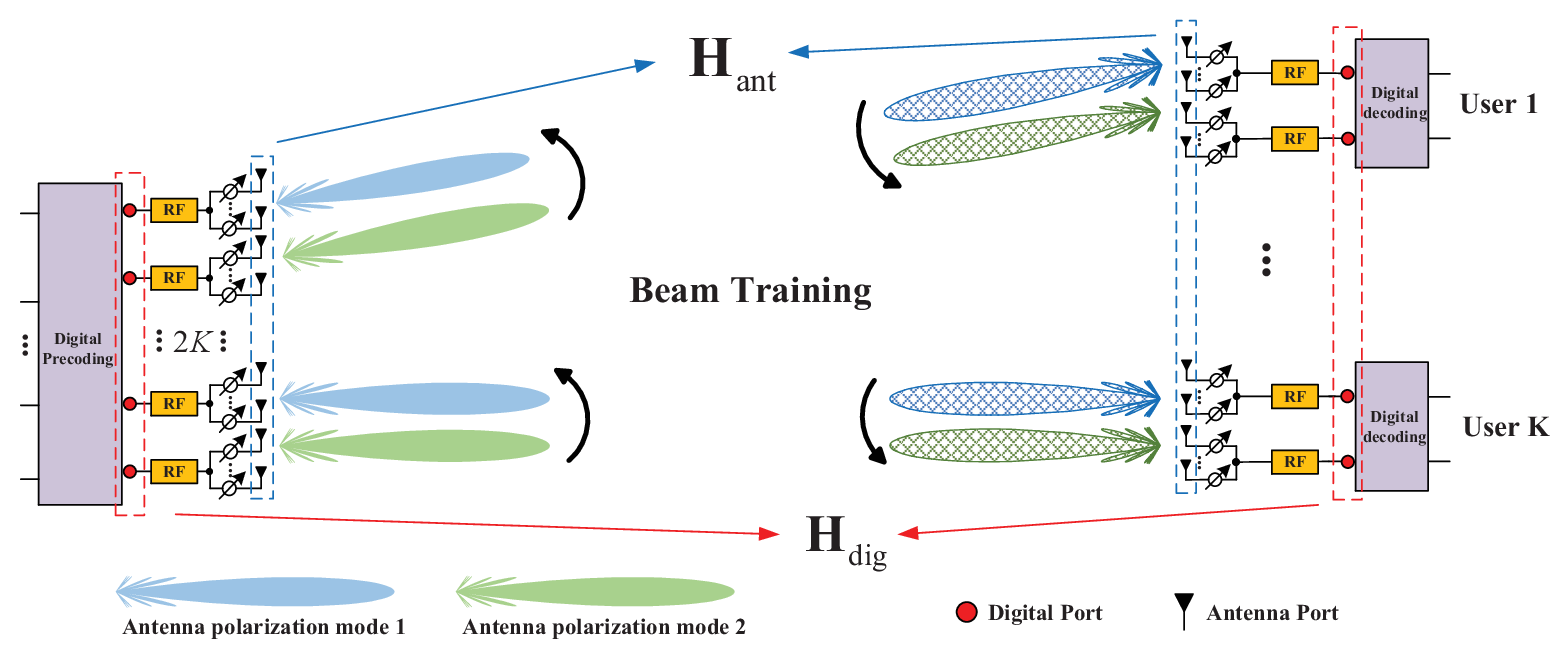}
\caption{The strategy of multi-user THz UM-MIMO: Apply beam training in the analog domain and applyprecoding in the digital domain.}\label{mumimo}
\vspace{-6pt}
\end{figure*}

By applying the above three phases, all the path angles can be found in IRS-assisted systems, which completes the THz IRS-assisted beam training. We would like to point out that the exhaustive search protocol has $N^2+N^4$ tests in IRS-assisted systems, whereas this training protocol has only $18N + 12{\log _3}N - 3$ tests \cite{tran4}. Next, we compare the performance of the beam training with the capacity bound of the schemes with optimal IRS phase shifts, random IRS phase shifts, and without IRS. The capacity bound, i.e., the performance of the optimal beam forming, can be obtained by assuming perfect CSI.\footnote{As we consider LoS channels, the optimal phase shifts ${\mathbf{\Theta }}$ can be obtained under perfect CSI, and the capacity bound is achievable by using SVD beamforming with water-filling power alloaction.} We consider a THz IRS-assisted UM-MIMO indoor scenario, in which a BS, a user, and an IRS are at the three vertices of a triangle with side length of $5$ m. The numbers of BS/user antennas and IRS elements are all $128$. The number of narrow beams in the bottom of the training codebook is $243$. The operating frequency is set to $140$ GHz with background noise power $-80$ dBm.  Fig. \ref{irs11} shows the performance of different schemes in THz LoS-blockage case, where ${\bf{H}}_{\rm{LoS}}={\bf{0}}$. In the non-IRS-assisted scheme, we treat IRS as an indoor wall whose first-order ray attenuation is randomly set to between $5.8$ dB and $19.3$ dB compared to the LoS \cite{wall}. It is observed that the performance gains of the IRS-assisted schemes are significant compared to the schemes with random IRS (phase shifts) and without IRS. Besides, the performance of THz IRS-assisted beam training scheme is close to the IRS-assisted capacity bound. Fig. \ref{irs22} shows the performance of different schemes in THz non-blockage case, where both the BS-user link and the BS-IRS-user link provide propagation paths. One can see that the performance gained by IRS-assisted schemes is still notable. This is because the IRS can provide additional aperture gains via controllable reflection, so as to increase the received power at the user. These results also validate the effectiveness of THz IRS-assisted beam training in the non-blockage scenario.

\section{Multi-User THz UM-MIMO Beamforming}
In a single-user MIMO system, the baseband digital beamforming needs linear processing on all antenna elements at the transmitter or receiver, whereas in a multi-user MIMO system, it is generally assumed that no coordination among antenna elements at different users, and thus the single-user beamforming techniques, e.g., SVD operation, are no longer feasible in multi-user case. \cite{mumimo}.  In light of this, there are two scenarios to be investigated in terms of multi-user beamforming, i.e., \emph{downlink beamforming}, where the BS transmits independent data streams to multiples users simultaneously, and \emph{uplink beamforming}, where a group of users transmit their to the BS simultaneously \cite{mumimo2}. The main interest for multi-user MIMO is to distinguish independent data streams in the spatial domain, which is known as space-division multiple access (SDMA)\cite{sdma}.

\subsection{Multi-User UM-MIMO Channels}
The two scenarios come with two kinds of multi-user MIMO channels, respectively. The downlink channel is referred to as a \emph{broadcast (BC) channel}, while the uplink channel is referred to as a \emph{multiple access (MA) channel}.  The sum-rate is a common performance metric for multi-user MIMO systems and there are some notable differences between BC and MA. In the former, the transmitted signal is a superposition of the data streams intended for all users, subject to the total transmit power constraint. By comparison, in the latter, the signal transmitted from each user is affected by other co-scheduled users, subject to individual power constraints. Uplink and downlink transmission policies depend on the accuracy of the channel state information at the transmitter (CSIT) and receiver (CSIR), and the optimal beamforming and power allocation between BC and MA channels have an implicit connection, known as duality \cite{duality}. In a multi-user UM-MIMO system with hybrid beamforming architecture as shown in Fig. \ref{mumimo}, the MIMO channel can be classified as an antenna-port channel (denoted by ${{\bf{H}}_{{\rm{ant}}}}$) and a digital-port channel (denoted by ${{\bf{H}}_{{\rm{dig}}}}$). The analog beamforming is the bridge between ${{\bf{H}}_{{\rm{ant}}}}$ and ${{\bf{H}}_{{\rm{dig}}}}$.

\subsection{THz UM-MIMO Beamforming Strategy}

In the THz UM-MIMO systems, the array at the transceivers has an extremely large number of antennas, which results in a large-dimensional matrix ${{\bf{H}}_{{\rm{ant}}}}$. Thus, the accurate channel estimation for ${{\bf{H}}_{{\rm{ant}}}}$ is time consuming and the multi-user precoding incurs high complexity. Here, we propose a cost-efficient solution which adopts the AoSA architecture and combines the beam training  (in the analog domain) and the precoding (in the digital domain).

Specifically, compared to digital beamforming, the AoSA can provide comparable performance in the low-rank channel using fewer RF chains. By performing the beam training in the analog domain, the digital-port channel is expected to have a high gain. Then,  the multi-user channel estimation and precoding are merely applied in the digital ports with a much smaller overhead. The beam training can be regarded as a compression process from ${{\bf{H}}_{{\rm{ant}}}}$ to ${{\bf{H}}_{{\rm{dig}}}}$. By applying the beam training, the system can find the LoS path in the multi-user UM-MIMO channel without the need of high-complexity calculations. Although this strategy sacrifices some DoFs in ${{\bf{H}}_{{\rm{ant}}}}$, its performance may not lose much given the fact that ${{\bf{H}}_{{\rm{ant}}}}$ is generally LoS dominant \cite{3DB2}. In particular, the LoS link can provide two DoFs for spatial multiplexing by using different antenna polarization modes  
\cite{polari}.  
\subsection{Linear Precoding Algorithms}
The goal of precoding in the digital domain is to convey independent data streams to different users. It was shown that Costa’s dirty-paper precoding can achieve the capacity region of the Gaussian BC multi-user MIMO channel\cite{costa}. However, it comes at the expense of non-linear processing. In comparison, linear precoding algorithms are more practical and well-explored in these years, which can be classified into the following three major categories: 

\subsubsection{Orthogonal Space-Division Multiplexing (OSDM)}
The core idea of this class is to completely eliminate the inter-user interference by dividing the space into multiple orthogonal sub-spaces and allocating them to different users or data streams. This method was first used in multi-user MISO scenarios, with the name channel-inversion or zero-forcing (ZF) \cite{ZF}. In the multiple-user MIMO scenarios, block-diagonalization (BD) \cite{BD} is used to create orthogonal sub-spaces by projecting each user's channel into the null-space of other users' channels. Based on ZF and BD, many extended precoding methods are studied to save the transmit power, reduce the computational complexity, or combat the effect of noise, e.g.,  iterative null space-directed SVD (Iterative Nu-SVD)\cite{NU-SVD}, eigen-based ZF (EZF)\cite{EZF}, coordinated BD (CBD)\cite{CBD}, regularized BD (RBD)\cite{RBD}, QR-Regularized BD (QR-RBD)\cite{QR-RBD}, complex lattice reduction RBD (CLR-RBD)\cite{CLR-RBD}.

\subsubsection{Successive Interference Elimination (SIE)}
The core idea of this class is to design each user's precoding matrix one by one and to make it lie in the null space of previous users' channels. As such, in a system with $K$ users, user $k$ optimizes its precoding to compensate for the interference received from users $1$, $2$, ..., $k-1$, and subject to the constraint that it does not interfere with any of those users. By leveraging the information of previous users' precoding matrices, one can eliminate the interference successively. Two specific algorithms are presented in \cite{BD} and \cite{SZF}, which are referred to as successive optimization (SO) and successive zero-forcing (SZF), respectively. However, these approaches generally have high-computational complexity since the solution depends on the design ordering of users. For a system of $K$ users, there are $K!$ sequentially optimized solutions, and are needs to find the best ordering to optimize the performance.

\subsubsection{Minimizing the Product of Mean Squared Error Determinants (PDetMSE)}
 The core idea of this class is to transform the classic sum-rate optimization into a tractable form, i.e., minimizing the PDetMSE. Specifically, by assuming that a minimum mean-squared-error (MMSE) decoder is applied at the receiver, the rate maximization problem with a sum power constraint is equivalent to a PDetMSE minimization problem under a sum power constraint, which can be solved by sequential quadratic programming (SQP) \cite{duality}. Based on this result, many approaches have been proposed to approximate or reformulate the objective function of PDetMSE, so as to simplify the difficulty. These approaches are known as minimizing the product of mean squared errors (PMSE)\cite{PMSE}, minimum total mean square error criterion (T-MMSE)\cite{T-MMSE}, and weighted sum minimum mean square error (WMMSE)\cite{WMMSE}. 

\begin{figure}[t]
\centering
\includegraphics[width=3.5in]{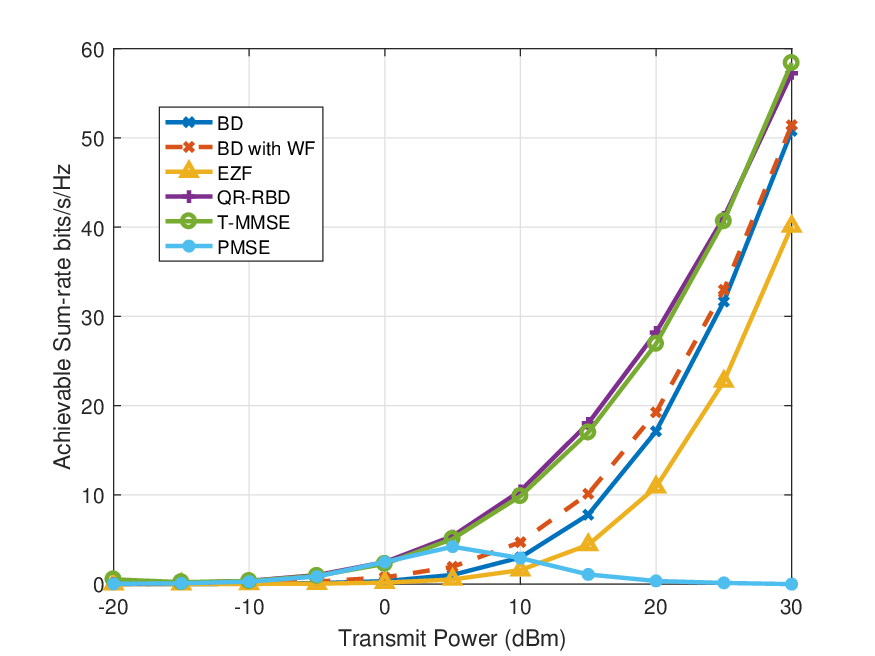}
\caption{The performance of different linear beamforming algorithms on THz UM-MIMO digital-port channel}\label{murate}\vspace{-2pt}
\end{figure}

The above algorithms have been proven to achieve decent performance under the Rayleigh channel, which models the antenna-port channel with small-scale arrays for frequency below $6$ GHz. In THz UM-MIMO systems, the precoder is optimized based on the digital-port channel ${{\bf{H}}_{{\rm{dig}}}}$ rather than the Rayleigh channel, and thus the performance of these algorithms needs to be re-evaluated. Now let us consider a THz UM-MIMO system, where a BS serves $8$ users randomly located $50$ meters away on a $140$ GHz carrier. The background noise power at the receiver is set to $-80$ dBm. The BS equips $16$ RF chains each connected to a subarray with $256$ antenna elements and each user equips $2$ RF chains each connected to a subarray with $16$ antenna elements. Assuming the beam alignment is successful in the analog domain with two antenna polarization modes, we select five low-complexity algorithms, i.e.,  BD\cite{BD}, EZF\cite{EZF},  QR-RBD\cite{QR-RBD}, PMSE\cite{PMSE}, and T-MMSE\cite{T-MMSE}, to seek solutions for the precoder to the digital-port channel. The averaged sum-rate over 1,000 random channel realizations is shown in Fig. \ref{murate}. It can be observed that EZF suffers notable performance loss and PMSE is no longer effective in a high-transmit power regime. BD with water-filling (WF) power allocation has a small improvement compared with BD between the power of $10$ and $20$dB. The performance of QR-RBD and T-MMSE are comparable, with the largest achievable sum-rate among the five algorithms. Thus, QR-RBD and T-MMSE might be the promising digital beamforming algorithms in THz multi-user UM-MIMO channels.

\section{Wideband Beamforming}\label{wb}
Based on the IFFT technologies, digital wideband beamforming can be realized by concurrently applying precodings for channels of different frequencies. In this section, we focus on the wideband beamforming in the analog domain, i.e., using the phased array.   There are two main effects, i.e., spatial-wideband effect and frequency-wideband effect, that need to be considered in the THz UM-MIMO systems. In sight of this, we introduce three solutions, true-time-delayer (TTD) beamformer, max-min beamformer, and beam split aggregation and multiplexing (BSAM) in the THz wideband communications.  

\subsection{Spatial-Wideband Effect}
Now, we consider a wideband communication with a very high symbol rate. When the plane wave arrives slantingly in a large array, the signal envelope received by different antenna elements might not belong to the same symbol and thus the phased array cannot effectively combine the signals, which is called the  \emph{spatial wideband effect} \cite{wb1,wb2,wb3}.

\begin{figure}[t]
\centering
\includegraphics[width=3.3in]{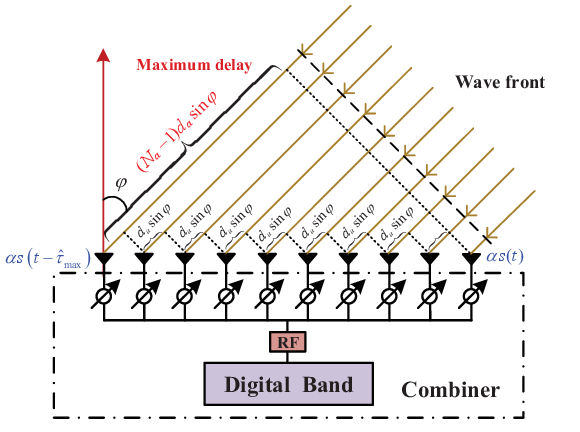}
\caption{Incoming signal with arrival angle $\varphi$ in large-scale ULA.}\label{wua}\vspace{-3pt}
\end{figure}

 Let $T_s$ represent the symbol period and the baseband signal can be expressed as $s(t) = \sum\nolimits_i {{\mathrm{sym}}[i]g(t - i{T_s})}$,
where ${\mathrm{sym}}[i]$ is the $i$-th symbol and $g(t)$ is the pulse shaping function, i.e., 
\begin{equation}
g(t)\left\{ {\begin{aligned}
&{1,\;\;0 \le t \le {T_s}}\\
&{0,\;\;{\mathrm{otherwise}}}
\end{aligned}} \right..
\end{equation}
For simplicity, we consider a SIMO ULA system and assume that the equivalent baseband signal received by the first antenna, i.e., the rightmost element shown in Fig. \ref{wua}, is $\alpha s(t)$, where $\alpha$ is the path loss. Since the time delay between adjacent elements is $\frac{{{d_a}\sin \varphi }}{c}$, the time delay at the $m$-th antenna element is 
\begin{equation}\label{71}
{\widehat \tau _m}=\frac{{(m - 1){d_a}\sin \varphi }}{c}.
\end{equation}
The phase difference at the $m$-th  antenna element is $2\pi  \cdot {{\hat \tau }_m} \cdot f = \frac{{2\pi (m - 1){d_a}\sin \varphi }}{\lambda }$. Thus, the equivalent baseband signal received by the $m$-th antenna element can be written as
\begin{equation}\label{73}
{y_m}(t) = \alpha s\left( {t - \frac{{(m - 1){d_a}\sin \varphi }}{c}} \right){e^{ - j\frac{{2\pi (m - 1){d_a}\sin \varphi }}{\lambda }}}.
\end{equation}
Let ${\widehat \tau _{\max }} = \frac{{({N_a} - 1){d_a}\sin \varphi }}{c}$ denote the maximum time delay. In the narrowband communication, $T_s$ is relatively large and we have ${\widehat \tau _{\max }} \ll {T_s}$. In this case, we can assume that $s\left( {t - \frac{{(m - 1){d_a}\sin \varphi }}{c}} \right) = s(t)$. Thus, the received signals are simplified as ${\mathbf{y}}(t) = \alpha {{\mathbf{a}}_r}(\varphi )s(t)$. As such, the multipath narrowband channel can be modeled as (\ref{chan}). This indicates that the received signal on each element has merely the phase difference and phase compensation can be used to provide coherent combining at the receiver. However, in the wideband communication, i.e., $T_s$ is close to or even less than ${\widehat \tau _{\max }}$, the antenna's first element and the last element may receive different symbols. Thus only using phase shifters for combination is infeasible. A larger number of elements $N_a$  yields possibly larger ${\widehat \tau _{\max }}$, leading to the spatial-wideband effect determined by the array size and symbol rate. 
To eliminate the spatial-wideband effect, we can adopt orthogonal frequency division multiplexing (OFDM) modulation and set its cyclic prefix duration larger than the maximum time delay.

\begin{figure}[t]
\centering
\includegraphics[width=3.5in]{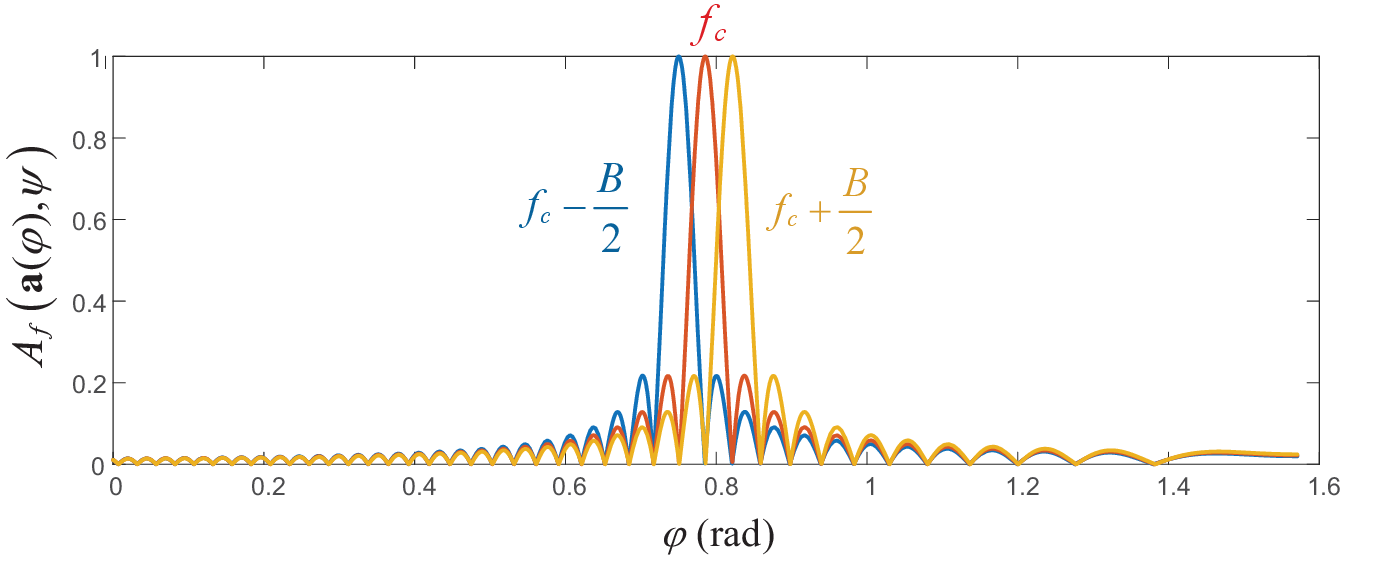}
\caption{Beam pattern received at different frequencies, where AoA $\psi =\pi/4$, $f_c=0.14$ THz, $B=10$ GHz, and $N_a=80$.}\label{squ}\vspace{-2pt}
\end{figure}
\subsection{Frequency-Wideband Effect}\label{fwe}
Even without the spatial-wideband effect, i.e., all antenna elements receive the same symbol, the signal components of different frequencies arrive at the array with different phase differences. Thus, using phased array that combines signals with the same phase compensation cannot simultaneously maximize the beam gains at all frequencies, which is called the \emph{frequency-wideband effect}. In other words, the beam pattern changes with the frequency of the signal, which is known as the phenomenon of \emph{beam squint} or \emph{beam split} \cite{bs1,bs2,bs3}.

The beam squint effect takes place at both the transmitter and the receiver. Without loss of generality, we analyze its effect in terms of a receive ULA. Let $B$ and $f_c$ denote the baseband bandwidth and carrier frequency, respectively.  The wideband response array vector can be expressed as
\begin{equation}\label{warv}
{{\mathbf{a}}}(\psi ,f) = \frac{1}{{\sqrt {{N_a}} }}{\left[ {1,{e^{ - j\frac{{2\pi f{d_a}\sin \psi }}{c}}},...,{e^{ - j\frac{{(N_a - 1)2\pi f{d_a}\sin \psi }}{c}}}} \right]^T},
\end{equation}
where $f \in \left[ {{f_c} - \frac{B}{2},{f_c} + \frac{B}{2}} \right]$. Commonly, the  combiner is set based on the carrier frequency, i.e., ${\mathbf{a}}(\varphi ) \buildrel \Delta \over = {\mathbf{a}}(\varphi ,{f_c})$. Thus, the normalized beam gain of the combiner ${{\mathbf{a}}}(\varphi )$ in the direction of $\psi$ at frequency $f$ can be expressed as
\begin{equation}\label{y78}
\begin{aligned}
{A_f}\left( {{\mathbf{a}}(\varphi ),\psi } \right)&= \left|\sqrt{N} {{\mathbf{a}}{{(\varphi )}^H}{\mathbf{a}}(\psi ,f)} \right|^2\\
 &=\left| {\frac{{\sin [\frac{{{N_a}\pi }}{2}(\xi \sin \psi  - \sin \varphi )]}}{{\sqrt{N_a}\sin [\frac{\pi }{2}(\xi \sin \psi  - \sin \varphi )]}}} \right|^2,
\end{aligned}
\end{equation}
where  $\xi  = \frac{f}{{{f_c}}}$. Consider a wideband incoming signals with AoA $\psi =\pi/4$, $f_c=0.14$ THz, $B=10$ GHz, $N_a=80$. Fig. \ref{squ} plots the normalized beam gains, i.e., beam patterns, at frequency ${f_c}$, ${f_c} - \frac{B}{2}$, and ${f_c} + \frac{B}{2}$. As we can see, compared to the beam pattern at ${f_c}$, the other patterns have angle squints. It is worth mentioning that the beam squint direction is different between the transmitter and the receiver. To be exact, as shown in Fig. \ref{squ}, the 
\emph{beam pattern received} with high frequency moves towards $90$ degrees while that with low-frequency moves towards $0$. On the contrary, we would like to point out that the\emph{ beam pattern transmitted} with high frequency moves towards $0$ while the low-frequency beam moves towards $90$ degrees.

Let us revisit the normalized beam gain given in (\ref{y78}). For incoming signals with AoA of $\psi$ at frequency $f$, the optimal combiner ${\mathbf{a}}(\varphi )$ satisfies 
\begin{equation}\label{y79}
\begin{aligned}
&\;\;\;\;\xi \sin \psi  - \sin \varphi  = 0\\
& \Rightarrow \;\varphi  = \arcsin \left( {\frac{f}{{{f_c}}}\sin \psi } \right).
\end{aligned}
\end{equation}
Thus, the beam squint angle is given by $|\varphi-\psi|$. Here,  we note some insights for the beam squint effects as below. 
\begin{itemize}
\item When $f = f_c$, there is no beam squint as $\varphi =\psi$.
\item Beam squint effect increases with $\psi$.
\item Beam squint effect increases with $\frac{B}{{{f_c}}}$.
\end{itemize}

Due to the frequency-wideband effect, the signal at edge frequency  suffers from performance loss even under a successful beam alignment. As a result, the minimum beam gain over $ \left[ {{f_c} - \frac{B}{2},{f_c} + \frac{B}{2}} \right]$ is given by 
\begin{equation}
{A_f}(\varphi ) = \left| {\frac{{\sin (\frac{{{N_a}\pi B}}{{4{f_c}}}\sin \varphi )}}{{\sqrt{N_a}\sin (\frac{{\pi B}}{{4{f_c}}}\sin \varphi )}}} \right|^2.
\end{equation}
Accordingly, the beam squint loss can be written as $1 - {A_f}({\varphi _{\max }})$, in which $\varphi _{\max }$ is the maximum beam deflection. When the beam patterns at the highest and the lowest frequencies are splitting, the minimum beam gain is zero.  

\begin{figure}[t]
\centering
\includegraphics[width=3.5in]{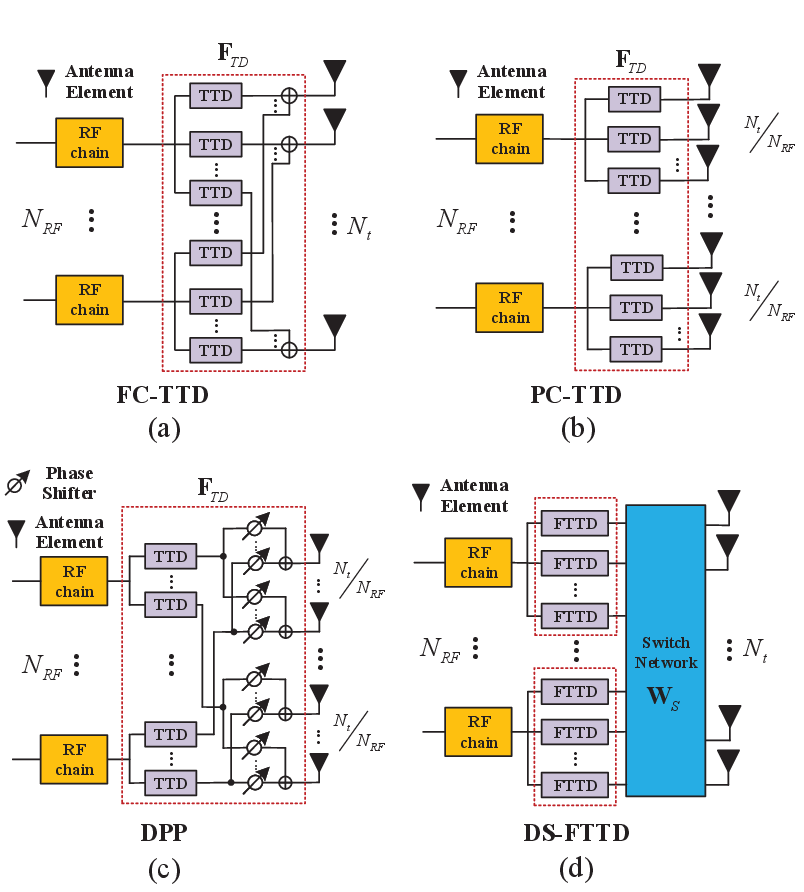}
\caption{(a) Full-connected TTD; (b) Partially-connected TTD; (c) Delay-phase precoding; (d) Dynamic-subarray with fixed TTD.}\label{widefig}
\vspace{-6pt}
\end{figure}

\subsection{TTD Beamformer}
The reason for both spatial-wideband and frequency-wideband effects is the time-delay spread over the antenna arrays, which cannot be eliminated by phased array.  Thus, the TTD beamformer can be used to address both effects, as the time-delay spread can be completely offset by TTD components\cite{ttd1,ttd2,ttd3,ttd4, ttd5, cmos13, pe1}.  
To realize the TTD, we can \emph{prolong the propagation length} or \emph{slow the wave velocity} of the electromagnetic signals in the precoder/combiner. The former approach has been mainly adopted in existing works \cite{ttd1,ttd2,ttd3}. Some emerging THz TTDs have been reported in \cite{ttd4, ttd5, cmos13, pe1}.   




There are four feasible TTD architectures in wideband beamforming, as presented in Fig. \ref{widefig}. In Fig. \ref{widefig} (a) and (b), full-connected TTD (FC-TTD) architecture and partially-connected TTD (PC-TTD) architecture replace the phase shift matrix in (\ref{hbfc}) and (\ref{hbpc}) with a time delay matrix, respectively. Although these architectures can directly apply the existing beamforming algorithms, THz time delay consumes more power than the conventional phase shifter \cite{sur381}, which is not cost-efficient when a large number of TTDs are required. To remedy this deficiency, delayed phase precoding (DPP) structures are used to provide a low-cost wideband effect compensation scheme, as shown in Fig. \ref{widefig} (c). The TTD layer inserted in the RF chains and phase shifters can support joint control of delay and phase to provide spatially aligned frequency-dependent beams over the entire bandwidth, thereby mitigating frequency-wideband effects while reducing hardware cost \cite{bs3}. Similarly, the authors in \cite{wbadd} proposed to combine the TTD devices and the conventional phase shifts to jointly design wideband beamforming with reduced implementation cost compared to the fully TTD arrays.
The infinite-precision TTD is impractical for hardware implementation. As a solution, a switching network is designed to flexibly select the appropriate time delay from a set of fixed TTDs, as shown in Fig. \ref{widefig} (d). The hardware complexity of dynamic-subarray FTTD (DS-FTTD) is lower than that of FC-TTD, PC-TTD, and DPP architectures, since there is no need to adjust the delay \cite{sur40,fttd}.

\subsection{Max-Min Beamformer}
The limitation of the phased array in wideband communication is that the phase shifts are flat over the signal band, which cannot cater to the wideband array response vector in (\ref{warv}) with $f \in \left[ {{f_c} - \frac{B}{2},{f_c} + \frac{B}{2}} \right]$. The conventional beamformer that maximally combines the signal power at $f_c$ might cause severe beam squint loss.  If we do not change the array architecture, i.e., still using the phased array,  a max-min beamformer is proposed to reduce the frequency-wideband effect. 

Specifically, in the wideband UM-MIMO systems,  the THz channel is frequency related and can be modelled as 
\begin{equation}
{{\bf{H}}_{{\rm{LoS}}}}(f) = \sqrt G {{\bf{a}}_r}(f){\bf{a}}_t^H(f),\;\;f \in \left[ {{f_c} - \frac{B}{2},{f_c} + \frac{B}{2}} \right].
\end{equation}
Thus, the performance metric in wideband communication systems should also be frequency-dependent and is denoted by ${\rm{U}}(f)$.  For example, the spectral efficiency of UM-MIMO can be expressed as 
\begin{equation}
{\rm{U}}(f) = {\log _2}\det \left| {{\bf{I}} + \frac{P}{{\sigma _n^2}}{{\bf{H}}_{{\rm{LoS}}}}(f){\bf{w}}{{\bf{w}}^H}{\bf{H}}_{{\rm{LoS}}}^H(f)} \right|.
\end{equation}
The design of the max-min beamformer ${\bf{w}}$ needs to cater to the metric over a wide band, which is formulated as
\begin{equation*}
{\rm{P}}(9):\;\;\mathop {{\rm{max}}}\limits_{\bf{w}} {\rm{ }}\mathop {{\rm{min}}}\limits_{f \in \left[ {{f_c} - B/2,{f_c} + B/2} \right]} {\rm{ U}}(f).
\end{equation*}
To the best of our knowledge, there has been no work considering the max-min beamformer design in THz UM-MIMO yet, which, however, is an effective solution for wideband communication and worthy of in-depth research in the future. To validate its effectiveness, we consider a beam gain metric and the problem is formulated as
\begin{figure}[t]
\centering
\includegraphics[width=3.5in]{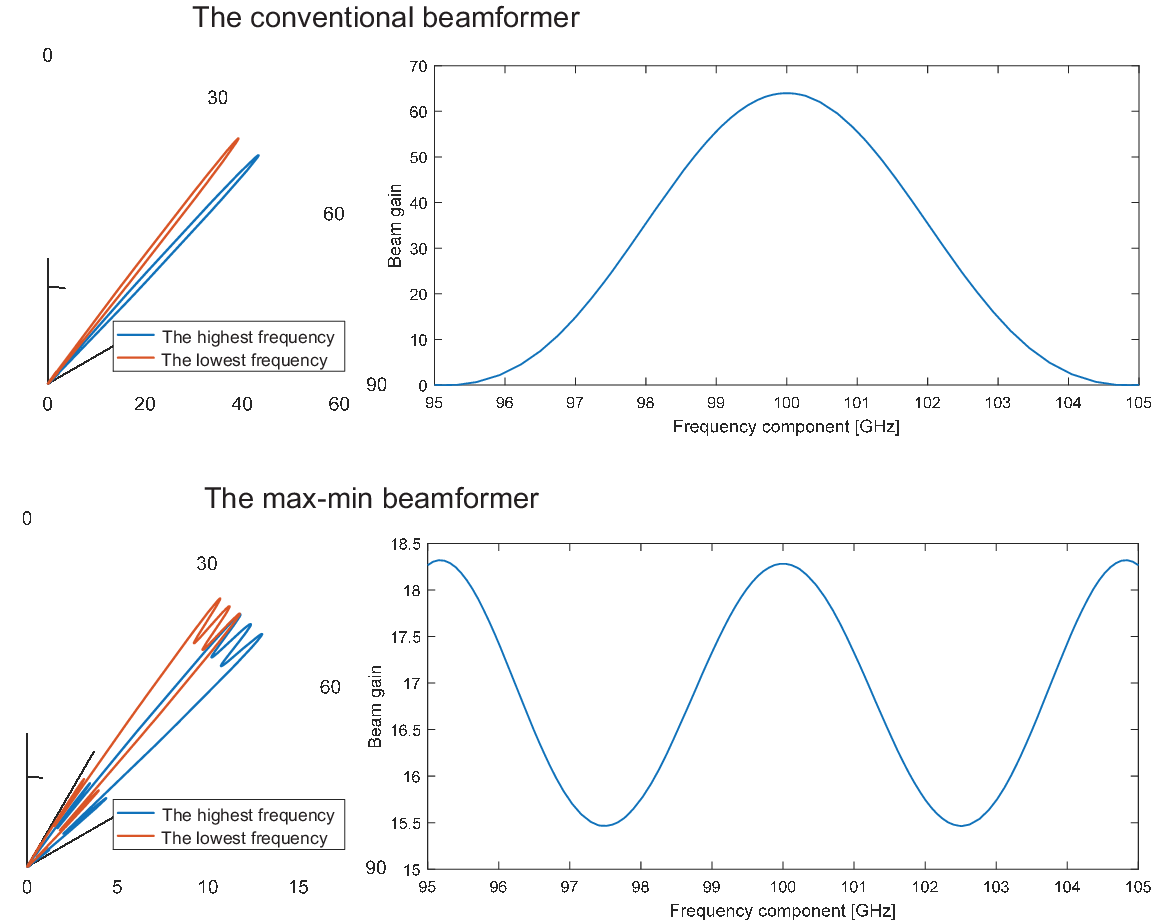}
\caption{Comparison of the beam patterns and beam gain between the conventional beamformer and the max-min beamformer.}\label{maxm}
\vspace{-6pt}
\end{figure}
\begin{figure*}[t]
\centering
\includegraphics[width=7in]{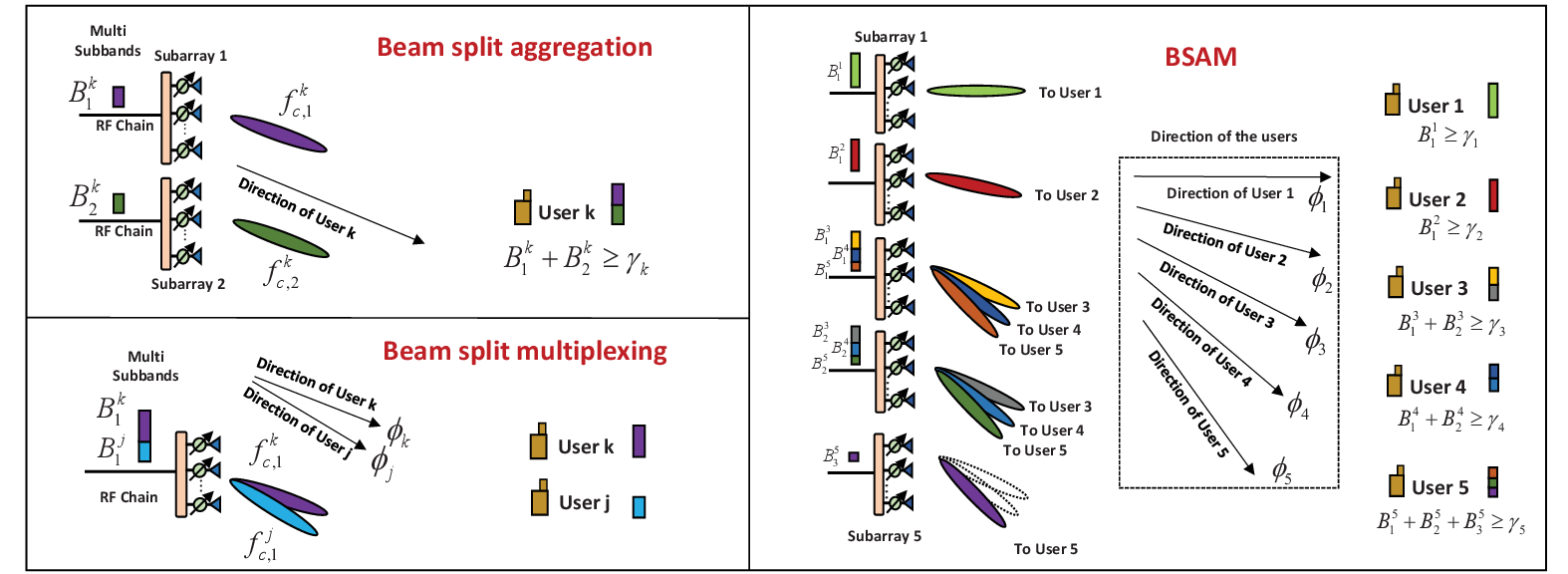}
\caption{Comparison of the beam patterns and beam gain between the conventional beamformer and the max-min beamformer.}\label{bsam}
\vspace{-6pt}
\end{figure*}

\begin{equation*}
\begin{aligned}
{\rm{P}}(10):\;\;&\mathop {{\rm{max}}}\limits_{\bf{w}} \mathop {{\rm{min}}}\limits_{f \in \left[ {{f_c} - B/2,{f_c} + B/2} \right]} {\rm{ }}\left| {{{\bf{w}}^H}{\bf{a}}(\psi ,f)} \right|^2\\
&{\rm{s}}{\rm{.t}}{\rm{. }}\;\;{\bf{w}}\left( i \right) = 1,\;\;i = 1,2,...,{N_a}.
\end{aligned}
\end{equation*}
Let us consider a system with $f_c=100$ GHz and $B=10$ GHz. By applying a proper algorithm to solve ${\rm{P}}(10)$ (for which the details are omitted), we design a max-min beamformer ${\bf{w}}$ with AoA $\psi=40^{\circ}$ and its beam pattern and beam gain over the band is shown in Fig. \ref{maxm}. It can be seen that the beams of the conventional beamformer are split and thus the wideband beam gain (i.e., the minimum beam gain over the band) is zero. In comparison, the max-min beamformer can form a wide beam, which avoids the beam misalignment under beam squint.

\subsection{BSAM}
An interesting concept of BSAM arises recently for realizing the wideband beamforming\cite{bsquint} with AoSA. The BSAM scheme is to use the conventional beamformer, i.e., array response vectors, on multiple phased arrays and judiciously optimize the frequency allocation for these arrays to simultaneously serve multiple users in wideband communications.

Specifically,  if user $k$ is not in the perpendicular direction of the array,  the effective signal band at user $k$ is limited due to the beam split effect. To satisfy the user's wideband demand, a \emph{beam split aggregation} scheme is proposed by using multiple subarrays to transmit multiple beams, respectively.  As shown in Fig. \ref{bsam} (left top), each subarray transmits a beam with an independent subband, where subarray 1 uses beamformer ${\bf{w}}_1 = {\bf{a}}(\phi_k ,f_{c,1}^k)$ and subarray 2 uses beamformer ${\bf{w}}_2 = {\bf{a}}(\phi_k, f_{c,2}^k)$, and the user aggregates the subbands to meet the wideband demand. Meanwhile,  a \emph{beam split multiplexing} scheme is proposed where each subarray may serve multiple users as shown in Fig. \ref{bsam} (left bottom). To be exact, if we use ${\bf{a}}(\phi_k ,f_{c,1}^k)$ to transmit the beam with central frequency $f_{c,1}^k$ to the user $k$, we define a beam split coverage (BSC) for this beamformer as
\begin{equation}
{\rm{BSC}}= [\psi _{k}^{\min },\psi _{k}^{\max }],
\end{equation}
where if ${\phi _k} \ge 0$ we have 
\begin{equation}
\begin{aligned}
&\psi _{k}^{\min } = \arcsin (\frac{{f_{c,1}^k}}{{{f_{\max }}}}\sin {\phi _k}),\;\\
&\psi _{k}^{\max } = \arcsin (\frac{{f_{c,1}^k}}{{{f_{\min }}}}\sin {\phi _k}),
\end{aligned}
\end{equation}
in which  $f_{\min}$ and $f_{\max}$ are the minimum and maximum frequency available in the systems, respectively.  If ${\phi _k} < 0$, we need to switch the values of $\psi _{k}^{\min }$ and $\psi _{k}^{\max }$. When the $j$-th user is in the  BSC, i.e., ${\phi _j} \in {\rm{BSC}}$, we can allocate a frequency band with bandwidth $B_1^j$ to user $j$, and the band range is given by
\begin{equation}
\Omega _1^j = \left[ {\frac{{f_{c,1}^k\sin {\phi _k}}}{{\sin {\phi _j}}} - \frac{{B_1^j}}{2},\frac{{f_{c,1}^k\sin {\phi _k}}}{{\sin {\phi _j}}} + \frac{{B_1^j}}{2}} \right],
\end{equation}
where the signals within the band $\Omega _1^j $ will be steered to user $j$ by the common beamformer.

The beam split aggregation scheme can satisfy the user's wideband demand but incurs a high cost since each subarray merely serves one user. On the other hand, the beam split multiplexing scheme enables one subarray to serve multiple users, but the effective bandwidth received by each user is limited. By combining their advantages, as shown in Fig. \ref{bsam} (right), the BSAM scheme can satisfy the user's wideband demand while reducing the number of subarrays used.

\section{Existing THz MIMO Arrays}
Currently, different fabrication techniques are considered to implement THz band antennas. These techniques can be roughly divided into three categories: electronic-based, photonics-based, and new materials-based. In the electronic approach, a variety of antenna types based on different materials and processes have been reported, including but not limited to horns \cite{manu1, manu2}, reflectors \cite{manu3, manu4}, and cavity-backed slot antenna arrays \cite{manu5}. At the same time, methods such as photo-conductive antennas \cite{manu6, manu7} and silicon-based lenses \cite{manu8, manu9, manu9-1} have been tried in the field of photonics. In addition, the latest option for developing antennas in the THz band is based on the new phase-change materials such as vanadium dioxide (VO2)\cite{table4}, graphene \cite{gra1} and liquid crystal (LC)\cite{lc1}. A detailed investigation of graphene THz antennas is provided in \cite{gra2}. 

Researches on individual THz band antennas reveal the possibility of making up the THz regions in the electromagnetic spectrum. However, the performance of individual THz antennas is mostly limited by the low transmit power or poor dynamic beam scanning capability. To this end, it is necessary to study the large-scale THz antenna array that supports high transmit power and adjustable direction. Fortunately, the very short wavelength of the THz band supports the integration of a large number of antenna arrays with a small footprint, which helps to achieve dynamic directional high-gain beamforming via antenna arrays. Although some methods have been tried for the implementation of THz antenna arrays \cite{manu10, manu11, manu12}, these arrays only enhance the directional gain and do not support dynamic beamforming. In the following, we will focus on array fabrication techniques that have the potential to support dynamic beamforming.

\begin{figure*}[t]
\centering
\includegraphics[width=7in]{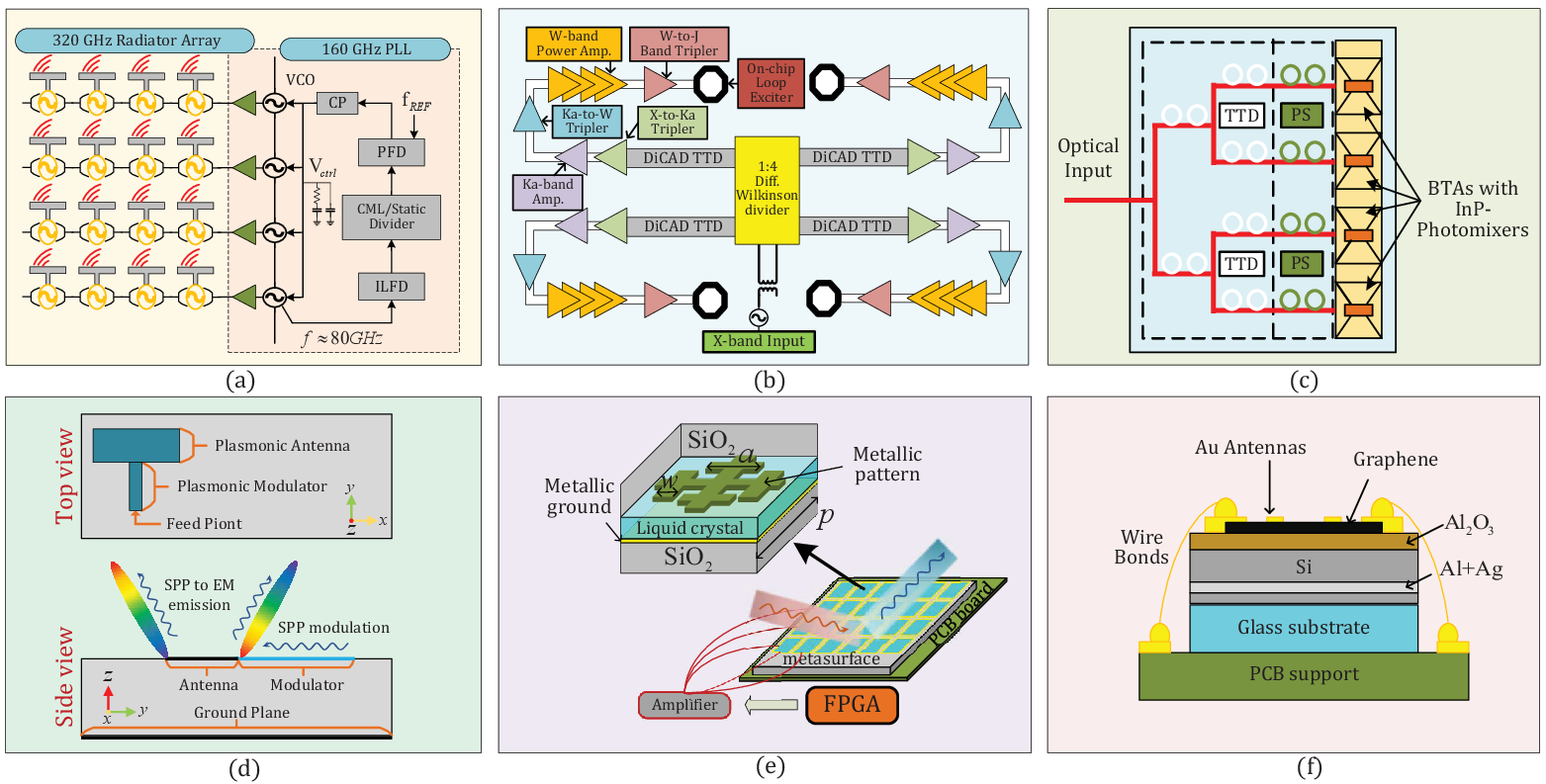}
\caption{Existing THz MIMO arrays: (a) on-chip phase-locking array; (b) chip-scale dielectric resonator antenna array; (c) digital controlled TTD antenna array; (d) graphene plasmonic array; (e) LC reflectsurface; (f) graphene reflectarray.}\label{mimoarray}\vspace{-2pt}
\end{figure*}

\subsection{Electronic and Photonic Approaches}
For supporting the integration of small-size on-chip antennas, silicon-based THz antennas, which are mainly supported by CMOS and silicon-germanium (SiGe) technologies, have been developing rapidly in recent years. Silicon-based technologies have the advantages of simple structure, easy array integration, small size, and potential low-cost \cite{sur320}. Its operating frequency is mostly limited to $0.5$ THz \cite{cmos2, meta3}, which can provide solutions in the lower frequency band of the THz spectrum. Due to the relatively mature integration capabilities of silicon-based, many small-scale phased arrays based on various silicon-based processes have been reported \cite{cmos4, cmos5, cmos6, cmos7, cmos8, cmos9, cmos10, cmos11}. An example is given in \cite{cmos12}, which shows a $4 \times 4$ URPA using $130$ nm SiGe bipolar-CMOS (BiCMOS) technology at $320$ GHz. As shown in Fig. \ref{mimoarray} (a), this array is made up of 16 elements and a fully-integrated $160$ GHz phase-locked loop (PLL), consisting of charge pump (CP), phase/frequency detector (PFD), current-mode logic (CML), and injection-locking frequency divider (ILFD).  The authors in \cite{cmos13} propose a 280 GHz $2 \times 2$ chip-scale dielectric resonator antenna array. As shown in Fig. \ref{mimoarray} (b), this array incorporates a balun for an X-band input signal, a $1:4$ Wilkinson divider, and four x27
active multipliers chains to drive the elements.

Despite the various progress that has been witnessed and is still ongoing in the field of silicon-based arrays, the drastic power drop associated with this technology is a major bottleneck \cite{sur320}. In the photonic approach, schemes for THz dynamic beam scanning are designed. For instance, frequency-scanning antennas can be used to control THz beam steering \cite{pe1}. 
In \cite{pe2}, the proposed photoelectric phase shifter can control $300$ GHz beam scanning within $50$ degree. Recently, the optical TTD phase shifters are also employed to offer stable time delay for wideband communications \cite{pe1}. Fig. \ref{mimoarray} (c) presents the schematic view of an optical TTD-based chip, wherein the input optical signals will be converted to 300 GHz frequency region by the InP photomixers and finally radiated by the $1 \times 4$ bow-tie antenna.  


\subsection{New Materials Approach}
In addition to the electronic and photonic processes, new materials provide another way to achieve high-performance THz antennas \cite{gra3, gra30, lc2}. Graphene, i.e., as a two-dimensional form of graphite, has attracted the attention of the scientific community due to its unique electronic and optical properties. Compared with conventional electronic materials, graphene is highly tunable, so it can be used to implement devices that support dynamic beamforming \cite{gra4, gra4-1}. 
Fig. \ref{mimoarray} (d) shows the working principle and design of the THz front end in \cite{gra4}.

Graphene antennas at the THz band with reconfigurable radiation patterns have been developed in \cite{gra5, gra6, gra7}. In addition to individual graphene antennas, small-scale graphene antenna arrays have been tried in \cite{gra8, gra9, gra10}, but the beam scanning range has not been piratically tested. Furthermore, \cite{gramimo1} proposed a reconfigurable MIMO antenna system for THz communications. \cite{gramimo2} envisages the use of graphene to implement UM-MIMO plasmonic nano-antenna arrays, which can implement a $1024  \times 1024$ UM-MIMO system at 1THz with arrays that occupy just $1$ mm$^2$. Liquid crystal and graphene also show application potential in reconfigurable reflectarrays \cite{reflc1, reflc2, refgr3, refgr2}. An example is given in \cite{reflc1}, which gives the theoretical analysis of the LC reflectarray at $0.67$ THz. As shown in Fig. \ref{mimoarray} (e), it consists of a $24$ element linear array, and each element is composed of $50$ rows and $2$ columns of unit cells with metal-insulator metal-resonator structure. In \cite{refgr3}, the graphene-based reconfigurable metasurface is designed to achieve beam control at $2$ THz. Authors in \cite{refgr2} experimentally demonstrate a $0.98$ THz graphene reflectarray metasurface that can achieve THz beam steering. As shown in Fig. \ref{mimoarray} (f), the reflectarray is mounted on a printed circuit board (PCB) and wire-bonded, where the substrate is comprising a reflective conductive ground plane and a dielectric spacer. It is worth pointing out that, so far, most new materials-based THz antenna arrays are still in the stage of theoretical design and analysis. The establishment of a complete array architecture for true THz frequencies with dynamic beamforming is not fulfilled. To sum up, Table \ref{ftta} shows the reported THz antenna arrays with dynamic beam scanning capability.

\begingroup
\renewcommand{\arraystretch}{1.2} 
\begin{table*}[]
\centering
\caption{Antenna arrays with dynamic beam scanning capability in THz band.} \label{ftta}
\vspace{-5pt}
\begin{threeparttable}  
\begin{tabular}{|l|l|l|l|l|l|l|}
\hline
Freq (Hz) & Size & Process             & Beam scan & Gain & Antenna type          & Reference    \\ \hline
$280$ G      & $4\times4$ arrays  & $45$ nm SOI CMOS   & $80$°/$80$° \tnote{1}  & $16$ dBi & on-chip               & \cite{cmos5}  \\ \hline
$140$ G      & $2 \times 4$ arrays & $65$ nm CMOS           & $40$°             & - & multi-chip            & \cite{cmos6}  \\ \hline
$0.53$ T     & $1 \times 4$ arrays & $40$ nm CMOS          & $60$°                    & $11.7$ dBi & patch          & \cite{cmos7} \\ \hline
$400$ G  & $1 \times 8$ arrays & $45$ nm SOI CMOS      & $75$°                    & $12$ dBi & patch        & \cite{cmos8} \\ \hline
$0.34$ T     & $2 \times 2$ arrays & $130$ nm SiGe BiCMOS & $128$°/$53$° \tnote{1}    & - & patch     & \cite{cmos9}  \\ \hline
$320$ G      & $1 \times 4$ arrays & $130$ nm SiGe BiCMOS & $24$°                    & $13$ dBi & patch  & \cite{cmos10} \\ \hline
$338$ G      & $4 \times 4$ arrays & $65$ nm CMOS     & $45$°/$50$° \tnote{1}
& $18$ dBi & microstrip  & \cite{cmos11} \\ \hline

$317$ G & $4 \times 4$ arrays & $130$ nm SiGe BiCMOS & - &   $17.3$ dBi & return-path gap coupler & \cite{cmos12} \\ \hline

$280$ G & $2 \times 2$ arrays & $65$ nm CMOS & $30$° & $12.5$ dBi & dielectric resonator  & \cite{cmos13} \\ \hline

$300$ G & $1 \times 4$ arrays & photonic &  $90$° & $10.6$ dBi & bow-tie antenna   & \cite{pe1} \\ \hline

$1.05$ T     & $4 \times 4$ arrays & graphene   & -    &  $13.9$ dBi & dipole  & \cite{gra8}    \\ \hline
$1.1$ T      & $2 \times 2$ arrays & graphene  &  $60$° 
&  $8.3$ dBi & patch  & \cite{gra9}    \\ \hline
 $220-320$ G     & $600$ elements & metallic  & $48$°
& $28.5$ dBi &  frequency scanning  & \cite{gra10}   
\\ \hline
$0.8$ T & $2 \times 2$ arrays & graphene & - &  - & photoconductive & \cite{gra11} \\ \hline

$1.3$ T & $25448$ elements & graphene & - &  $29.3$ dBi & reflectarray & \cite{gra3} \\ \hline

$220-320$ G & $8 \times 8$ arrays & brass sheets & $50$°/$45$° \tnote{1}  &  $17$ dBi & frequency scanning & \cite{table1} \\ \hline

$100$ G & $54 \times 52$ cells & liquid crystals & $55$° &  $15$ dBi & reflectarray & \cite{table2} \\ \hline

$345$ G & - & liquid crystals & $20$° &  $35$ dBi & reflectarray & \cite{table3} \\ \hline

$100$ G & - & VO2 & $44$°/$44$° &  - & metasurfaces & \cite{table4} \\ \hline

$115$ G & $39 \times 39$ cells & liquid crystals & $20$° &  $16.55$ dBi & reflectarray & \cite{table5} \\ \hline

\end{tabular}
\begin{tablenotes}    
\footnotesize               
\item[1] In both azimuth and elevation
\end{tablenotes}    
\end{threeparttable}  
\vspace{-2pt}      
\end{table*} 
\endgroup

To sum up, some promising fabrication techniques for implementing UM-MIMO antenna arrays have been developed in the THz range, and future research may focus on seeking potential solutions for better beam flexibility as well as a larger array size, e.g., i) improving the dynamic beamforming capabilities of the THz arrays, including adjustment accuracy and scanning range, ii) increasing the size of the antenna array and pushing it to the level of thousands of elements, iii) and mutual coupling effects caused by large-scale integration.  

\section{Emerging Applications}

\begin{figure*}[t]
\centering
\includegraphics[width=7in]{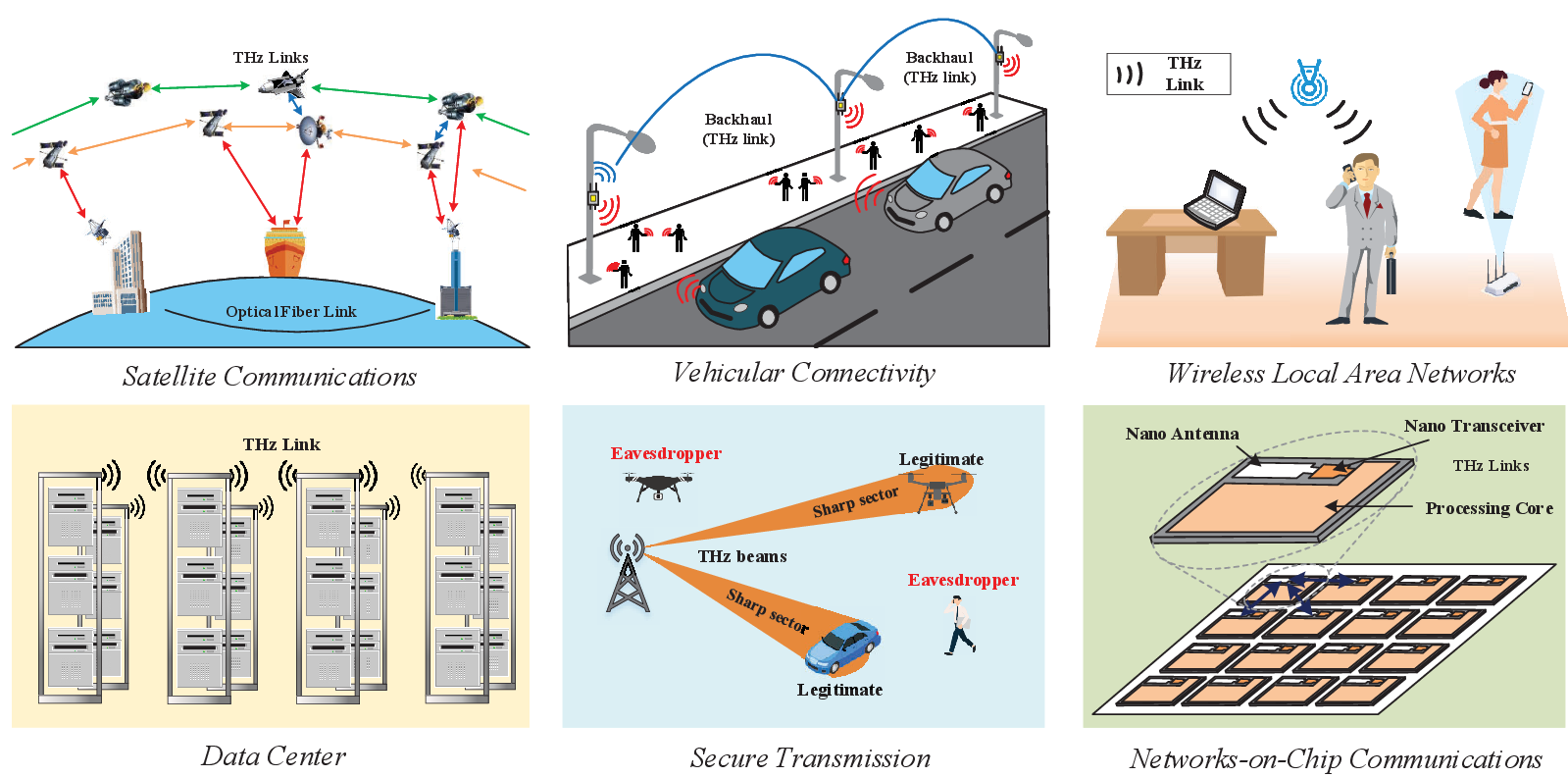}
\caption{THz communication with beamforming technologies in six different scenarios.}\label{appli} \vspace{-2pt}
\end{figure*}

In this section, we introduce some applications of beamforming technologies for THz communication. Specifically, we discuss six major scenarios also shown in Fig. \ref{appli} and their envisioned benefits.

\subsection{Satellite Communications}

In the past, free space optical communication has been extensively studied for realizing satellite networks since it allows high-bandwidth data transmissions, which are difficult to be detected and intercepted \cite{sc1}. Recently, research progress has shown the prospect of using THz communication to build satellite networks, thanks to its high bandwidth and negligible molecular absorption in this scenario. In addition to the advantages of high transmission rate and security of directional optical communication, THz communication has higher energy efficiency and easier beam control. In the future, the miniaturized THz communication systems can be applied to the high-speed inter-satellite communication of satellite clusters and satellite-to-ground communication to realize the integrated space-air-ground communication envisaged by 6G \cite{sc2}.

\subsection{Vehicular Connectivity}
To realize the concept of  intelligent transportation system, vehicle communication network has been extensively studied, which roughly includes three types of scenarios: vehicle to vehicle (V2V), vehicle to infrastructure (V2I), and vehicle to anything (V2X) \cite{sur22, vc2}. Vehicle networks require high data rates, low latency, and reliable communications. For instance, real-time traffic demands a data rate of $50$ Mbps, and auxiliary communication for autonomous driving requires a delay  less than $10$ ms and reliability of $99.999$\% \cite{sur33}. For the era of wireless interconnected smart cars, THz communication is anticipated to provide ultra-capacity, low-latency, high-mobility, and high-reliability for safe autonomous driving and intelligent transportation system (ITS) \cite{sur32, vc5}. It has been shown that THz wireless communication can support high-speed mobile railway application scenarios through dynamic beam tracking technology \cite{vc6}. The channel propagation characteristics of the $300$ GHz carrier in the V2I scenario have been explored in \cite{vc7}, which can be used to support the link-level and system-level designs for future THz vehicular communications.

\subsection{Wireless Local Area Networks}
The capability of THz communication of achieving short-distance high-speed coverage in indoor scenarios can be leveraged to enhance the WLAN performance. Considering that the transmission capacity of optical fiber links is much higher than that of conventional WLAN with mmWave band,  THz WLAN can enable seamless ultra-high-speed connections between high-speed wired networks and user mobile devices.  THz beamforming provides the ability to implement bandwidth-intensive applications, such as virtual reality, high-definition holographic video conferencing, and multi-user concurrent high-speed data transmission \cite{sur15}.
At present, the achievable communication distance of THz is small, which drives the community to propose the idea of ``information shower", or ``kiosk application", to maximize the capabilities of T-WLAN \cite{kiosk}. The kiosk application recommends deploying THz access points (APs) in specific high hotspot areas (e.g. public building entrances, shopping mall halls, etc.) to provide local high-rate transmissions \cite{st1}.

\subsection{Wireless Data Center}
As the demand for cloud service applications increases steadily, data centers have become an important part of modern Internet architecture. However, wired data centers result in high complexity, power consumption, maintenance costs, and large space occupied by cables. \cite{dc1} proposes to introduce a wireless link in the data center to achieve higher reconfigurability and dynamic operation. THz communication technologies have the potential to provide dense and extremely fast wireless connections for data centers with high flexibility \cite{dc2, dc3}. Some researchers also point out that combining wireless communications with existing data center architectures can help reduce costs \cite{dc7}. There have been some investigations on the THz channel modeling for wireless channel environment in data center\cite{dc4, dc5, dc6}. 

\subsection{Secure Transmission}
The THz beamforming via ultra-massive antenna elements to create highly directional beams brings many benefits to secure transmission applications, especially in commercial scenarios and secure payment scenarios \cite{st1,st2}.  
To be exact, the directional beams result in fewer scattering components and space sparsity, which prevents any node located outside the beam direction from eavesdropping the information \cite{st1}. In the meanwhile, as the transmit and receive THz beams are aligned within specific angle pairs, interference from  other angles can be considerably reduced.

Despite the above benefits, information security is still imperfect in the vicinity of the THz beam propagation path. Note that it is usually the case that the eavesdropper’s antenna is located in the signal propagation path to monitor the signal. In particular, \cite{nature} shows that eavesdroppers can place small (compared to the beam size) scatterers or even beam splitters on the legal receiving path to radiate legal signals to undesired positions.  To tackle this challenge, \cite{st3} controls the coverage distance of the beam to improve the security of THz communication by optimizing carrier frequencies, power allocation, and rate distribution on each sub-band. In addition, it has been validated that integrating IRS to THz communication can further enhance the physical-layer security in the  wiretap channel \cite{st4, st5, st6}. 


\subsection{Networks-on-Chip Communications}
Typical computer cores must constantly interact with each other to share common data and synchronize their activities. However, with the increase in the number of cores on the chip, the conventional wired topology is not enough to ensure high-speed interaction under the predicament of complex multi-core wiring \cite{st1}. \cite{ncc2} proposes to use graphene-based nano-antennas to replace on-chip wiring to realize on-chip wireless communication. The chip-to-chip link can only be enabled with a smaller transceiver (sub-millimeter level). By using planar THz nano-antennas to create ultra-high-speed links, it is expected to meet the stringent requirements of on-chip scenarios with limited area and dense communications \cite{sur15}.

\begin{center}
\begin{table*}[h]
\caption{The characteristics of wireless communication frequency bands.} 
\label{table_rivalry}
\begin{tabular}{|>{\columncolor[HTML]{C0C0C0}}m{2.5cm}|m{0.2\textwidth}<{\centering}|m{0.25\textwidth}<{\centering}|m{0.31\textwidth}<{\centering}|}
\hline
 & \textbf{sub-6 GHz}  & \textbf{mmWave}& \textbf{THz }
\\\hline

\textbf{Data rates}  & Low (below $1$ Gbps)  & Medium (up to $10$ Gbps)   & High (up to $100$ Gbps) 
\\\hline
\textbf{Bandwidth} & Small (around 100 MHz)  & Wide (less than $10$ GHz)   & Ultra-wide (up to $100$ GHz)   
\\ \hline
\textbf{Multipath}   & Rich scattering   & Few path  & LoS-dominant
\\\hline
\textbf{Path loss} & Mainly spreading loss & Spreading loss and slight molecules absorption (tens of dB/km) & Spreading loss and severe molecular absorption ($100$ dB/km) 
\\\hline
\textbf{Angular spread} & All around  & Sparse ($30$ degree)  & Very sparse ($15$ degree)
\\\hline
\textbf{Noise}  & Thermal noise & Thermal noise  & Thermal noise and molecular reradiation noise 
\\\hline
\textbf{Blockage  effect}  & almost uninterruptible & Easily blocked  & Severe blockage problem
\\\hline
\textbf{Beamwidth} & Wide to medium & Narrow & Extremely narrow  
\\\hline
\textbf{Distance} & Big range & Medium range $\le200$ m &  Short range $\le50$ m  
\\ \hline
\textbf{\shortstack[l]{Number of \\ antennas}} & Limited number of elements   & Small size with massive elements   & Tiny size with ultra-massive elements
\\\hline

\textbf{\shortstack[l]{ Hardware\\ constraint}} & Normal  & High  & Very high
\\\hline
\textbf{Beamforming architecture} & Digital beamforming  & Hybrid beamforming  & Analog beamforming, TTD beamforming, IRS-assisted beamforming
\\\hline          
\end{tabular}
\end{table*}
\end{center}

\section{Open Challenges}
In this section, we outline some open challenges of the THz beamforming in UM-MIMO systems. 

\subsection{Channel Modeling and Measurement}
In Section {\ref{sm}}, we consider modifying and tailoring the conventional S-V channel model to characterize the THz wave propagation. As the spatial superposition of ultra-massive  single-input single-output (SISO) channels is complicated, accurate THz channel modeling is still lacking for UM-MIMO systems. Various factors need to be considered, including modeling in static and time-varying environments, channel space-time correlation under large-scale antenna arrays, near-field effects caused by the expansion of array scale, and modeling mutual coupling effects \cite{sur25}. In particular, the large array size in UM-MIMO leads to inaccuracy of planar-wave channel model (PWM), while an increased channel matrix dimension leads to excessive parameters in applying spherical-wave channel model (SWM). Thus, a hybrid spherical- and planar-wave channel model (HSPM) can be adopted to balance the accuracy and overhead of channel modeling\cite{hspw}.  Channel measurement is an important way to verify and improve the channel model. However, channel measurement in the THz band requires high-precision equipment, diverse test scenarios, and long iteration cycles, which is associated with high costs. At present, apart from the lower THz frequencies \cite{sur29}, the measurements for about $1$ THz is still limited \cite{umcm9, umcm10}. 

\subsection{THz Transceiver Device}
Wideband THz beamforming requires effective excitation of precise THz waves. However, the difficulty of exciting THz signals in a wide bandwidth is owing to the particularity of the THz frequency, which is too low for optical devices and exceeds the upper limit of conventional electronic oscillators.  In the past ten years, some emerging solutions, e.g.,  SiGe, gallium nitride (GaN), indium phosphide (InP), quantum cascade laser (QCL), and graphene have been proposed to achieve efficient generation and  detection of THz waves. Nevertheless, there are still many challenges in hardware devices \cite{sur320}. SiGe-based devices have only limited power gain and it is difficult to operate above 500 GHz. GaN high electron mobility transistor (HEMT) faces lower breakdown voltage in some scenarios. QCL can only work in a low-temperature laboratory environment, and face the dilemma of miniaturization \cite{sur28}. Graphene can be used to design compact THz transceivers due to its high conductivity and support the propagation of THz surface plasmon polariton (SPP) waves. However, graphene is far less mature than the above-mentioned technologies due to the lack of research \cite{sur15}.

\subsection{Low-Resolution Hardware}
Typically, the RF chain consists of analog-to-digital conversion/digital-to-analog conversion (ADC/DAC), demodulator, up/down-converter, low noise amplifier (LNA), mixers, automatic gain control (AGC), variable gain amplifier (VGA), and some filters. The existing signal processing algorithms developed for ideal components with infinite resolution require high-cost hardware and high energy consumption. Besides, the overall impact of non-ideal components may seriously undermine the theoretically expected performance. Thus, the investigation of low-resolution RF components is a very important topic.  For example, some works have investigated the influence of nonlinear distortions,  inflicted by low-resolution ADCs \cite{adc1,adc2,adc3} and quantized phase shifts \cite{ps1,ps2,ps3}, on the beamforming design.

\subsection{Large-Scale THz Array Design}

Thanks to the coherent superposition of electromagnetic waves, ultra-massive antenna elements are essential to compensate for the severe propagation loss of THz waves. Large-scale integrated phased arrays can increase the antenna gain while maintaining the advantages of miniaturization and flexible beam control. However, the practical design of a large-scale THz array is not an easy task. Challenges include the complicated signal distribution network design, the heat dissipation layout design of the dense array, and the low direct current conversion efficiency issue.  As an emerging solution, graphene-based large-scale antenna arrays are expected to overcome the above issues \cite{sur28}. The major challenge for graphene antennas is to characterize the interaction and coupling effects among adjacent elements \cite{sur15}. Current works on graphene-based antennas still focus on theoretical analysis but lack experimental validation. 

\subsection{Wideband Beamforming} 
To reduce/eliminate the frequency wideband effect as discussed in Sec. \ref{fwe},  there are several reported approaches to enhance the wideband beam gain, such as using the TTD beamformer \cite{bs6}, redesign the codebook \cite{bs4}, using subprocessors for different frequency groups \cite{wb1}, etc. However, these technolgies are of high energy consumption and require sophisticated hardware, which may not be practical in UM-MIMO systems. The overhead can be reduced by using the architecture of hybrid TTDs and phase shifters,  where the joint design of the parameters on TTD and phase shifts are worthy of study in THz wideband beamforming\cite{bs3,wbadd}. In addition, as for the beam squint effect for analog beamformer, we can use the AoSA architecture to realize wideband beamforming based on the beam split aggregation and multiplexing (BSAM) \cite{bsquint}, which is a promising low-cost solution.  For a single analog array,  to enable the user to receive a wideband signal under the beam squint effect, customized wideband codebooks can be designed, where each beam serves a wide zone instead of a narrow zone as in its narrowband counterpart, so as to avoid the misalignment with the AoA due to beam squint.

\subsection{3D Beamforming} 
3D beamforming technology supported by the THz UM-MIMO system allows higher transmission gain and stronger directivity for THz communication to provide large-capacity and less-interference multiple access. This feature will help the connection management of a large number of air targets (e.g. cellular-connected drones) in the future air-ground integrated wide-area coverage network \cite{3DB1}. However, compared with 2D beamforming, 3D schemes bring more challenges to codebook design and beam management. 
On one hand, the 3D codebook directly extended from the 2D version through the Kronecker product will yield irregular coverage. It is necessary to construct a new 3D codebook to ensure the maximum coverage in both azimuth and elevation dimensions. On the other hand, extremely sharp THz beam pairs need to be continuously adjusted in high-speed mobile scenarios, which requires more efficient and fast 3D beam management.  A grid-based unified beam training and beam tracking procedure can support such a fast 3D beam management 
\cite{3DB2}.

\subsection{Blockage Problem}
The THz wave has limited diffraction and the LoS path is extremely easy to be blocked by obstacles \cite{cov1}. When the THz waves are transmitted in the non-LoS (NLoS) channel, the data rate drops considerably due to reflection and/or scattering losses on rough surfaces \cite{cov2}.  Thus, one possible solution is to add APs to expand the signal coverage. However, coverage planning requires a detailed 3D model of the geographic environment to properly deploy APs, but the dense cells greatly increase the complexity of AP location planning. On the other hand, new challenges need to be fully considered before obtaining the diversity gain brought by the coordinated transmission of multiple APs, including the control signal overhead between APs and the timeliness of finding the best joint transmission strategy \cite{cov31}. Another solution is adding IRSs to reconstruct the wireless environment with THz propagation\cite{cov4}. Some key concerns include how to locate the IRS to enhance the signal coverage \cite{cov5} and how to realize the optimal routine \cite{cov6}.  Besides, in high-mobility outdoor scenarios, the beam misalignment caused by Doppler expansion needs to be seriously considered \cite{cov3}.

\section{CONCLUSION}
The THz frequency band is recently becoming a hot spot in the  communication community. Table \ref{table_rivalry} summarizes the features that distinguish sub-6 GHz, mmWave, and THz communications. In this paper, the model of the THz UM-MIMO systems was first established, in which the channel parameters, antenna geometry, and transceiver architecture are specified. We highlighted the basic principles of beamforming and presented the schemes of beam training. Then, we proceeded to address the on-trend THz beamforming topics, i.e.,  IRS-assisted joint beamforming and multi-user beamforming. For the former, we studied the model of IRS-assisted systems and provided a cooperative beam training scheme. For the latter, we provided a beamforming strategy for multi-user UM-MIMO THz systems and evaluated different linear beamforming algorithms for it.  Next, we discussed the spatial-wideband and frequency-wideband effects, along with the feasible solutions in wideband beamforming. Then, the existing THz MIMO arrays were classified based on the fabrication techniques, including the electronic approach, photonic approach, and new materials approach. In the end, the emerging applications and some open challenges were elaborated. It is expected that these open problems will motivate the researchers and engineers to endeavor for more efficient solutions to future THz UM-MIMO systems.

\begin{IEEEbiography}[{\includegraphics[width=1in,clip]{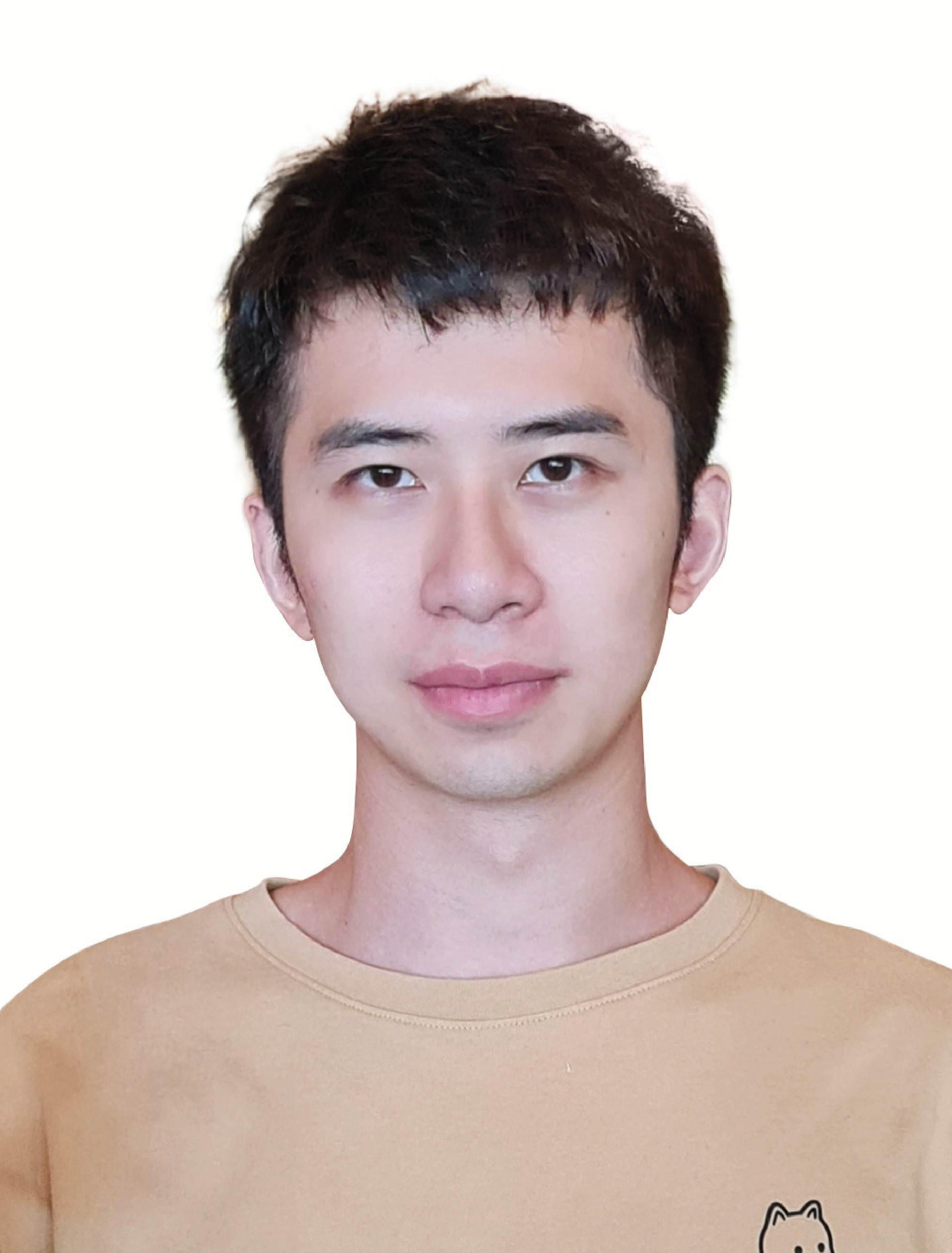}}]{Boyu Ning} received the B.E. degree in Communication Engineering, along with the Certification of the Talent Program in Yingcai Honors College, from the University of Electronic Science and Technology of China (UESTC), Chengdu, China, in 2018.  He is currently pursuing the Ph.D. degree with the National Key Laboratory of Science and Technology on Communications, UESTC.   He was a recipient of the Honor of Outstanding Students of UESTC.  His research interests include Terahertz communication, intelligent reflecting surface, massive MIMO, physical-layer security, and convex optimization.
\end{IEEEbiography}

\begin{IEEEbiography}
[{\includegraphics[width=1in,clip]{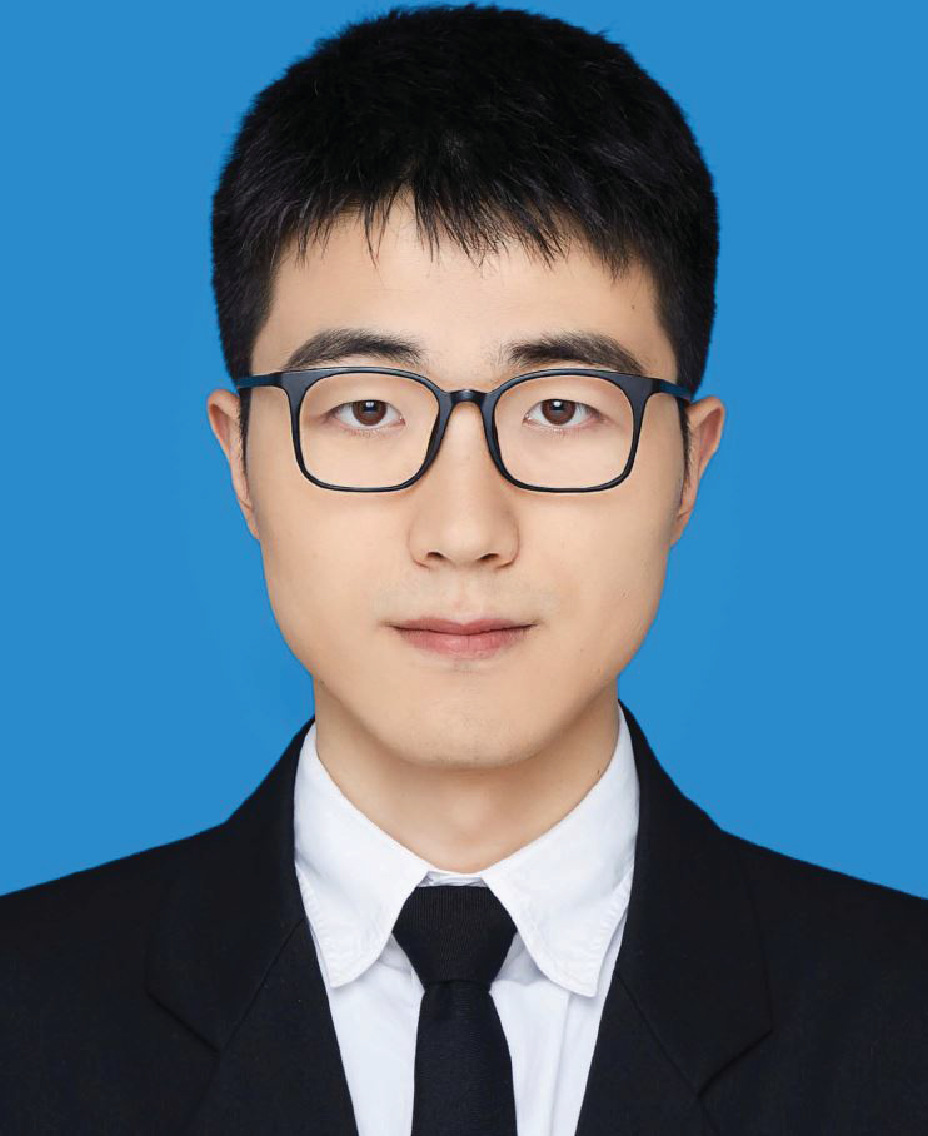}}]{Zhongbao Tian } received the B.E. degree in communication engineering from the University of Electronic Science and Technology of China (UESTC) in 2019. He is currently working toward the M.E. degree with the National Key Laboratory of Science and Technology on Communications, UESTC. He research and study interests include Terahertz communication and 3D beam forming.
\end{IEEEbiography}

\begin{IEEEbiography}[{\includegraphics[width=1in,clip]{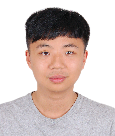}}]{Weidong Mei}
received the B.Eng. degree in communication engineering and the M.Eng. degree in communication and information systems from the University of Electronic Science and Technology of China, Chengdu, China, in 2014 and 2017, respectively, and the Ph.D. degree from the NUS Graduate School, National University of Singapore, in 2021 under the Integrative Sciences and Engineering Programme (ISEP) Scholarship. 

From 2021 to 2022, he was a Research Fellow with the Department of Electrical and Computer Engineering, National University of Singapore. He is currently a Professor with the University of Electronic Science and Technology of China. His research interests include intelligent reflecting surface, wireless drone communications, physical-layer security, and convex optimization techniques. He was a recipient of the Outstanding Master’s Thesis Award from the Chinese Institute of Electronics in 2017, and the Best Paper Award from the IEEE International Conference on Communications in 2021. He was honored as the Exemplary Reviewer of the \textsc{IEEE Transactions on Communications} from 2019 to 2020, the \textsc{IEEE Wireless Communications Letters} in 2019 and 2021, and the \textsc{IEEE Communications Letters} from 2021 to 2022.
\end{IEEEbiography}

\begin{IEEEbiography}
[{\includegraphics[width=1in,clip]{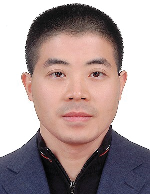}}]{Zhi Chen} received B. Eng, M. Eng., and Ph.D. degree in Electrical Engineering from University of Electronic Science and Technology of China (UESTC), in 1997, 2000, 2006, respectively. On April 2006, he joined the National Key Lab of Science and Technology on Communications (NCL), UESTC, and worked as a professor in this lab from August 2013. He was a visiting scholar at University of California, Riverside during 2010-2011. He is also the deputy director of Key Laboratory of Terahertz Technology, Ministry of Education. His current research interests include Terahertz communication, 5G/6G mobile communications. 
\end{IEEEbiography}

\begin{IEEEbiography}
[{\includegraphics[width=1in,clip]{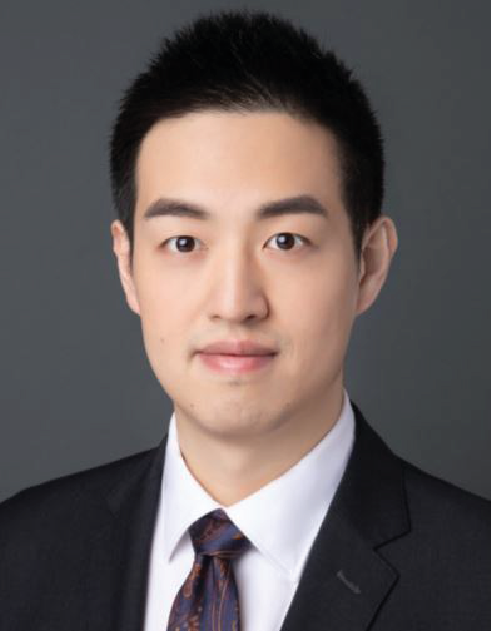}}]{Chong Han} is with Shanghai Jiao Tong University, Shanghai, China, where he is currently a John WU \& Jane Sun Endowed Associate Professor, and the Director of the Terahertz Wireless Communications (TWC) Laboratory. He obtained the Ph.D. degree in Electrical and Computer Engineering from Georgia Institute of Technology, Atlanta, GA, USA, in 2016. He received 2019–2021 Distinguished TPC Member Award, IEEE International Conference on Computer Communications (INFOCOM) and 2018 Elsevier Nano Communication Network Journal Young Investigator Award, 2018 Shanghai Chenguang Funding Award, and 2017 Shanghai Yangfan Funding Award. He is a (guest) editor of IEEE JSAC, JSTSP, Open Journal of Vehicular Technology, etc. He has published 6 book chapters, 50+ journal articles, and 60+ conference papers, most of which, if not all, are related to THz communications. He is a TPC Co-Chair or General Co-Chair for the 1st–6th International Workshop on Terahertz Communications, in conjunction with IEEE ICC/Globecom. Furthermore, he will be the symposium chair of SAC THz Communications with Globecom’2023. He is serving as a Vice Chair of IEEE ComSoc RCC Special Interest Group (SIG) on THz Communications.

\end{IEEEbiography}

\begin{IEEEbiography}
[{\includegraphics[width=1in,clip]{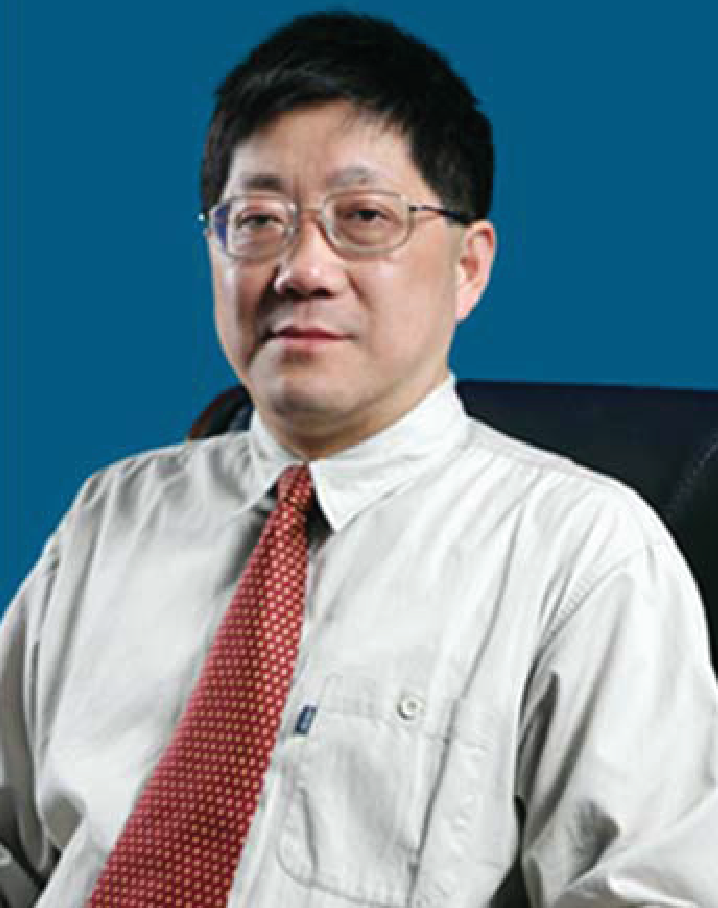}}]{Shaoqian Li} (Fellow, IEEE) received the B.E. degree in communication technology from Northwest Institute of Telecommunication Engineering (currently Xidian University), Xian, China, in 1981, and the M.E. degree in information and communication systems from the University of Electronic Science and Technology of China (UESTC), Chengdu, China, in 1984. He joined UESTC as an Academic Member in 1984, where he became a Professor of information and communication systems in 1997, and a Ph.D. Supervisor in 2000. He is currently the Director of the National Key Laboratory of Communications, UESTC. He has authored hundreds of journal or conference papers, and published several books. His research topics cover a broad range, including multiple-antenna signal processing technologies for mobile communications, cognitive radios, coding and modulation for next generation mobile broadband communications systems, wireless and mobile communications, anti-jamming technologies, and signal processing for communications. He has been a member of the Communication Expert Group of the National 863 Plan since 1998 and a member of The Future Project since 2005. He was the TPC Co-Chair of the IEEE International Conference on Communications, Circuits, and Systems in 2005, 2006, and 2008. He is currently a member of the Board of Communications and Information Systems of Academic Degrees Committee, State Council, China. He is also a member of the Editorial Board of the \emph{Chinese Science Bulletin} and the \emph{Chinese Journal of Radio Science}.
\end{IEEEbiography}

\begin{IEEEbiography}
[{\includegraphics[width=1in,clip]{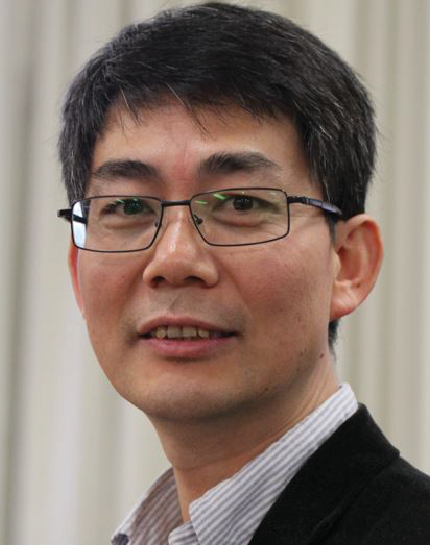}}]{Jinhong Yuan}(M'02--SM'11--F'16) received the B.E. and Ph.D. degrees in electronics engineering from the Beijing Institute of Technology, Beijing, China, in 1991 and 1997, respectively. From 1997 to 1999, he was a Research Fellow with the School of Electrical Engineering, University of Sydney, Sydney, Australia. In 2000, he joined the School of Electrical Engineering and Telecommunications, University of New South Wales, Sydney, Australia, where he is currently a Professor and Head of Telecommunication Group with the School. He has published two books, five book chapters, over 300 papers in telecommunications journals and conference proceedings, and 50 industrial reports. He is a co-inventor of one patent on MIMO systems and four patents on low-density-parity-check codes. He has co-authored four Best Paper Awards and one Best Poster Award, including the Best Paper Award from the IEEE International Conference on Communications, Kansas City, USA, in 2018, the Best Paper Award from IEEE Wireless Communications and Networking Conference, Cancun, Mexico, in 2011, and the Best Paper Award from the IEEE International Symposium on Wireless Communications Systems, Trondheim, Norway, in 2007. He is an IEEE Fellow and currently serving as an Associate Editor for the IEEE Transactions on Wireless Communications and IEEE Transactions on Communications. He served as the IEEE NSW Chapter Chair of Joint Communications/Signal Processions/Ocean Engineering Chapter during 2011-2014 and served as an Associate Editor for the IEEE Transactions on Communications during 2012-2017. His current research interests include error control coding and information theory, communication theory, and wireless communications.
\end{IEEEbiography}

\begin{IEEEbiography} 
[{\includegraphics[width=1in,clip]{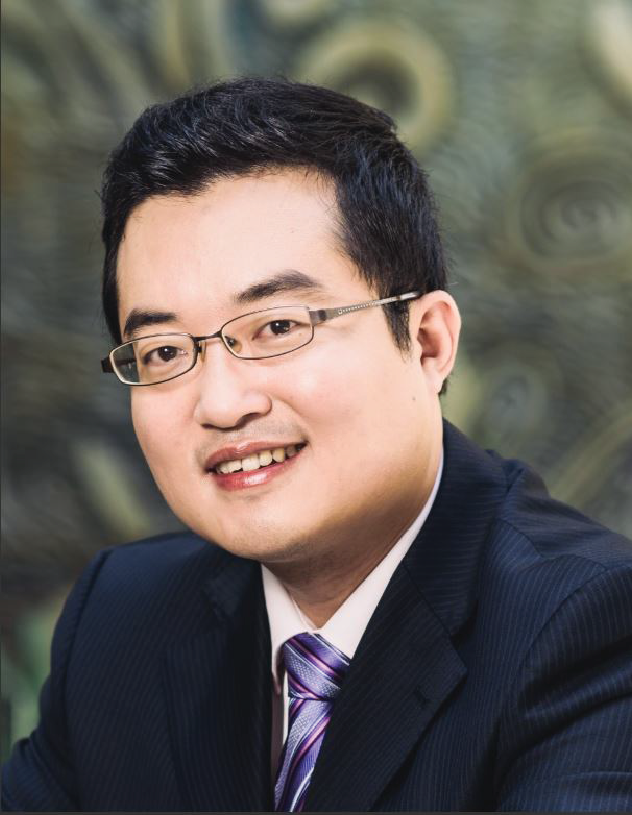}}]
{Rui Zhang} (S'00-M'07-SM'15-F'17) received the B.Eng. (first-class Hons.) and M.Eng. degrees from the National University of Singapore, Singapore, and the Ph.D. degree from the Stanford University, Stanford, CA, USA, all in electrical engineering.

From 2007 to 2009, he worked as a researcher at the Institute for Infocomm Research, ASTAR, Singapore. In 2010, he joined the Department of Electrical and Computer Engineering of National University of Singapore, where he was appointed as a Provost’s Chair Professor in 2020. Since 2022, he has joined the School of Science and Engineering, The Chinese University of Hong Kong, Shenzhen, as a Principal’s Diligence Chair Professor. He has published over 300 journal papers and over 200 conference papers. He has been listed as a Highly Cited Researcher by Thomson Reuters/Clarivate Analytics since 2015. His current research interests include UAV/satellite communications, wireless power transfer, intelligent reflecting surface, reconfigurable MIMO, radio mapping and optimization methods.

He was the recipient of the 6th IEEE Communications Society Asia-Pacific Region Best Young Researcher Award in 2011, the Young Researcher Award of National University of Singapore in 2015, the Wireless Communications Technical Committee Recognition Award in 2020, and the IEEE Signal Processing and Computing for Communications (SPCC) Technical Recognition Award in 2020. He received 14 IEEE Best Paper Awards, including the IEEE Marconi Prize Paper Award in Wireless Communications in 2015 and 2020, the IEEE Signal Processing Society Best Paper Award in 2016, the IEEE Communications Society Heinrich Hertz Prize Paper Award in 2017, 2020 and 2022, the IEEE Communications Society Stephen O. Rice Prize in 2021, etc. He served for over 30 international conferences as the TPC co-chair or an organizing committee member. He was an elected member of the IEEE Signal Processing Society SPCOM Technical Committee from 2012 to 2017 and SAM Technical Committee from 2013 to 2015, and served as the Vice Chair of the IEEE Communications Society Asia-Pacific Board Technical Affairs Committee from 2014 to 2015. He was a Distinguished Lecturer of IEEE Signal Processing Society and IEEE Communications Society from 2019 to 2020. He served as an Editor for the IEEE TRANSACTIONS ON WIRELESS COMMUNICATIONS from 2012 to 2016, the IEEE JOURNAL ON SELECTED AREAS IN COMMUNICATIONS: Green Communications and Networking Series from 2015 to 2016, the IEEE TRANSACTIONS ON SIGNAL PROCESSING from 2013 to 2017, the IEEE TRANSACTIONS ON GREEN COMMUNICATIONS AND NETWORKING from 2016 to 2020, and the IEEE TRANSACTIONS ON COMMUNICATIONS from 2017 to 2022. He served as a member of the Steering Committee of the IEEE Wireless Communications Letters from 2018 to 2021. He is a Fellow of the Academy of Engineering Singapore. \end{IEEEbiography}


\begin{thebibliography}{1}
  \bibitem{5g}
  J. G. Andrews \emph{et al.}, ``What will 5G be?'' \emph{IEEE J. Sel. Areas Commun.}, vol. 32, no. 6, pp. 1065-1082, June 2014.
  
  \bibitem{5g2}
  S. Mumtaz \emph{et al.}, ``Introduction to mmWave massive MIMO,'' in \emph{mmWave Massive MIMO: A Paradigm for 5G}, London, U.K.: Academic Press, 2017, pp. 1-18.
  
  \bibitem{lte}
  A. Ghosh \emph{et al.}, ``LTE-advanced: next-generation wireless broadband technology [Invited Paper],'' \emph{IEEE Wireless Commun.}, vol. 17, no. 3, pp. 10-22, June 2010.
  
  \bibitem{wpan}
  E. Callaway \emph{et al.}, ``Home networking with IEEE 802.15.4: a developing standard for low-rate wireless personal area networks,'' \emph{IEEE Commun. Mag.}, vol. 40, no. 8, pp. 70-77, Aug. 2002.
  
  \bibitem{wihd}
  WirelessHD Consortium, ``Wireless HD specification version 1.1 overview,'' \emph{Tech. Rep.}, CA, USA , May 2010.
  
  \bibitem{8015}
  T. Baykas \emph{et al.}, ``IEEE 802.15.3c: the first IEEE wireless standard for data rates over 1 Gb/s,'' \emph{IEEE Commun. Mag.}, vol. 49, no. 7, pp. 114-121, July 2011.
  
  \bibitem{wlan}
  B. P. Crow \emph{et al.}, ``IEEE 802.11 wireless local area networks,'' \emph{IEEE Commun. Mag.}, vol. 35, no. 9, pp. 116-126, Sept. 1997.
  
  \bibitem{edh}
  S. Cherry, ``Edholm’s law of bandwidth,'' \emph{IEEE Spectr.}, vol. 41, no. 7, pp. 58-60, July 2004.
  
  
  \bibitem{mimo}
  R. W. Heath \emph{et al.}, ``An overview of signal processing techniques for millimeter wave MIMO systems,'' \emph{IEEE J. Sel. Topics Signal Process.}, vol. 10, no. 3, pp. 436-453, Apr. 2016.
  
  \bibitem{comp}
  S. Bassoy \emph{et al.}, ``Coordinated multi-point clustering schemes: A survey,'' \emph{IEEE Commun. Surv. \& Tuts.}, vol. 19, no. 2, pp. 743-764, 2nd Quart. 2017.
  
  \bibitem{caf}
  M. Iwamura \emph{et al.}, ``Carrier aggregation framework in 3GPP LTE-advanced [WiMAX/LTE update],'' \emph{IEEE Commun. Mag.}, vol. 48, no. 8, pp. 60-67, Aug. 2010.
  
  \bibitem{het}
  N. Zhang \emph{et al.}, ``Cloud assisted HetNets toward 5G wireless networks,'' \emph{IEEE Commun. Mag.}, vol. 53, no. 6, pp. 59-65, June 2015.
  
  \bibitem{het2}
  J. An \emph{et al.}, ``Achieving sustainable ultra-dense heterogeneous networks for 5G,'' \emph{IEEE Commun. Mag.}, vol. 55, no. 12, pp. 84-90, Dec. 2017.
  
  \bibitem{otfs1}
  R. Hadani \emph{et al.}, ``Orthogonal time frequency space modulation,'' in \emph{2017 IEEE Wireless Commun. Networking Conf. (WCNC)}, 2017, pp. 1-6.
  
  
  \bibitem{otfs2}
  Z. Wei \emph{et al.}, ``Transmitter and receiver window designs for orthogonal time-frequency space modulation,'' \emph{IEEE Trans. Commun.}, vol. 69, no. 4, pp. 2207-2223, April 2021.
  
  \bibitem{rates1}
  S. Yang \emph{et al.}, ``Degrees of freedom of time correlated MISO broadcast channel with delayed CSIT,'' \emph{IEEE Trans. Inf. Theory}, vol. 59, no. 1, pp. 315-328, Jan. 2013.
  
  \bibitem{rates2}
  M. Dai, \emph{et al.}, ``A rate splitting strategy for massive MIMO with imperfect CSIT,'' \emph{IEEE Trans. Wireless Commun.}, vol. 15, no. 7, pp. 4611-4624, July 2016.
  
  \bibitem{noma1}
  L. Dai \emph{et al.}, ``Nonorthogonal multiple access for 5G: Solutions, challenges, opportunities, and future research trends,'' \emph{IEEE Commun. Mag.}, vol. 53, no. 9, pp. 74-81, Sept. 2015
  
  \bibitem{noma2}
  Z. Ding \emph{et al.}, ``A survey on non-orthogonal multiple access for 5G networks: research challenges and future trends,'' \emph{IEEE J. Select. Areas Commun.}, vol. 35, no. 10, pp. 2181-2195, Oct. 2017.
  
  \bibitem{sur320}
  H. Elayan \emph{et al.}, ``Terahertz band: The last piece of RF spectrum puzzle for communication systems,'' \emph{IEEE Open J. Commun. Soc.}, vol. 1, pp. 1-32, 2020.
  
  
  \bibitem{mmimo0}
J. Hoydis, S. Ten Brink, and M. Debbah, ``Massive mimo: How many
antennas do we need?'' in  \emph{Proc. of 49th Annual Allerton Conference on
Communication, Control, and Computing. IEEE,} 2011, pp. 545-550.
  
  \bibitem{gramimo2} 
I. F. Akyildiz and J. M. Jornet, ``Realizing ultra-massive MIMO (1024 $\times$ 1024) communication in the (0.06-10) terahertz band,'' \emph{Nano Commun. Networks}, vol. 8, pp. 46-54, Nov. 2016.
  
  \bibitem{sur1}
 P. H. Siegel, ``Terahertz technology,'' \emph{IEEE Trans. Microw. Theory Tech.}, vol. 50, no. 3, pp. 910-928, Mar. 2002.
  
  \bibitem{sur2}
  M. J. Fitch and R. Osiander, ``Terahertz waves for communications and sensing,'' \emph{Johns Hopkins APL Techn. Dig.}, vol. 25, no. 4, pp. 348-355, 2004.
  
  \bibitem{sur3}
  R. Piesiewicz \emph{et al.}, ``Short-range ultra-broadband terahertz communications: Concepts and Perspectives,'' \emph{IEEE Antennas Propagat. Mag.}, vol. 49, no. 6, pp. 24-39, Dec. 2007.
  
  \bibitem{sur4}
  I. Hosako \emph{et al.}, ``At the dawn of a new era in terahertz technology,'' in \emph{Proc. IEEE}, vol. 95, no. 8, pp. 1611-1623, Aug. 2007.
  
  \bibitem{sur5}
  M. Tonouchi, ``Cutting-edge terahertz technology,'' \emph{Nat. photon.}, vol. 1, no. 2, pp. 97-105, 2007.
  
  \bibitem{sur6}
  J. Federici and L. Moeller, ``Review of terahertz and subterahertz wireless communications,'' \emph{J. Appl. Phys.}, vol. 107, no. 11, p. 6, 2010.
  
  \bibitem{sur7}
  K. C. Huang and Z. Wang, ``Terahertz terabit wireless communication,'' \emph{IEEE Microw. Mag.}, vol. 12, no. 4, pp. 108-116, June 2011.
  
  \bibitem{sur8}
  T. Kleine-Ostmann and T. Nagatsuma, ``A review on terahertz communications research,'' \emph{J. Infrared Millimeter Terahertz Waves}, vol. 32, no. 2, pp. 143-171, 2011.
  
  \bibitem{sur9}
  T. Nagatsuma, ``Terahertz technologies: present and future,'' \emph{IEICE Electron. Exp.}, vol. 8, no. 14, pp. 1127-1142, July 2011.
  
  \bibitem{sur10}
  H. J. Song and T. Nagatsuma, ``Present and future of terahertz communications,'' \emph{IEEE Trans. THz Sci. Technol.}, vol. 1, no. 1, pp. 256-263, Sept. 2011.
  
  \bibitem{sur11}
  K. Wu \emph{et al.}, ``Substrate-integrated millimeter-wave and terahertz antenna technology,'' in \emph{Proc. IEEE}, vol. 100, no. 7, pp. 2219-2232, July 2012.
  
  \bibitem{sur12}
  T. Kurner, ``Towards future terahertz communications systems,'' \emph{Terahertz Sci. Technol.}, vol. 5, pp. 11-17, Jan. 2012.
  
  
  \bibitem{sur13}
  T. Nagatsuma \emph{et al.}, ``Terahertz wireless communications based on photonics technologies,'' \emph{Opt. Exp.}, vol. 21, no. 20, pp. 23736-23747, 2013.
  
  \bibitem{sur131}
  k. Jha and G.Singh, ``Terahertz planar antennas for future wireless communication: A technical review,'' \emph{Infrared Phys. $\&$ Tech.}, vol. 60, pp. 71-80, Sept. 2013.
  
  \bibitem{sur14}
  I. F. Akyildiz, J. M. Jornet, and C. Han, ``Terahertz band: Next frontier for wireless communications,'' \emph{Phys. Commun.}, vol. 12, pp. 16-32, Sept. 2014.
  
  \bibitem{sur15} 
  T. Kurner and S. Priebe, ``Towards THz communications-status in research, standardization and regulation,'' \emph{J. Infrared Millimeter Terahertz Waves}, vol. 35, no. 1, pp. 53-62, Jan. 2014.
  
  \bibitem{sur16}
  I. F. Akyildiz, J. M. Jornet, and C. Han, ``TeraNets: ultra-broadband communication networks in the terahertz band,'' \emph{IEEE Wireless Commun.}, vol. 21, no. 4, pp. 130-135, Aug. 2014.
  
  \bibitem{sur17}
  A. Hirata and M. Yaita, ``Ultrafast terahertz wireless communications technologies,'' \emph{IEEE Trans. THz Sci. Technol.}, vol. 5, no. 6, pp. 1128-1132, Nov. 2015.
  
  \bibitem{sur18}
  C. Lin and G. Y. Li, ``Terahertz communications: An array-of-subarrays solution,'' \emph{IEEE Commun. Mag.}, vol. 54, no. 12, pp. 124-131, Dec. 2016.
  
  \bibitem{sur19}
  M. Hasan \emph{et al.}, ``Graphene terahertz devices for communications applications,'' \emph{Nano Commun. Netw.}, vol. 10, pp. 68-78, Dec. 2016.
  
  \bibitem{sur20}
  Nagatsuma \emph{et al.}, ``Advances in terahertz communications accelerated by photonics'' \emph{Nature Photon.} vol. 10, pp. 371-379, 2016.
  
  \bibitem{sur21}
  J. F. Federici, J. Ma, and L. Moeller, ``Review of weather impact on outdoor terahertz wireless communication links,'' \emph{Nano Commun. Netw.}, vol. 10, pp. 13-26, Dec. 2016.
  
  \bibitem{sur22}
  S. Mumtaz \emph{et al.}, ``Terahertz communication for vehicular networks,'' \emph{IEEE Trans. Veh. Technol.}, vol. 66, no. 7, pp. 5617-5625, July 2017.
  
  \bibitem{sur23}
  V. Petrov \emph{et al.}, ``Last meter indoor terahertz wireless access: Performance insights and implementation roadmap,'' \emph{IEEE Commun. Mag.}, vol. 56, no. 6, pp. 158-165, June 2018.
  
  \bibitem{sur24}
  A. A. A. Boulogeorgos \emph{et al.}, ``Terahertz technologies to deliver optical network quality of experience in wireless systems beyond 5G,'' \emph{IEEE Commun. Mag.}, vol. 56, no. 6, pp. 144-151, June 2018.
  
  \bibitem{sur25}
  C. Han and Y. Chen, ``Propagation modeling for wireless communications in the terahertz band,'' \emph{IEEE Commun. Mag.}, vol. 56, no. 6, pp. 96-101, June 2018.
  
  \bibitem{sur251}
  I. F. Akyildiz \emph{et al.}, ``Combating the distance problem in the millimeter wave and terahertz frequency bands,'' \emph{IEEE Commun. Mag.}, vol. 56, no. 6, pp. 102-108, June 2018.
  
  \bibitem{sur26}
  N. Khalid, T. Yilmaz, and O. B. Akan, ``Energy-efficient modulation and physical layer design for low terahertz band communication channel in 5G femtocell Internet of Things,'' \emph{Ad Hoc Netw.}, vol. 79, pp. 63-71, Oct. 2018.
  
  \bibitem{sur261}
  D. Headland \emph{et al.} , ``Tutorial: Terahertz beamforming, from concepts to realizations,'' \emph{APL Photon.}, vol. 3, pp. 051101, 2018.
  
  \bibitem{sur27}
  Z. Chen \emph{et al.}, ``A survey on terahertz communications,'' \emph{China Commun.}, vol. 16, no. 2, pp. 1-35, Feb. 2019.
  
  \bibitem{sur28}
  K. Tekbıyık \emph{et al.}, ``Terahertz band communication systems: Challenges, novelties and standardization efforts,'' \emph{Phys. Commun.}, vol. 35, Aug. 2019. 
  
  \bibitem{sur29}
  T. S. Rappaport \emph{et al.}, ``Wireless communications and applications above 100 GHz: Opportunities and challenges for 6G and beyond,'' \emph{IEEE Access}, vol. 7, pp. 78729-78757, 2019.
  
  
  \bibitem{sur31}
  K. K. O \emph{et al.}, ``Opening terahertz for everyday applications,'' \emph{IEEE Commun. Mag.}, vol. 57, no. 8, pp. 70-76, Aug. 2019.
  
  \bibitem{sur32}
  K. M. S. Huq \emph{et al.}, ``Terahertz-enabled wireless system for beyond-5G ultra-fast networks: A brief survey,'' \emph{IEEE Netw.}, vol. 33, no. 4, pp. 89-95, July 2019.
  
  \bibitem{sur321}
  X. Fu \emph{et al.}, ``Terahertz beam steering technologies: From phased arrays to field-programmable metasurfaces,'' \emph{Adv. Optical Mater.}, vol. 8, pp. 1900628, 2020.
  
  \bibitem{sur33}
  H. Sarieddeen \emph{et al.}, ``Next generation terahertz communications: A rendezvous of sensing, imaging, and localization,'' \emph{IEEE Commun. Mag.}, vol. 58, no. 5, pp. 69-75, May 2020.
  
  \bibitem{sur34}
  L. Zhang \emph{et al.}, ``Beyond 100 Gb/s optoelectronic terahertz communications: Key technologies and directions,'' \emph{IEEE Commun. Mag.}, vol. 58, no. 11, pp. 34-40, Nov. 2020.
  
  \bibitem{sur35}
  M. A. Jamshed \emph{et al.}, ``Antenna selection and designing for THz applications: suitability and performance evaluation: A survey,'' \emph{IEEE Access}, vol. 8, pp. 113246-113261, 2020.
  
  \bibitem{sur36}
  S. Ghafoor \emph{et al.}, ``MAC protocols for terahertz communication: A comprehensive survey,'' \emph{IEEE Commun. Surv. \& Tuts.}, vol. 22, no. 4, pp. 2236-2282, 4th Quart. 2020.
  
  \bibitem{sur37}
  C. X. Wang \emph{et al.}, ``6G wireless channel measurements and models: Trends and challenges,'' \emph{IEEE Veh. Technol. Mag.}, vol. 15, no. 4, pp. 22-32, Dec. 2020.
  
  \bibitem{sur38}
  A. Faisal \emph{et al.}, ``Ultramassive MIMO systems at terahertz bands: Prospects and challenges,'' \emph{IEEE Veh. Technol. Mag.}, vol. 15, no. 4, pp. 33-42, Dec. 2020.
  
  \bibitem{sur381}
  J. Tan \emph{et al.}, ``THz precoding for 6G: Applications, challenges, solutions, and opportunities,'' [Online]. Available: https://arxiv.org/abs/2005.10752
  
  \bibitem{sur39}
  F. Lemic \emph{et al.}, ``Survey on terahertz nanocommunication and networking: A top-down perspective,'' \emph{IEEE J. Sel. Areas Commun. }, vol. 39, no. 6, pp. 1506-1543, June 2021.
  
  \bibitem{sur40}
  C. Han, L. Yan and J. Yuan, ``Hybrid beamforming for terahertz wireless communications: Challenges, architectures, and open problems,''  \emph{IEEE Wireless Commun.}, vol. 28, no. 4, pp. 198-204, Aug. 2021.
  
  \bibitem{sur41}
  H. Sarieddeen, M. S. Alouini and T. Y. Al-Naffouri, ``An overview of signal processing techniques for terahertz communications,'' \emph{Proceedings of the IEEE}, vol. 109, no. 10, pp. 1628-1665, Oct. 2021.
  
  
  
  
  \bibitem{sur44}
  H. J. Song, ``Terahertz wireless communications: Recent developments including a prototype system for short-range data downloading,'' \emph{IEEE Microw. Mag.}, vol. 22, no. 5, pp. 88-99, May 2021.
  
  \bibitem{sur45}
  H. Do \emph{et al.}, ``Terahertz line-of-sight MIMO communication: Theory and practical challenges,''  \emph{IEEE Commun. Mag.},  vol. 59, no. 3, pp. 104-109, May 2021.
  
  \bibitem{sur46}
  Z. Chen \emph{et al.}, ``Terahertz wireless communications for 2030 and beyond: A cutting-edge frontier,'' \emph{IEEE Commun. Mag.}, vol. 59, no. 11, pp. 66-72, Nov. 2021.
  
  \bibitem{sur47}
  Z. Chen \emph{et al.}, ``Intelligent reflecting surface assisted terahertz communications toward 6G,'' \emph{IEEE Wireless Commun.}, vol. 28, no. 6, pp. 110-117, Dec. 2021.
  
  
  \bibitem{sur42}
  C. Chaccour, M. N. Soorki, W. Saad, M. Bennis, P. Popovski and M. Debbah, ``Seven Defining Features of Terahertz (THz) Wireless Systems: A Fellowship of Communication and Sensing,'' \emph{IEEE Commun. Surv. \& Tuts.}, vol. 24, no. 2, pp. 967-993, secondquarter 2022.
  
  \bibitem{holis}
  C. Han \emph{et al.},  ``Terahertz Wireless Channels: A Holistic Survey on Measurement, Modeling, and Analysis,'' \emph{IEEE Commun. Surv. \& Tuts.}, vol. 24, no. 3, pp. 1670-1707, thirdquarter 2022.
  
  \bibitem{umm}
  H. Sarieddeen, M. S. Alouini and T. Y. Al-Naffouri, ``Terahertz-band ultra-massive spatial modulation MIMO,'' \emph{IEEE J. Sel. Areas Commun.}, vol. 37, no. 9, pp. 2040-2052, Sept. 2019. 
  
  \bibitem{mod2}
  S. Priebe and T. Kürner, ``Stochastic modeling of THz indoor radio channels,'' \emph{IEEE Trans. Wireless Commun.}, vol. 12, no. 9, pp. 4445-4455, Sept. 2013.
  
  \bibitem{mod3}
  J. M. Jornet and I. F. Akyildiz, ``Channel modeling and capacity analysis for electromagnetic wireless nanonetworks in the Terahertz band,'' \emph{IEEE Trans. Wireless Commun.}, vol. 10, no. 10, pp. 3211-3221, Oct. 2011.
  
  \bibitem{mpm}
  H. J. Liebe, ``MPM-an atmospheric millimeter-wave propagation model'', \emph{Int. J. Infrared Millimeter Waves}, vol. 10, pp. 631-650, 1989.
  
  \bibitem{am}
  Paine S.(2012). \emph{AM atmospheric model}[Online]. Available: https://www.cfa.harvard.edu/~spaine/am/
  
  \bibitem{itu}
  \emph{Attenuation by atmospheric gases}, ITU Rec. ITU-R P.676-10, ITU, Sept. 2013.
  
  \bibitem{umcm1}
  C. Han and I.F. Akyildiz, ``Three-dimensional end-to-end modeling and analysis for graphene-enabled terahertz band communications,'' \emph{IEEE Trans. Veh. Tech.}, vol. 66, no. 7, pp. 5626-34, July 2017.
  
  \bibitem{umcm2}
  Young-Seek Chung \emph{et al.}, ``FDTD analysis of propagation characteristics of terahertz electromagnetic pulses,'' \emph{IEEE Trans. Magn.}, vol. 36, no. 4, pp. 951-955, July 2000.
  
  \bibitem{umcm3}
  M. R. Akdeniz \emph{et al.}, ``Millimeter wave channel modeling and cellular capacity evaluation,'' \emph{IEEE J. Select. Areas Commun.}, vol. 32, no. 6, pp. 1164-1179, June 2014.
  
  \bibitem{umcm4}
  A.F. Molisch, ``A generic model for MIMO wireless propagation channels in macro- and microcells,'' \emph{IEEE Trans. Signal Process.}, vol. 52, no. 1, pp. 61-71, Jan. 2004.
  
  \bibitem{umcm5}
  S. Kim and A. Zajić, ``Statistical modeling and simulation of shortrange device-to-device communication channels at sub-THz frequencies,'' \emph{IEEE Trans. Wireless Commun.}, vol. 15, no. 9, pp. 6423-6433, Sept. 2016.
  
  
  \bibitem{sv}
  A. Saleh and R. Valenzuela, ``A statistical model for indoor multipath propagation,'' \emph{IEEE J. Sel. Areas Commun.}, vol. 5, no. 2, pp. 128-137, May 1987.
  
  \bibitem{mod1}
  C. Han \emph{et al.}, ``Multi-ray channel modeling and wideband characterization for wireless communications in the terahertz band,'' \emph{IEEE Trans. Wireless Commun.}, vol. 14, no. 5, pp. 2402-2412, May 2015.
  
  \bibitem{mod4}
  C. Lin and G. Y. Li, ``Indoor terahertz communications: How many antenna arrays are needed?'' \emph{IEEE Trans. Wireless Commun.}, vol. 14, no. 6, pp. 3097-3107, June 2015.
  
  \bibitem{mod5}
  C. Han \emph{et al.}, ``Ultra-massive MIMO channel modeling for graphene-enabled terahertz-band communications,'' in \emph{IEEE Veh. Technol. Conf. (VTC Spring)}, Porto, pp. 1-5, June 2018.
  
  \bibitem{ang1}
  Q. Spencer, M. Rice, B. Jeffs, and M. Jensen, ``A statistical model for
  angle of arrival in indoor multipath propagation,'' in  \emph{Proc. IEEE 47th Veh. Technol. Conf. (VTC)}, vol. 3. May 1997, pp. 1415-1419.
  
  \bibitem{ang2}
  S. Priebe, M. Jacob, and T. Kuerner, ``AoA, AoD and ToA characteristics
  of scattered multipath clusters for THz indoor channel modeling,''
  in \emph{Proc. Eur. Wireless Conf. (EW)}, Apr. 2011, pp. 1-9.
  
  \bibitem{af}
  A. F. Molisch, \emph{Wireless Communications}, vol. 15. Hoboken, NJ, USA: Wiley, 2010.
  
  \bibitem{hitran}
  L. Rothman \emph{et al.}, ``The hitran2012 molecular spectroscopic database,'' \emph{J. Quant. Spectrosc. Radiat. Transf.}, vol. 130, pp. 4-50, July 2013.
  
  \bibitem{pat}
  W. L. Stutzman and G. A. Thiele, \emph{Antenna theory and design.} New York: Wiley, 1998.
  
  \bibitem{rs}
  R. S. Elliott, ``Beamwidth and directivity of large scanning arrays,'' \emph{The Microw. J.}, pp. 74-82, Jan. 1964.
  
  \bibitem{sun}
  S. Sun \emph{et al.}, ``Propagation models and performance evaluation for 5G millimeter-wave bands,'' \emph{IEEE Trans. Veh. Technol.}, vol. 67, no. 9, pp. 8422-8439, Sept. 2018.
  
  \bibitem{bcnoise}
  L. Kish, ``Stealth communication: Zero-power classical communication, zero-quantum quantum communication and environmental-noise communication,'' \emph{Appl. Phys. Lett.}, vol. 87, 2005.
  
  \bibitem{abnoise}
  J. M. Jornet and I. F. Akyildiz, ``Femtosecond-long pulse-based modulation for terahertz band communication in nanonetworks,'' \emph{IEEE Trans. Commun.}, vol. 62, no. 5, pp. 1742-1754, May 2014.
  
  \bibitem{abnoise2}
  V. Petrov \emph{et al.}, ``On the efficiency of spatial channel reuse in ultra-dense THz networks,'' in \emph{2015 IEEE Global Commun. Conf. (GLOBECOM)}, San Diego, CA, USA, 2015, pp. 1-7.
  
  \bibitem{abnoise3}
  S. A. Hoseini \emph{et al.}, ``Massive MIMO performance comparison of beamforming and multiplexing in the terahertz band,'' in \emph{ IEEE Globecom Workshops (GC Wkshps)}, Singapore, 2017, pp. 1-6.
  
  
  \bibitem{array2}
  Q. Ding \emph{et al.}, ``Hybrid precoding for mmWave massive MIMO systems with different antenna arrays,'' \emph{China Commun.}, vol. 16, no. 10, pp. 45-55, Oct. 2019.
  
  \bibitem{array1}
  W. Tan \emph{et al.}, ``Analysis of different planar antenna arrays for mmWave massive MIMO systems,'' in \emph{2017 IEEE 85th Veh. Technol. Conf. (VTC Spring)}, Sydney, NSW, 2017, pp. 1-5.
  
  \bibitem{archi1}
  T. Rappaport, \emph{Millimeter Wave Wireless Communications}, Prentice Hall, 2014.
  
  \bibitem{archi2}
  S. Kutty and D. Sen, ``Beamforming for millimeter wave communications: An inclusive survey,'' \emph{IEEE Commun. Surv. \& Tuts.}, vol. 18, no. 2, pp. 949-973, 2nd Quart. 2016.
  
  \bibitem{archi3}
  I. Ahmed \emph{et al.}, ``Resource allocation for transmit hybrid beamforming in decoupled millimeter wave multiuser-MIMO downlink,'' \emph{IEEE Access}, vol. 5, pp. 170-182, 2017.
  
  \bibitem{archi4}
  Xinying Zhang \emph{et al.}, ``Variable-phase-shift-based RF-baseband codesign for MIMO antenna selection,'' \emph{IEEE Trans. Signal Process.}, vol. 53, no. 11, pp. 4091-4103, Nov. 2005.
  
  \bibitem{archi6}
  A. F. Molisch \emph{et al.}, ``Hybrid beamforming for massive MIMO: A survey,'' \emph{IEEE Commun. Mag.}, vol. 55, no. 9, pp. 134-141, Sept. 2017.
  
  \bibitem{archi61}
  E. Zhang and C. Huang, ``On achieving optimal rate of digital precoder by RF-baseband codesign for MIMO systems,'' in \emph{2014 IEEE 80th Veh. Technolo. Conf. (VTC-Fall)}, 2014, pp. 1-5.
  
  \bibitem{archi7}
  X. Yu \emph{et al.}, ``Alternating minimization algorithms for hybrid precoding in millimeter wave MIMO systems,'' \emph{IEEE J. Sel. Topics Signal Process.}, vol. 10, no. 3, pp. 485-500, Apr. 2016.
  
  \bibitem{archi8}
  I. Ahmed \emph{et al.}, ``A survey on hybrid beamforming techniques in 5G: Architecture and system model perspectives,'' \emph{IEEE Commun. Surv. \& Tuts.}, vol. 20, no. 4, pp. 3060-3097, 4th Quart. 2018.
  
  \bibitem{archi9}
  S. Han \emph{et al.}, ``Large-scale antenna systems with hybrid analog and digital beamforming for millimeter wave 5G,'' \emph{IEEE Commun. Mag.}, vol. 53, no. 1, pp. 186-194, Jan. 2015.
  
  \bibitem{archi10}
  X. Gao \emph{et al.}, ``Energy-efficient hybrid analog and digital precoding for mmWave MIMO systems with large antenna arrays,'' \emph{IEEE J. Sel. Areas Commun.}, vol. 34, no. 4, pp. 998-1009, Apr. 2016.
  
  \bibitem{archadd1}
  C. Lin and G. Y. Li, ``Adaptive beamforming with resource allocation for distance-aware multi-user indoor terahertz communications,'' \emph{IEEE Trans. Commun.}, vol. 63, no. 8, pp. 2985-2995, Aug. 2015.
  
  \bibitem{archadd2}
  C. Lin \emph{et al.}, ``Subarray-based coordinated beamforming training for mmWave and sub-THz communications,'' \emph{IEEE J. Sel. Areas Commun.}, vol. 35, no. 9, pp. 2115-2126, Sept. 2017. 
  
  \bibitem{archadd3}
  H. Yuan \emph{et al.}, ``Hybrid beamforming for terahertz multi-carrier systems over frequency selective fading,'' \emph{IEEE Trans. Commun.}, vol. 68, no. 10, pp. 6186-6199, Oct. 2020.
  
  \bibitem{archi11}
  S. Park \emph{et al.}, ``Dynamic subarrays for hybrid precoding in wideband mmWave MIMO systems,'' \emph{IEEE Trans. Wireless Commun.}, vol. 16, no. 5, pp. 2907-2920, May 2017.
  
  \bibitem{archi12}
  D. Zhang \emph{et al.}, ``Hybridly connected structure for hybrid beamforming in mmWave massive MIMO systems,'' \emph{IEEE Trans. Commun.}, vol. 66, no. 2, pp. 662-674, Feb. 2018.
  
  \bibitem{archi13}
  J. Jin \emph{et al.}, ``Channel-statistics-based hybrid precoding for millimeter-wave MIMO systems with dynamic subarrays,'' \emph{IEEE Trans. Commun.}, vol. 67, no. 6, pp. 3991-4003, June 2019.
  
  \bibitem{archi14}
  F. Yang \emph{et al.}, ``A partially dynamic subarrays structure for wideband mmWave MIMO systems,'' \emph{IEEE Trans. Commun.}, vol. 68, no. 12, pp. 7578-7592, Dec. 2020.
  
  \bibitem{archi15}
  L. Yan \emph{et al.}, ``A dynamic array-of-subarrays architecture and hybrid precoding algorithms for terahertz wireless communications,'' \emph{IEEE J. Sel. Areas Commun.}, vol. 38, no. 9, pp. 2041-2056, Sept. 2020.
  
  
  \bibitem{vhj}
  Visser, Hubregt J., \emph{Array and phased array antenna basics.}, Chichester, UK: Wiley, 2005.
  
  \bibitem{mrcode}
  S. Noh \emph{et al.}, ``Multi-resolution codebook and adaptive beamforming sequence design for millimeter wave beam alignment,'' \emph{IEEE Trans. Wireless Commun.}, vol. 16, no. 9, pp. 5689-5701, Sept. 2017.
  
  \bibitem{ca}
  C. A. Balanis, \emph{Antenna Theory: Analysis and Design.} Hoboken, NJ, USA: Wiley, 2005.
  
  \bibitem{ERD}
  M. Cui \emph{et al.}, ``Near-field wideband beamforming for extremely large antenna array,''  \emph{arXiv preprint arXiv: 2109.10054},  Sep. 2021.
  
  \bibitem{tran1}
  Z. Xiao \emph{et al.}, ``Hierarchical codebook design for beamforming training in millimeter-wave communication,'' \emph{IEEE Trans. Wireless Commun.}, vol. 15, no. 5, pp. 3380-3392, May 2016.
  
  \bibitem{tran2}
  R. Zhang \emph{et al.}, ``Subarray-based simultaneous beam training for multiuser mmWave massive MIMO systems,'' \emph{IEEE Wireless Commun. Lett.}, vol. 8, no. 4, pp. 976-979, Aug. 2019.
  
  \bibitem{tran3}
  H. Yu \emph{et al.}, ``An improved beam training scheme under hierarchical codebook,'' \emph{IEEE Access}, vol. 8, pp. 53627-53635, 2020.
  
  \bibitem{steer1}
  O. E. Ayach \emph{et al.}, ``The capacity optimality of beam steering in large millimeter wave MIMO systems,'' in \emph{2012 IEEE 13th Int. Workshop Signal Process. Advances Wireless Commun. (SPAWC)}, Cesme, 2012, pp. 100-104.
  
  \bibitem{steer2}
  S. Hur \emph{et al.}, ``Millimeter wave beamforming for wireless backhaul and access in small cell networks,'' \emph{IEEE Trans. Commun.}, vol. 61, no. 10, pp. 4391-4403, Oct. 2013.
  
  \bibitem{steer3}
  B. Ning \emph{et al.}, ``Optimal beam steering design for large-scale mmWave MIMO wiretap channel,'' in \emph{2018 IEEE Global Commun. Conf. (GLOBECOM)}, Abu Dhabi, United Arab Emirates, 2018, pp. 1-6.
  
\bibitem{tran4}
  B. Ning \emph{et al.}, ``Terahertz multi-user massive MIMO with intelligent reflecting surface: Beam training and hybrid beamforming,'' \emph{IEEE Trans. Veh. Technol.}, vol. 70, no. 2, pp. 1376-1393, Feb. 2021.
  
\bibitem{3DB2}
  B. Ning \emph{et al.}, ``A unified 3D beam training and tracking procedure for terahertz communication,'' \emph{IEEE Trans. Wireless Commun.}, vol. 21, no. 4, pp. 2445-2461, Apr. 2022.
  
  \bibitem{al1}
  M. Li \emph{et al.}, ``Explore and eliminate: Optimized two-stage search for millimeter-wave beam alignment,'' \emph{IEEE Trans. Wireless Commun.}, vol. 18, no. 9, pp. 4379-4393, Sept. 2019.
  
  \bibitem{al2}
  J. Zhang \emph{et al.}, ``Codebook design for beam alignment in millimeter wave communication systems,'' \emph{IEEE Trans. Commun.}, vol. 65, no. 11, pp. 4980-4995, Nov. 2017.
  
  \bibitem{al3}
  C. Liu \emph{et al.}, ``Millimeter wave beam alignment: Large deviations analysis and design insights,'' \emph{IEEE J. Sel. Areas Commun.}, vol. 35, no. 7, pp. 1619-1631, July 2017.
  
  \bibitem{hbt2}
  T. Nitsche \emph{et al.}, ``IEEE 802.11ad: Directional 60 GHz communication for multi-gigabit-per-second Wi-Fi,'' \emph{IEEE Commun. Mag.}, vol. 52, no. 12, pp. 132-141, Dec. 2014.
  
  \bibitem{hbt3}
  H. Yan and D. Liu, ``Multiple RF chains assisted parallel beam search for mmWave hybrid beamforming systems,'' in \emph{IEEE Globecom Workshops (GC Wkshps)}, Abu Dhabi, United Arab Emirates, 2018, pp. 1-6.
  
  \bibitem{hbt1}
  A. Alkhateeb \emph{et al.}, ``Channel estimation and hybrid precoding for millimeter wave cellular systems,'' \emph{IEEE J. Sel. Topics Signal Process.,} vol. 8, no. 5, pp. 831-846, Oct. 2014.
  
    \bibitem{ps1}
  K. Chen \emph{et al.}, ``Two-step codeword design for millimeter wave massive MIMO systems with quantized phase shifters,'' \emph{IEEE Trans. Signal Process.}, vol. 68, pp. 170-180, 2020.
  
  
  \bibitem{SNB}
  K. Xu \emph{et al.}, ``Fast beam training
  for FDD multi-user massive MIMO systems with finite phase shifter
  resolution,'' \emph{IEEE Trans. Veh. Technol.}, vol. 70, no. 1, pp. 459-473, Jan. 2021.
  
   \bibitem{wbeam1}
  B. Ning and Z. Chen, ``An optimization-based wide-beam design for THz MIMO,''  in \emph{Global Commun. Conf.}, Rio de Janeiro, Brazil, 2022, pp. 1-6.
  
   \bibitem{wbeam2}
B. Ning \emph{et al.}, ``Wide-beam designs for terahertz massive MIMO: SCA-ATP and S-SARV,''  early access in \emph{IEEE Internet of Things J.}, 2023.  doi: 10.1109/JIOT.2023.3241224.
  
  
  \bibitem{deact}
  T. He and Z. Xiao, ``Suboptimal beam search algorithm and codebook
  design for millimeter-wave communications,'' \emph{Mobile Netw. Appl.}, vol. 20, no. 1, pp. 86-97, Jan. 2015.
  
  \bibitem{hbt2a}
  B. Peng \emph{et al.}, ``Fast beam searching concept for indoor Terahertz communications,'' in \emph{8th European Conf. Antennas and Propagation (EuCAP 2014)}, The Hague, Netherlands, 2014, pp. 639-643.
  
  \bibitem{len1}
  J. Brady \emph{et al.}, ``Beamspace MIMO for millimeter-wave communications: System architecture, modeling, analysis, and measurements,'' \emph{IEEE Trans. Antennas Propag.}, vol. 61, no. 7, pp. 3814-3827, July 2013.
  
  \bibitem{len2}
  Y. J. Cho \emph{et al.}, ``RF lens-embedded antenna array for mmWave MIMO: Design and performance,'' \emph{IEEE Commun. Mag.}, vol. 56, no. 7, pp. 42-48, July 2018.
  \bibitem{len3}
  G. H. Song \emph{et al.}, ``Beamspace MIMO transceivers for low-complexity and near-optimal communication at mm-wave frequencies,'' in \emph{IEEE Int. Conf. Acoust. Speech Signal Process.}, 2013, pp. 4394-4398.
  
  \bibitem{len4}
  Y. Zeng and R. Zhang, ``Millimeter wave MIMO with lens antenna array: A new path division multiplexing paradigm,'' \emph{IEEE Trans. Commun.}, vol. 64, no. 4, pp. 1557-1571, Apr. 2016.
  
  \bibitem{l38}
  Z. Popovic and A. Mortazawi, ``Quasi-optical transmit/receive front ends,'' \emph{IEEE Trans. Microw. Theory Tech.}, vol. 46, no. 11, pp. 1964-1975, Nov. 1998.
  
  \bibitem{l39}
  Z. P. S. Hollung and A. Cox, ``A bi-directional quasi-optical lens amplifier,'' \emph{IEEE Trans. Microw. Theory Tech.}, vol. 45, no. 12, pp. 1964-1975, Dec. 1997.
  
  \bibitem{l40}
  B. Bares \emph{et al.}, ``A new accurate design method for millimeter-wave homogeneous dielectric substrate lens antennas of arbitrary shape,'' \emph{IEEE Trans. Antennas Propag.}, vol. 53, no. 3, pp. 1069-1082, Mar. 2005.
  
  \bibitem{l41}
  P.Y. Lau \emph{et al.}, ``Electromagnetic field distribution of lens antennas,'' in \emph{Asia-Pac. Conf. Antennas Propag.}, Aug. 2013.
  
  \bibitem{l42}
  M. A. Al-Joumayly and N. Behdad, ``Wideband planar microwave lenses using sub-wavelength spatial phase shifters,'' \emph{IEEE Trans. Antennas Propag.}, vol. 59, no. 12, pp. 4542-4552, Dec. 2011.
  
  \bibitem{l43}
  M. Li \emph{et al.}, ``Broadband true-time-delay microwave lenses based on miniaturized element frequency selective surfaces,'' \emph{IEEE Trans. Antennas Propag.}, vol. 61, no. 3, pp. 1166-1179, Mar. 2013.
  
  \bibitem{prot}
  J. Brady \emph{et al.}, ``Multi-beam MIMO prototype for real-time multiuser communication at 28 GHz,'' in \emph{2016 IEEE Globecom Workshops (GC Wkshps)}, Washington, DC, 2016, pp. 1-6.
  
  \bibitem{l44}
  T. L. Marzetta, ``Noncooperative cellular wireless with unlimited numbers of base station antennas,'' \emph{IEEE Trans. Wireless Commun.}, vol. 9, no. 11, pp. 3590-3600, Nov. 2010.
  
  \bibitem{l45}
  F. Rusek \emph{et al.}, ``Scaling up MIMO: Opportunities and challenges with very large arrays,'' \emph{IEEE Signal Process. Mag.}, vol. 30, no. 1, pp. 40-60, Jan. 2013.
  
  \bibitem{l46}
  L. Lu \emph{et al.}, ``An overview of massive MIMO: Benefits and challenges,'' \emph{IEEE J. Sel. Topics Signal Process.}, vol. 8, no. 5, pp. 742-758, Oct. 2014.
  
  \bibitem{l47}
  Y. Zeng, R. Zhang, and Z. N. Chen, ``Electromagnetic lens-focusing antenna enabled massive MIMO: performance improvement and cost reduction,'' \emph{IEEE J. Sel. Areas Commun.}, vol. 32, no. 6, pp. 1194-1206, June, 2014.
  
  \bibitem{dft}
  A. M. Sayeed, ``Deconstructing multiantenna fading channels,'' \emph{IEEE Trans. Signal Process.}, vol. 50, no. 10, pp. 2563-2579, Oct. 2002.
  
  \bibitem{sel}
  P. V. Amadori and C. Masouros, ``Low RF-complexity millimeter-wave beamspace-MIMO systems by beam selection,'' \emph{IEEE Trans. Commun.}, vol. 63, no. 6, pp. 2212-2222, June 2015.
  
  \bibitem{I1}
  E. Bjornson \emph{et al.}, ``Massive MIMO has unlimited capacity,'' \emph{IEEE Trans. Wireless Commun.}, vol. 17, no. 1, pp. 574-590, Jan. 2018.
  
  \bibitem{I2}
  D. W. K. Ng \emph{et al.}, ``Energy-efficient resource allocation in OFDMA systems with large numbers of base station antennas,'' \emph{IEEE Trans. Wireless Commun.}, vol. 11, no. 9, pp. 3292-3304, Sept. 2012.
  
  \bibitem{csa}
  C. Liaskos \emph{et al.}, ``A new wireless communication paradigm through software controlled metasurfaces,'' \emph{IEEE Commun. Mag.,} vol. 56, no. 9, pp. 162-169, Sept. 2018.
  
  \bibitem{cometa}
  T. J. Cui \emph{et al.}, ``Coding metamaterials, digital metamaterials and programmable metamaterials,'' \emph{Light Sci. Applicat.}, vol. 3, no. 10, pp. e218, Oct. 2014.
  
  \bibitem{basar}
  E. Basar \emph{et al.}, ``Wireless communications through reconfigurable intelligent surfaces,'' \emph{IEEE Access}, vol. 7, pp. 116753-116773, 2019.
  
  \bibitem{IRSAWN}
  Q. Wu and R. Zhang, ``Towards smart and reconfigurable environment: intelligent reflecting surface aided wireless network,'' \emph{IEEE Commun. Mag.}, vol. 58, no. 1, pp. 106-112, Jan. 2020.
  
  \bibitem{IRSAWC}
  Q. Wu \emph{et al.}, ``Intelligent reflecting surface aided wireless communications: a tutorial,'' \emph{IEEE Trans. Commun.}, vol. 69, no. 5, pp. 3313-3351, May 2021.
  \bibitem{shu}
  S. Hu \emph{et al.}, ``Beyond massive MIMO: The potential of data transmission with large intelligent surfaces,'' \emph{IEEE Trans. Signal Process.}, vol. 66, no. 10, pp. 2746-2758, May 2018.
  
  \bibitem{act}
  A. C. Tasolamprou \emph{et al.}, ``Exploration of intercell wireless millimeter-wave communication in the landscape of intelligent metasurfaces,'' \emph{IEEE Access}, vol. 7, pp. 122931-122948, 2019.
  
  \bibitem{qinte} 
  Q. Wu and R. Zhang, ``Intelligent reflecting surface-enhanced wireless network via joint active and passive beamforming,''  \emph{IEEE Trans. Wireless Commun.}, vol. 18, no. 11, pp. 5394-5409, Nov. 2019.
  
  \bibitem{ghy}
  H. Guo \emph{et al.}, ``Weighted sum-rate maximization for reconfigurable intelligent surface aided wireless networks,'' \emph {IEEE Trans. Wireless Commun.} vol. 19, no. 5, pp. 3064-3076, May 2020.
  
  \bibitem{huangchi}
  C. Huang \emph{et al.},  ``Reconfigurable intelligent surfaces for energy efficiency in wireless communication,'' \emph{IEEE Trans. Wireless Commun.}, vol. 18, no. 8, pp. 4157-4170, Aug. 2019.
  
  \bibitem{multicast}
  G. Zhou \emph{et al.}, ``Intelligent reflecting surface aided multigroup multicast MISO communication systems,'' \emph{IEEE Trans. Signal Process.}, vol. 68, pp. 3236-3251, 2020.
  
  \bibitem{latency}
  T. Bai \emph{et al.}, ``Latency minimization for intelligent reflecting surface aided mobile edge computing,'' \emph{IEEE J. Sel. Areas Commun.}, vol. 38, no. 11, pp. 2666-2682, Nov. 2020.
  \bibitem{nby}
  B. Ning \emph{et al.}, ``Beamforming optimization for intelligent reflecting surface assisted MIMO: A sum-path-gain maximization approach,'' \emph{IEEE Wireless Commun. lett.} vol. 9, no. 7, pp. 1105-1109, July 2020.
  
  \bibitem{szhang}
  S. Zhang and R. Zhang, ``Capacity characterization for intelligent reflecting surface aided MIMO communication,'' \emph{IEEE J. Sel. Areas Commun.}, vol. 38, no. 8, pp. 1823-1838, Aug. 2020,
  
  \bibitem{pwang}
  P. Wang \emph{et al.}, ``Joint transceiver and large intelligent surface design for massive MIMO mmWave systems,'' \emph{IEEE Trans. Wireless Commun.}, vol. 20, no. 2, pp. 1052-1064, Feb. 2021.
  
  \bibitem{ch1}
  P. Wang \emph{et al.}, ``Compressed channel estimation for intelligent reflecting surface-assisted millimeter wave systems,'' \emph{IEEE Signal Process. Lett.}, vol. 27, pp. 905-909, 2020.
  
  \bibitem{ch2}
  Z.-Q. He and X. Yuan, ``Cascaded channel estimation for large intelligent metasurface assisted massive MIMO,'' \emph{IEEE Wireless Commun. Lett.}, vol. 9, no. 2, pp. 210-214, Feb. 2020.
  
  \bibitem{ch3}
  Z. Wang \emph{et al.}, ``Channel estimation for intelligent reflecting surface assisted multiuser communications: Framework, algorithms, and analysis,'' \emph{IEEE Trans. Wireless Commun.}, vol. 19, no. 10, pp. 6607-6620, Oct. 2020.
  
  \bibitem{ch4}
  H. Liu \emph{et al.}, ``Matrix-calibration-based cascaded channel estimation for reconfigurable intelligent surface assisted multiuser MIMO,'' \emph{IEEE J. Sel. Areas Commun.}, vol. 38, no. 11, pp. 2621-2636, Nov. 2020.
   
  \bibitem{ch5}
  Z. Wan \emph{et al.}, ``Broadband channel estimation for intelligent reflecting surface aided mmWave massive MIMO systems,'' in \emph{2020 IEEE Int. Conf. Commun. (ICC)}, Dublin, Ireland, 2020, pp. 1-6.
  
  \bibitem{op1}
  C. Pan \emph{et al.}, ``Multicell MIMO communications relying on intelligent reflecting surfaces,'' \emph{IEEE Trans. Wireless Commun.} vol. 19, no. 8, pp. 5218-5233, Aug. 2020.
  
  \bibitem{op2}
  C. Pan \emph{et al.}, ``Intelligent reflecting surface aided MIMO broadcasting for simultaneous wireless information and power transfer,'' \emph{IEEE J. Sel. Areas Commun.}, vol. 38, no. 8, pp. 1719-1734, Aug. 2020.
  
  \bibitem{op3}
  L. Zhang \emph{et al.}, ``Intelligent reflecting surface aided MIMO cognitive radio systems,'' \emph{IEEE Trans. Veh. Technol.}, vol. 69, no. 10, pp. 11445-11457, Oct. 2020.
  
  \bibitem{op4}
  S. Hong \emph{et al.}, ``Artificial-noise-aided secure MIMO wireless communications via intelligent reflecting surface,'' \emph{IEEE Trans. Commun.}, vol. 68, no. 12, pp. 7851-7866, Dec. 2020.
  
  \bibitem{op5}
  B. Di \emph{et al.}, ``Hybrid beamforming for reconfigurable intelligent surface based multi-user communications: Achievable rates with limited discrete phase shifts,'' \emph{IEEE J. Sel. Areas Commun.}, vol. 38, no. 8, pp. 1809-1822, Aug. 2020.
  
  \bibitem{op6}
  H. Zhang \emph{et al.}, ``Reconfigurable intelligent surfaces assisted communications with limited phase shifts: How many phase shifts are enough?,'' \emph{IEEE Trans. Veh. Technol.}, vol. 69, no. 4, pp. 4498-4502, Apr. 2020.
  
  \bibitem{op7}
  B. Ning, T. Wang, P. Wang, Z. Chen and J. Fang, ``Space-orthogonal scheme for IRSs-aided multi-user MIMO in mmWave/THz communications,'' in \emph{IEEE Int. Conf. Commun. (ICC)}, Seoul, South Korea, pp. 3478-3483, 2022.
  \bibitem{irstran1}
  C. You \emph{et al.}, ``Fast beam training for IRS-assisted multiuser communications,'' \emph{IEEE Wireless Commun. Lett.}, vol. 9, no. 11, pp. 1845-1849, Nov. 2020.
  
  \bibitem{irstran2}
  C. Jia \emph{et al.}, ``Machine learning empowered beam management for intelligent reflecting surface assisted MmWave networks,'' \emph{China Commun.}, vol. 17, no. 10, pp. 100-114, Oct. 2020.
  
  \bibitem{irstran3}
  B. Ning \emph{et al.}, ``Channel estimation and transmission for intelligent reflecting surface assisted THz communications,'' in \emph{IEEE Int. Conf. Commun. (ICC)}, Dublin, Ireland, June 2020, pp. 1-7. 
  
  \bibitem{wall}
  S. Priebe \emph{et al.}, ``Ultra broadband indoor channel measurements and calibrated ray tracing propagation modeling at THz frequencies,'' \emph{IEEE J. Commun. Netw.}, vol. 15, no. 6, pp. 547-558, Dec. 2013.
  
  
  \bibitem{mumimo}
  Q. H. Spencer \emph{et al.}, ``An introduction to the multi-user MIMO downlink,''  \emph{IEEE Commun. Mag.}, vol. 42, no. 10, pp. 60-67, Oct. 2004
  
  \bibitem{mumimo2}
  E. Castañeda \emph{et al.}, ``An overview on resource allocation techniques for multi-user MIMO systems,'' \emph{IEEE Commun. Surv. \& Tuts.}, vol. 19, no. 1, pp. 239-284, 1st Quart., 2017.
  
  \bibitem{sdma}
  J. Mietzner \emph{et al.}, ``Multiple-antenna techniques for wireless communications-A comprehensive literature survey,'' \emph{IEEE Commun. Surv. \& Tuts.}, vol. 11, no. 2, pp. 87-105, 2nd Quart., 2009.
  
  \bibitem{duality}
  S. Vishwanath, N. Jindal, and A. Goldsmith, ``Duality, achievable rates, and sum-rate capacity of Gaussian MIMO broadcast channels,'' \emph{IEEE Trans. Inf. Theory}, vol. 49, pp. 2658-2668, Oct. 2003.
  
%
%
  
  
  \bibitem{polari}
  V. Eiceg, H. Sampath and S. Catreux-Erceg, ``Dual-polarization versus single-polarization MIMO channel measurement results and modeling,'' \emph{IEEE Trans. Wireless Commun.}, vol. 5, no. 1, pp. 28-33, Jan. 2006,
  
  \bibitem{costa}
  H. Weingarten, Y. Steinberg, and S. Shamai, ``The capacity region of the Gaussian multiple-input multiple-output broadcast channel,'' \emph{IEEE Trans. Inf. Theory}, vol. 52, no. 9, pp. 3936-3964, Sept. 2006.
  
  \bibitem{ZF}
  T. Yoo and A. Goldsmith, ``On the optimality of multiantenna broadcast scheduling using zero-forcing beamforming,'' \emph{IEEE J. Sel. Areas Commun.}, vol. 24, no. 3, pp. 528-541, Mar. 2006.
  
  \bibitem{BD}
  Q. Spencer, A. Swindlehurst, and M. Haardt, ``Zero-forcing methods for downlink spatial multiplexing in multiuser MIMO channels,'' \emph{IEEE Trans. Signal Process.}, vol. 52, no. 2, pp. 461-471, Feb. 2004.
  
  \bibitem{NU-SVD}
  Z. Pan, K. K. Wong, and T. S. Ng, ``Generalized multiuser orthogonal space-division multiplexing,'' \emph{IEEE Trans. Wireless Commun.}, vol. 3, no. 6, pp. 1969-1973, Nov. 2004.
  
  \bibitem{EZF}
  L. Sun and M. R. McKay, ``Eigen-based transceivers for the MIMO broadcast channel with semi-orthogonal user selection,''  \emph{IEEE Trans. Signal Process.}, vol. 58, no. 10, pp. 5246-5261, Oct. 2010.
  
  \bibitem{CBD}
  J. H. Chang, L. Tassiulas, and F. Rashid-Farrokhi, ``Joint transmitter receiver diversity for efficient space division multiaccess,'' \emph{IEEE Trans. Wireless Commun.}, vol. 1, pp. 16-27, Jan. 2002.
  
  \bibitem{RBD}
  V. Stankovic and M. Haardt, ``Generalized design of multiuser MIMO precoding matrices,'' \emph{IEEE Trans. Wireless Commun.}, vol. 7, pp. 953-961, Mar. 2008.
  
  \bibitem{QR-RBD}
  H. Wang \emph{et al.}, ``A linear precoding scheme for downlink multiuser MIMO precoding systems,'' \emph{IEEE Commun. Lett.}, vol. 15, no. 6, pp. 653-655, June 2011.
  
  \bibitem{CLR-RBD}
  C. Windpassinger and R. Fischer, ``Low-complexity near-maximum likelihood detection and precoding for MIMO systems using lattice reduction,'' in \emph{Proc. 2003 IEEE Inf. Theory Workshop}, pp. 345-348.
  
  \bibitem{SZF}
  A. D. Dabbagh and D. J. Love, ``Precoding for multiple antenna Gaussian broadcast channels with successive zero-forcing,''  \emph{IEEE Trans. Signal Process.}, vol. 55, no. 7, pp. 3837-3850, July 2007.
  
  \bibitem{PMSE}
  A. Tenenbaum and R. Adve, ``Linear processing and sum throughput in the multiuser MIMO downlink''\emph{IEEE Trans. Wireless Commun.}, vol. 8, no. 5, pp. 2652-2661, May 2009.
  
  \bibitem{T-MMSE}
  J. Zhang \emph{et al.}, ``Joint linear transmitter and receiver design for the downlink of multiuser MIMO systems,'' \emph{IEEE Commun. Lett.}, vol. 9, pp. 991-993, Nov. 2005.
  
  \bibitem{WMMSE}
  Q. Shi \emph{et al.}, ``An iteratively weighted MMSE approach to distributed sum-utility maximization for a MIMO interfering broadcast channel,'' \emph{IEEE Trans. Signal Process.}, vol. 59, pp. 4331-4340, Sept. 2011.
  
    \bibitem{wb1}
  J. Brady and A. Sayeed, ``Wideband communication with high-dimensional arrays: New results and transceiver architectures,'' in \emph{2015 IEEE Int. Conf. Commun. Workshop (ICCW)}, London, 2015, pp. 1042-1047.
  
  \bibitem{wb2}
  H. Hashemi \emph{et al.}, ``Integrated true-time-delay-based ultra-wideband array processing,'' \emph{IEEE Commun. Mag.}, vol. 46, no. 9, pp. 162-172, Sept. 2008.
  
  \bibitem{wb3}
  B. Wang \emph{et al.}, ``Spatial- and frequency-wideband effects in millimeter-wave massive MIMO systems,'' \emph{IEEE Trans. Signal Process.}, vol. 66, no. 13, pp. 3393-3406, July 2018.
  
  \bibitem{bs1}
  M. Cai \emph{et al.}, ``Effect of wideband beam squint on codebook design in phased-array wireless systems,'' in \emph{2016 IEEE GLOBECOM}, Washington, DC, USA, 2016, pp. 1-6.
  
  \bibitem{bs2}
  B. Wang \emph{et al.}, ``Spatial-wideband effect in massive MIMO with application in mmWave systems,'' \emph{IEEE Commun. Mag.}, vol. 56, no. 12, pp. 134-141, Dec. 2018.
  
  \bibitem{bs3}
  L. Dai, J. Tan, Z. Chen, and H. V. Poor, ``Delay-Phase Precoding for Wideband THz Massive MIMO,'' \emph{IEEE Trans. Wireless Commun.} , vol. 21, no. 9, pp. 7271-7286, Sept. 2022.
  
  \bibitem{ttd1}
  J. Roderick \emph{et al.}, ``An UWB beamformer with 4ps true time delay resolution,'' in \emph{IEEE 2005 Custom Integrated Circuit Design Conf.}, San Jose, CA, 2005, pp. 805-808.
  
  \bibitem{ttd2}
  T. Chu \emph{et al.}, ``A 4-channel UWB beam-former in 0.13$\mu$m CMOS using a path-sharing true-time-delay architecture,'' in \emph{2007 IEEE Int. Solid-State Circuits Conf.}, San Francisco, CA, 2007, pp. 426-613.
  
  \bibitem{ttd3}
  T. Chu and H. Hashemi, ``A CMOS UWB camera with 7×7 simultaneous active pixels,'' in \emph{2008 IEEE Int. Solid-State Circuits Conf.}, San Francisco, CA, 2008, pp. 120-600.
  
  \bibitem{ttd4}
  F. Miyamaru \emph{et al.}, `` Ultrafast optical control of group delay of narrow-band terahertz waves,'' \emph{Sci. Rep.}, vol. 4, pp. 4346, 2014.
  
  \bibitem{ttd5}
  Wei Jia \emph{et al.}, ``Active control and large group delay in graphene-based terahertz metamaterials,'' \emph{The J. Phys. Chem. C}, vol. 123, no. 23, pp. 18560-18564. 
  
  \bibitem{cmos13}
  N. Buadana, S. Jameson and E. Socher, ``A 280-GHz digitally controlled four port chip-scale dielectric resonator antenna transmitter with DiCAD true time delay,'' \emph{IEEE Solid-State Circuits Lett.}, vol. 3, pp. 454-457, 2020.
  
  \bibitem{pe1}
  P. Lu \emph{et al.}, ``Photonic assisted beam steering for millimeter-wave and THz antennas,'' in \emph{2018 IEEE Conf. Antenna Measurements \& Applicat. (CAMA)}, Vasteras, 2018, pp. 1-4.
  
  \bibitem{wbadd}
  B. Zhai \emph{et al.}, ``THzPrism: frequency-based beam spreading for terahertz communication systems,'' \emph{IEEE Wireless Commun. Lett.}, vol. 9, no. 6, pp. 897-900, June 2020.
  
  \bibitem{fttd}
  L. Yan, C. Han, and J. Yuan, ``Energy-efficient dynamic-subarray with fixed true-time-delay design for terahertz wideband hybrid beamforming'', \emph{IEEE J. Sel. Areas Commun.}, vol. 40, no. 10, pp. 2840-2854, Oct. 2022.
  \bibitem{manu1}
  B. Aqlan \emph{et al.}, ``Sub-THz circularly polarized horn antenna using wire electrical discharge machining for 6G wireless communications,'' \emph{IEEE Access}, vol. 8, pp. 117245-117252, 2020.
  
  \bibitem{manu2} 
  N. Chahat \emph{et al.}, ``1.9-THz multiflare angle horn optimization for space instruments,'' \emph{IEEE Trans. Terahertz Sci. Technol.}, vol. 5, no. 6, pp. 914-921, Nov. 2015.
  
  \bibitem{manu3}
  K. Fan \emph{et al.}, ``Development of a high gain 325-500 GHz antenna using quasi-planar reflectors,'' \emph{IEEE Trans. Antennas Propag.}, vol. 65, no. 7, pp. 3384-3391, July 2017.
  
  \bibitem{manu4}
  H. Wang \emph{et al.}, ``Terahertz high-gain offset reflector antennas ssing SiC and CFRP material,'' \emph{IEEE Trans. Antennas Propag.}, vol. 65, no. 9, pp. 4443-4451, Sept. 2017.
  \bibitem{manu5} 
  X. Deng \emph{et al.}, ``340-GHz SIW cavity-backed magnetic rectangular slot loop antennas and arrays in silicon technology,'' \emph{IEEE Trans. Antennas Propag.}, vol. 63, no. 12, pp. 5272-5279, Dec. 2015.
  
  \bibitem{manu6}
  R. Mendis \emph{et al.}, ``Spectral characterization of broadband THz antennas by photoconductive mixing: toward optimal antenna design,'' \emph{IEEE Antennas Wireless Propag. Lett.}, vol. 4, pp. 85-88, 2005.
  
  \bibitem{manu7}
  N. Zhu and R. W. Ziolkowski, ``Photoconductive THz antenna designs with high radiation efficiency, high directivity, and high aperture efficiency,'' \emph{IEEE Trans. Terahertz Sci. Technol.}, vol. 3, no. 6, pp. 721-730, Nov. 2013.
  \bibitem{manu8} 
  M. Alonso-delPino \emph{et al.}, ``Development of silicon micromachined microlens antennas at 1.9 THz,'' \emph{IEEE Trans. Terahertz Sci. Technol.}, vol. 7, no. 2, pp. 191-198, Mar. 2017.
  
  \bibitem{manu9}
  N. Llombart \emph{et al.}, ``Silicon micromachined lens antenna for THz integrated heterodyne arrays,'' \emph{IEEE Trans. Terahertz Sci. Technol.}, vol. 3, no. 5, pp. 515-523, Sept. 2013.
  
  \bibitem{manu9-1}
  N. Llombart \emph{et al.}, ``Novel terahertz antenna based on a silicon lens fed by a leaky wave enhanced waveguide,'' \emph{IEEE Trans. Antennas Propag.}, vol. 59, no. 6, pp. 2160-2168, June 2011.
  
  \bibitem{table4} 
  Mohammed Reza M. Hashemi \emph{et al.}, ``Electronically-controlled beam-steering through vanadium dioxide metasurfaces,'' \emph{Sci. Rep.}, vol. 6, pp. 35439, 2016.
  
  \bibitem{gra1}
  J. M. Jornet and I. F. Akyildiz, ``Graphene-based plasmonic nano-antenna for terahertz band communication in nanonetworks,'' \emph{IEEE J. Sel. Areas Commun.}, vol. 31, no. 12, pp. 685-694, Dec. 2013.
  
  \bibitem{lc1} %
  B. Scherger \emph{et al.}, ``Discrete terahertz beam steering with an electrically controlled liquid crystal device,'' \emph{J. Infrared Milli. Terahz. Waves}, vol. 33, pp. 1117-1122, 2012.
  
  \bibitem{gra2}
  Correas-Serrano \emph{et al.} (2017), ``Graphene-based antennas for terahertz systems: A review,'' [Online]. Available: https://arxiv.org/abs/1704.00371v1
  
  \bibitem{manu10}
  C. Lee \emph{et al.}, ``Terahertz antenna arrays with silicon micromachined-based microlens antenna and corrugated horns,'' in \emph{Proc. Int. Workshop Antenna Technol. (iWAT)}, Seoul, 2015, pp. 70-73.
  
  \bibitem{manu11}
  E. García-Muñoz \emph{et al.}, ``Photonic-based integrated sources and antenna arrays for broadband wireless links in terahertz communications,'' \emph{Semicond. Sci. Technol.}, vol. 34, no. 5, May 2019.
  
  \bibitem{manu12}
  K. M. Luk \emph{et al.}, ``A microfabricated low-profile wideband antenna array for terahertz communications,'' \emph{Sci. Rep.}, vol. 7, no. 1, pp. 1268, Dec. 2017.
  
  \bibitem{meta3}
  K. Sengupta, T. Nagatsuma, and D. M. Mittleman, ``Terahertz integrated electronic and hybrid electronic photonic systems,'' \emph{Nature Electron.}, vol. 1, no. 12, pp. 622-635, Dec. 2018.
  
  \bibitem{cmos2}
  Z. Wang \emph{et al.}, ``A CMOS 210-GHz fundamental transceiver with OOK modulation,'' \emph{IEEE J. Solid-State Circuits}, vol. 49, no. 3, pp. 564-580, Mar. 2014.
  
  \bibitem{cmos4}
  P. Reynaert \emph{et al.}, ``THz arrays in CMOS,'' in \emph{2020 3rd Int. Workshop on Mobile Terahertz Syst. (IWMTS)}, Essen, Germany, 2020, pp. 1-5.
  
  \bibitem{cmos5}
  K. Sengupta and A. Hajimiri, ``A 0.28 THz power-generation and beam-steering array in CMOS based on distributed active radiators,'' \emph{IEEE J. Solid-State Circuits}, vol. 47, no. 12, pp. 3013-3031, Dec. 2012.
  
  \bibitem{cmos6}
  A. Tang \emph{et al.}, ``A 65nm CMOS 140 GHz 27.3 dBm EIRP transmit array with membrane antenna for highly scalable multi-chip phase arrays,'' in \emph{2014 IEEE MTT-S Int. Microw. Symp. (IMS2014)}, Tampa, FL, 2014, pp. 1-3.
  
  \bibitem{cmos7}
  K. Guo \emph{et al.}, ``A 0.53-THz subharmonic injection-locked phased array with 63- $\mu W $ radiated power in 40-nm CMOS,'' \emph{IEEE J. Solid-State Circuits}, vol. 54, no. 2, pp. 380-391, Feb. 2019.
  
  \bibitem{cmos8}
  Y. Yang \emph{et al.}, ``An eight-element 370-410-GHz phased-array transmitter in 45-nm CMOS SOI with peak EIRP of 8-8.5 dBm,'' \emph{IEEE Trans. Microw. Theory Techn.}, vol. 64, no. 12, pp. 4241-4249, Dec. 2016.
  
  \bibitem{cmos9}
  H. Jalili and O. Momeni, ``A 0.34-THz wideband wide-angle 2-D steering phased array in 0.13- $\mu$ m SiGe BiCMOS,'' \emph{IEEE J. Solid-State Circuits}, vol. 54, no. 9, pp. 2449-2461, Sept. 2019.
  
  \bibitem{cmos10}
  X. Deng \emph{et al.}, ``A 320-GHz 1$\times$4 fully integrated phased array transmitter using 0.13-$\mu$m SiGe BiCMOS technology,'' \emph{IEEE Trans. Terahertz Sci. Technol.}, vol. 5, no. 6, pp. 930-940, Nov. 2015.
  
  \bibitem{cmos11}
  Y. Tousi and E. Afshari, ``A high-power and scalable 2-D phased array for terahertz CMOS integrated systems,'' \emph{IEEE J. Solid-State Circuits}, vol. 50, no. 2, pp. 597-609, Feb. 2015.
  
  \bibitem{cmos12}
  R. Han \emph{et al.}, ``A SiGe terahertz heterodyne imaging transmitter with 3.3 mW radiated power and fully-integrated phase-locked loop,'' \emph{IEEE J. Solid-State Circuits}, vol. 50, no. 12, pp. 2935-2947, Dec. 2015.
  
  \bibitem{pe2}
  M. Che \emph{et al.}, ``Optoelectronic THz-wave beam steering by arrayed photomixers with integrated antennas,'' \emph{IEEE Photon. Technol. Lett.}, vol. 32, no. 16, pp. 979-982, Aug. 2020.
  
  \bibitem{gra3}
  E. Carrasco and J. Perruisseau-Carrier, ``Reflectarray antenna at terahertz using graphene,'' \emph{IEEE Antennas Wireless Propag. Lett.}, vol. 12, pp. 253-256, 2013.
  
  \bibitem{gra30}
  Y. Luo \emph{et al.}, ``Graphene-based multi-beam reconfigurable THz antennas,'' \emph{IEEE Access}, vol. 7, pp. 30802-30808, 2019.
  
  \bibitem{lc2}
  B. Vasić \emph{et al.}, ``Tunable beam steering at terahertz frequencies using reconfigurable metasurfaces coupled with liquid crystals,'' \emph{IEEE J. Sel. Areas Quant. Electron.}, vol. 26, no. 5, pp. 1-9, Sept.-Oct. 2020.
  
  \bibitem{gra4} 
  A. Singh \emph{et al.}, ``Design and operation of a graphene-based plasmonic nano-antenna array for communication in the terahertz band,''  \emph{IEEE J. Sel. Areas Commun.}, vol. 38, no. 9, pp. 2104-2117, Sept. 2020.
  
  \bibitem{gra4-1}
  M. Esquius-Morote \emph{et al.}, ``Sinusoidally modulated graphene leaky-wave antenna for electronic beamscanning at THz,'' \emph{IEEE Trans. Terahertz Sci. Technol.}, vol. 4, no. 1, pp. 116-122, Jan. 2014.
  
  \bibitem{gra5} 
  Y. Dong \emph{et al.}, ``Dual-band reconfigurable terahertz patch antenna with graphene-stack-based backing cavity,'' \emph{IEEE Antennas Wireless Propag. Lett.}, vol. 15, pp. 1541-1544, 2016.
  
  \bibitem{gra6} 
  W. Fuscaldo \emph{et al.}, ``A reconfigurable substrate-superstrate graphene-based leaky-wave THz antenna,'' \emph{IEEE Antennas Wireless Propag. Lett.}, vol. 15, pp. 1545-1548, 2016.
  
  \bibitem{gra7} 
  Z. Chang \emph{et al.}, ``A reconfigurable graphene reflectarray for generation of vortex THz waves,'' \emph{IEEE Antennas Wireless Propag. Lett.}, vol. 15, pp. 1537-1540, 2016.
  
  \bibitem{gra8} 
  M. Dragoman \emph{et al.}, ``Terahertz antenna based on graphene,'' \emph{J. Appl. Phys.}, vol. 107, no. 10, pp. 104313, 2010.
  
  \bibitem{gra9} 
  Esfandiyari \emph{et al.}, ``Channel capacity enhancement by adjustable graphene-based MIMO antenna in THz band,'' \emph{Opt. Quant. Electron.}, vol. 51, no. 5, pp. 137, 2019. 
  
  \bibitem{gra10} 
  K. Sarabandi \emph{et al.}, ``A novel frequency beam-steering antenna array for submillimeter-wave applications,'' \emph{IEEE Trans. Terahertz Sci. Technol.}, vol. 8, no. 6, pp. 654-665, Nov. 2018.
  
  \bibitem{gramimo1} 
  Z. Xu \emph{et al.}, ``Design of a reconfigurable MIMO system for THz communications based on graphene antennas,'' \emph{IEEE Trans. Terahertz Sci. Technol.}, vol. 4, no. 5, pp. 609-617, Sept. 2014.
  

  
  
  
  \bibitem{reflc1} 
  Wu, Jingbo \emph{et al.}. ``Liquid crystal programmable metasurface for terahertz beam steering,'' \emph{Appl. Phys. Lett.}, vol. 116, no. 13, pp. 131104, 2020.
  
  \bibitem{reflc2} 
  X. Fu and T. J. Cui, ``Recent progress on metamaterials: From effective medium model to real-time information processing system,'' \emph{Progr. Quantum Electron.}, vol. 67, Sept. 2019.
  
  \bibitem{refgr3} 
  S. E. Hosseininejad \emph{et al.}, ``Digital metasurface based on graphene: An application to beam steering in terahertz plasmonic antennas,'' \emph{IEEE Trans. Nanotechnology}, vol. 18, pp. 734-746, 2019.
  
  \bibitem{refgr2} 
  Tamagnone, Michele \emph{et al.} (2018). ``Graphene reflectarray metasurface for terahertz beam steering and phase modulation,'' [Online]. Available: https://arxiv.org/abs/1806.02202
  
  \bibitem{gra11} 
  F. Zangeneh-Nejad and R. Safian, ``A tunable high-impedance THz antenna array,'' in \emph{2015 23rd Iranian Conf. on Elect. Eng.}, Tehran, 2015, pp. 402-405.
  
  \bibitem{table1} 
  R. Camblor \emph{et al.}, ``Full 2-D submillimeter-wave frequency scanning array,'' \emph{IEEE Trans. Antennas Propag.}, vol. 65, no. 9, pp. 4486-4494, Sept. 2017.
  
  \bibitem{table2} 
  G. Perez-Palomino \emph{et al.}, ``Design and demonstration of an electronically scanned reflectarray antenna at 100 GHz using multiresonant cells based on liquid crystals,'' \emph{IEEE Trans. Antennas Propag.}, vol. 63, no. 8, pp. 3722-3727, Aug. 2015.
  
  \bibitem{table3} 
  G. Perez-Palomino \emph{et al.}, ``Preliminary design of a liquid crystal-based reflectarray antenna for beam-scanning in THz,'' in \emph{2013 IEEE Antennas Propagation Soc. Int. Symp. (APSURSI)}, Orlando, FL, USA, 2013, pp. 2277-2278.
  
  \bibitem{table5} 
  Jun Yang \emph{et al.}, ``A novel electronically controlled two-dimensional terahertz beam-scanning reflectarray antenna based on liquid crystals,'' \emph{Front. Phys.}, vol. 8, Oct. 2020.
  
  
  \bibitem{sc1}
  V. W. S. Chan, ``Optical satellite networks,'' \emph{J. Lightw. Technol.}, vol. 21, no. 11, pp. 2811-2827, Nov. 2003.
  
  \bibitem{sc2}
  You X \emph{et al.}, ``Towards 6G wireless communication networks: Vision, enabling technologies, and new paradigm shifts,'' \emph{Sci. China Inf. Sci.}, vol. 64, issue 1, pp. 110301-, 2021.
  
  \bibitem{vc2}
  F. Jameel \emph{et al.}, ``Propagation channels for mmWave vehicular communications: State-of-the-art and future research directions,'' \emph{IEEE Wireless Commun.}, vol. 26, no. 1, pp. 144-150, Feb. 2019.
  
  \bibitem{vc5}
  S. A. Busari \emph{et al.}, ``Generalized hybrid beamforming for vehicular connectivity using THz massive MIMO,'' \emph{IEEE Trans. Veh. Technol.}, vol. 68, no. 9, pp. 8372-8383, Sept. 2019.
  
  \bibitem{vc6}
  K. Guan \emph{et al.}, ``On millimeter wave and THz mobile radio channel for smart rail mobility,'' \emph{IEEE Trans. Veh. Technol.}, vol. 66, no. 7, pp. 5658-5674, July 2017.
  
  \bibitem{vc7}
  H. Yi \emph{et al.}, ``Characterization for the vehicle-to-infrastructure channel in urban and highway scenarios at the terahertz band,'' \emph{IEEE Access}, vol. 7, pp. 166984-166996, 2019.
  
  \bibitem{kiosk}
  H. Song \emph{et al.}, ``Prototype of KIOSK data downloading system at 300 GHz: Design, technical feasibility, and results,'' \emph{IEEE Commun. Mag.}, vol. 56, no. 6, pp. 130-136, June 2018.
  
  \bibitem{st1}
  V. Petrov \emph{et al.}, ``Terahertz band communications: Applications, research challenges, and standardization activities,'' in \emph{2016 8th Int. Congr. Ultra Modern Telecommun. and Control Syst. and Workshops (ICUMT)}, Lisbon, 2016, pp. 183-190.
  
  \bibitem{dc1}
  Wu, K. \emph{et al.}, ``Rethinking the architecture design of data center networks,'' \emph{Front. Comput. Sci.}, vol. 6, pp. 596-603, 2012.
  
  \bibitem{dc2}
  Y. Cui \emph{et al.}, ``Wireless data center networking,'' \emph{IEEE Wireless Commun.}, vol. 18, no. 6, pp. 46-53, Dec. 2011.
  
  \bibitem{dc3}
  D. Halperin \emph{et al.}, ``Augmenting data center networks with multi-gigabit wireless links,'' \emph{SIGCOMM Comput. Commun. Rev.}, vol. 41, no. 4, pp. 38-49, Aug. 2011.
  
  \bibitem{dc7}
  L. Bariah \emph{et al.}, ``A prospective look: Key enabling technologies, applications and open research topics in 6G networks,'' \emph{IEEE Access}, vol. 8, pp. 174792-174820, 2020.
  
  \bibitem{dc4}
  B. Peng and T. Kürner, ``A stochastic channel model for future wireless THz data centers,'' in \emph{2015 Int. Symp. Wireless Commun. Syst. (ISWCS)}, Brussels, 2015, pp. 741-745.
  
  \bibitem{dc5}
  J. M. Eckhardt \emph{et al.}, ``Measurements in a real data centre at 300 GHz and recent results,'' in \emph{2019 13th European Conf. Antennas and Propag. (EuCAP)}, Krakow, Poland, 2019, pp. 1-5.
  
  \bibitem{dc6}
  Chia-Lin Cheng \emph{et al.}, ``THz cluster-based modeling and propagation characterization in a data center environment,'' \emph{IEEE Access}, vol. 8, pp. 56544-56558, 2020.
  
  \bibitem{st2}
  \emph{IEEE standard for high data rate wireless multi-media networks-amendment 2: 100 Gb/s wireless switched point-to-point physical,} standard 802.15.3d-2017, July 2017.
  
  \bibitem{nature}
  J. Ma \emph{et al.}, ``Security and eavesdropping in terahertz wireless links,'' \emph{Nature}, vol. 563, no. 7729, pp. 89-93, 2018.
  
  \bibitem{st3}
  W. Gao \emph{et al.}, ``Distance-adaptive absorption peak modulation (DA-APM) for terahertz covert communications,'' \emph{IEEE Trans. Wireless Commun.}, vol. 20, no. 3, pp. 2064-2077, Mar. 2021.
  
  \bibitem{st4}
  B. Ning \emph{et al.}, ``Improving security of THz communication with intelligent reflecting surface,'' \emph{IEEE Globecom Workshops(GC Wkshps)}, Waikoloa, HI, USA, pp. 1-6, 2019.
  
  \bibitem{st5}
  J. Qiao and M. S. Alouini, ``Secure transmission for intelligent reflecting surface-assisted mmWave and terahertz systems,'' \emph{IEEE Wireless Commun. Lett.}, vol. 9, no. 10, pp. 1743-1747, Oct. 2020.
  
  \bibitem{st6}
  B. Ning, Z. Chen, Z. Tian and S. Li, ``Optimization for IRS-assisted systems with both multicast and confidential messages,''  \emph{IEEE Global Communications Conference (GLOBECOM)}, Madrid, Spain, pp. 1-6, 2021.
  
  \bibitem{ncc2}
  S. Abadal \emph{et al.}, ``Graphene-enabled wireless communication for massive multicore architectures,'' \emph{IEEE Commun. Mag.}, vol. 51, no. 11, pp. 137-143, Nov. 2013.
  
  
  \bibitem{hspw}
  Y. Chen, L. Yan and C. Han, ``Hybrid Spherical- and Planar-Wave Modeling and DCNN-Powered Estimation of Terahertz Ultra-Massive MIMO Channels,'' \emph{IEEE Trans. Commun.}, vol. 69, no. 10, pp. 7063-7076, Oct. 2021.
  
  \bibitem{umcm9}
  B. Peng \emph{et al.}, ``Channel modeling and system concepts for future terahertz communications: Getting ready for advances beyond 5G,'' \emph{IEEE Veh. Technol. Mag.}, vol. 15, no. 2, pp. 136-143, June 2020.
  
  \bibitem{umcm10}
  Y. Yang \emph{et al.}, ``Generative-adversarial-network-based wireless channel modeling: Challenges and opportunities,'' \emph{IEEE Commun. Mag.}, vol. 57, no. 3, pp. 22-27, Mar. 2019.
  
  
  \bibitem{adc1}
  J. Zhang \emph{et al.}, ``On low-resolution ADCs in practical 5G millimeter-wave massive MIMO systems,'' \emph{IEEE Commun. Mag.}, vol. 56, no. 7, pp. 205-211, July 2018.
  
  \bibitem{adc2}
  J. Mo, P. Schniter, and R. W. Heath, ``Channel estimation in broadband millimeter wave MIMO systems with few-bit ADCs,'' \emph{IEEE Trans. Signal Process.}, vol. 66, no. 5, pp. 1141-1154, Mar. 2018.
  
  \bibitem{adc3}
  J. Zhang \emph{et al.}, ``On the spectral efficiency of massive MIMO systems with low-resolution ADCs,'' \emph{IEEE Commun. Lett.}, vol. 20, no. 5, pp. 842-845, May 2016.
  
  
  \bibitem{ps2}
  J. Chen, ``Hybrid beamforming with discrete phase shifters for millimeterwave massive MIMO systems,'' \emph{IEEE Trans. Veh. Technol.}, vol. 66, no. 8, pp. 7604-7608, Aug. 2017.
  
  \bibitem{ps3}
  Y. Lin, ``On the quantization of phase shifters for hybrid precoding systems,'' \emph{IEEE Trans. Signal Process.}, vol. 65, no. 9, pp. 2237-2246, May 2017.
  
    \bibitem{bs6}
  M. Cho, I. Song, and J. D. Cressler, ``A true time delay-based SiGe bidirectional T/R chipset for large-scale wideband timed array antennas,'' in \emph{IEEE Radio Freq. Integr. Circuits Symp. (RFIC)}, Philadelphia, PA, USA, June 2018, pp. 272-275.
  
  
  \bibitem{bs4}
  M. Cai, J. N. Laneman, and B. Hochwald, ``Beamforming codebook compensation for beam squint with channel capacity constraint,'' in \emph{IEEE Int. Symp. Inf. Theory}, June 2017, pp. 76-80.
  
  
  \bibitem{bsquint}
  B. Ning, L. Li, W. Chen, and Z. Chen, ``Wideband Terahertz Communications with AoSA: Beam Split Aggregation and Multiplexing,''  in \emph{Global Commun. Conf.}, Rio de Janeiro, Brazil, 2022, pp. 1-6.
  
  \bibitem{3DB1}
  L. Liu \emph{et al.}, ``Multi-beam UAV communication in cellular uplink: Cooperative interference cancellation and sum-rate maximization,'' \emph{IEEE Trans. Wireless Commun.}, vol. 18, no. 10, pp. 4679-4691, Oct. 2019.
  
  \bibitem{cov1}
  C. Jansen \emph{et al.}, ``The impact of reflections from stratified building materials on the wave propagation in future indoor terahertz communication systems,'' \emph{IEEE Trans. Antennas Propag.}, vol. 56, no. 5, pp. 1413-1419, May 2008.
  
  \bibitem{cov2}
  C. Jansen \emph{et al.}, ``Diffuse scattering from rough surfaces in THz communication channels,'' \emph{IEEE Trans. THz Sci. Technol.}, vol. 1, no. 2, pp. 462-472, Nov. 2011.
  
  
  \bibitem{cov31}
  L. You \emph{et al.}, ``Network massive MIMO transmission over millimeter-wave and terahertz bands: Mobility enhancement and blockage mitigation,'' \emph{IEEE J. Sel. Areas Commun.}, vol. 38, no. 12, pp. 2946-2960, Dec. 2020.
  
  \bibitem{cov4}
  B. Ning, P. Wang, L. Li, Z. Chen and J. Fang, ``Multi-IRS-Aided Multi-User MIMO in mmWave/THz Communications: A Space-Orthogonal Scheme,'' \emph{IEEE Trans.  Commun.}, vol. 70, no. 12, pp. 8138-8152, Dec. 2022
  
  \bibitem{cov5}
  S. Zeng \emph{et al.}, ``Reconfigurable intelligent surface (RIS) assisted wireless coverage extension: RIS orientation and location optimization,'' \emph{IEEE Commun. Lett.},  vol. 25, no. 1, pp. 269-273, Jan. 2021.
  
  \bibitem{cov6}
  W. Mei and R. Zhang, ``Cooperative beam routing for multi-IRS aided communication,'' \emph{IEEE Wireless Commun. Lett.}, vol. 10, no. 2, pp. 426-430, Feb. 2021.
  
  \bibitem{cov3}
  L. You \emph{et al.}, ``BDMA for millimeter-wave/terahertz massive MIMO transmission with per-beam synchronization,'' \emph{IEEE J. Sel. Areas Commun.}, vol. 35, no. 7, pp. 1550-1563, July 2017.


\end{thebibliography}
\end{document}